\documentclass[fleqn,usenatbib]{mnras}

\usepackage{mathptmx}
\usepackage[T1]{fontenc}

\DeclareRobustCommand{\VAN}[3]{#2}
\let\VANthebibliography\thebibliography
\def\thebibliography{\DeclareRobustCommand{\VAN}[3]{##3}\VANthebibliography}

\usepackage{graphicx}	% Including figure files
\usepackage{amsmath}	% Advanced maths commands
\usepackage{amssymb}	% Extra maths symbols
\usepackage{tabularx}  % Remove if this breaks table 2
\usepackage{subfigure}  % added by author 
\usepackage{caption}    % added by author

\title[Investigating the structure of SFRs using INDICATE]{Investigating the structure of star-forming regions using INDICATE}

\author[G. A. Blaylock-Squibbs, R. J. Parker, Anne S.M. Buckner and Manuel Güdel]{
George A. Blaylock-Squibbs$^{1}$\thanks{E-mail: gablaylock-squibbs1@sheffield.ac.uk}, Richard J. Parker$^{1}$, Anne S.M. Buckner$^{2}$ and Manuel Güdel$^{3}$
\\
$^{1}$Department of Physics and Astronomy,
The University of Sheffield, Hounsfield Road,
Sheffield,
S3 7RH\\
$^{2}$Department of Physics and Astronomy, University of Exeter, North Park Road, Exeter, EX4 4QL\\
$^{3}$Department of Astrophysics, University of Vienna, Türkenschanzstr. 17, 1180 Vienna, Austria
}

\date{Accepted XXX. Received YYY; in original form ZZZ}

\pubyear{2021}

\begin{document}
\label{firstpage}
\pagerange{\pageref{firstpage}--\pageref{lastpage}}
\maketitle

\begin{abstract}
The ability to make meaningful comparisons between theoretical and observational data of star-forming regions is key to understanding the star formation process. In this paper we test the performance of INDICATE, a new method to quantify the clustering tendencies of individual stars in a region, on synthetic star-forming regions with sub-structured, and smooth, centrally concentrated distributions. INDICATE quantifies the amount of stellar affiliation of each individual star, and also determines whether this affiliation is above random expectation for the star-forming region in question. We show that INDICATE cannot be used to quantify the overall structure of a region due to a degeneracy when applied to regions with different geometries. We test the ability of INDICATE to detect differences in the local stellar surface density and its ability to detect and quantify mass segregation. We then compare it to other methods such as the mass segregation ratio $\Lambda_{\rm{MSR}}$, the local stellar surface density ratio $\Sigma_{\rm{LDR}}$ and the cumulative distribution of stellar positions. INDICATE detects significant differences in the clustering tendencies of the most massive stars when they are at the centre of a smooth, centrally concentrated distribution, corresponding to areas of greater stellar surface density. When applied to a subset of the 50 most massive stars we show INDICATE can detect signals of mass segregation. We apply INDICATE to the following nearby star-forming regions: Taurus, ONC, NGC 1333, IC 348 and $\rho$~Ophiuchi and find a diverse range of clustering tendencies in these regions.

\end{abstract}


%%%%%%%%%%%%%%%%%%%%%%%%%%%%%%%%%%%%%%%%%%%%%%%%%%

%%%%%%%%%%%%%%%%% BODY OF PAPER %%%%%%%%%%%%%%%%%%
%This is a simple template for authors to write new MNRAS papers.
%See \texttt{mnras\_sample.tex} for a more complex example, and \texttt{mnras\_guide.tex}
%for a full user guide.

%All papers should start with an Introduction section, which sets the work
%in context, cites relevant earlier studies in the field by \citet{bate_interpreting_1998},
%and describes the problem the authors aim to solve \citep[e.g.][]{bate_interpreting_1998}.
%Multiple citations can be joined in a simple way like \citet{buckner_spatial_2019, buckner_spatial_2020}.
\section{Introduction}
Star formation is observed to occur in giant molecular clouds where the stars form in embedded groups \citep{lada_embedded_2003}. These embedded groups are often part of star-forming regions that have a range of different morphologies (i.e. smooth centrally concentrated spherical distributions or more complex substructured distributions) and densities \citep[][]{bressert_spatial_2010,kruijssen_fraction_2012}. Quantifying the amount of spatial (and kinematic) substructure is key to determining whether star formation is universal (i.e. the same everywhere) or whether it is dependant on local environmental factors. \par
Young star-forming regions are often observed to be substructured and subvirial, but this substructure can be erased over a very short time period (the order of a few crossing times within substructured regions) due to dynamical interactions. These interactions lead to dynamical mass segregation \citep[e.g.][]{mcmillan_dynamical_2007, allison_using_2009, allison_early_2010, moeckel_limits_2009, parker_dynamical_2014, dominguez_how_2017} where the most massive stars migrate to the centre of the region over the order of the crossing time scale \citep[][]{bonnell_mass_1998}. Consequently, the observed locations and spatial arrangement of massive stars are not necessarily identical to those when they formed.\par

To test theories of star formation the overall structure of such regions and the distribution of the massive stars inside them must be detected and quantified. Methods that can accurately detect and quantify structure and mass segregation are therefore needed. The mean surface density of companions (two-point or auto correlation function) has been used in \citet{gomez_spatial_1993}, \citet{larson_star_1995} and \citet{gouliermis_complex_2014} to quantify the distributions of stars. This method looks at the excess number of pairs as a function of the separation compared to a random distribution of stars \citep[][]{gomez_spatial_1993,simon_clustering_1997, bate_interpreting_1998,kraus_spatial_2008}. \par 
\citet{cartwright_statistical_2004} introduced the $Q$-Parameter which uses minimum spanning trees (MSTs) to determine the overall structure of a star-forming region (see also \citet{schmeja_evolving_2006}, \citet{cartwright_measuring_2009}, \citet{bastian_spatial_2009}, \citet{sanchez_spatial_2009}, \citet{lomax_statistical_2011} and \citet{jaffa_mathcal_2017}). The $Q$-Parameter can be used as a proxy for the dynamical age, with lower $Q$ values (substructured distributions) corresponding to dynamically younger regions and higher $Q$ values (smooth, centrally concentrated distributions) corresponding to dynamically older regions \citep{parker_dynamics_2014}. Using the $Q$-parameter as a proxy for dynamical age in combination with $\Sigma$ (local stellar surface density), \citet{parker_characterizing_2012} and \citet{parker_dynamical_2014} showed that the current dynamical state of a star-forming region could be estimated. \par
\citet{allison_using_2009} developed the mass segregation ratio ($\Lambda_{\rm{MSR}}$) method which detects and quantifies the amount of mass segregation present in a star-forming region (again using MSTs). The $\Lambda_{\rm{MSR}}$ method allows the degree of mass segregation to be found using a plot of $\Lambda_{\rm{MSR}}$ against the number of stars in the minimum spanning tree (see $\S$~\ref{sec:mass_segregation_ratio}). \par
To mitigate the effects of outlying datapoints on $\Lambda_{\rm{MSR}}$, \citet{maschberger_global_2011} introduced the local stellar surface density ratio, defined as the ratio between the median surface densities of a chosen subset and all stars in the region. This method determines if the most massive stars are located in areas of higher than average surface density. \par
The results obtained from these methods must be interpreted with care, especially when determining whether a star-forming region is mass segregated (see \citet{parker_comparisons_2015}). Using these methods, mass segregation has been defined in two main ways: i) the most massive stars are located in areas of higher than average local stellar surface density (as measured by $\Sigma_{\rm{LDR}}$) and ii) the most massive stars are centrally concentrated in the region (as measured by $\Lambda_{\rm{MSR}}$). \par
INDICATE is a new method proposed in \citet{buckner_spatial_2019} to quantify the clustering tendencies of points in a distribution (e.g. stars in a star-forming region), which they used to characterise the spatial behaviours of stars in the Carina Nebula (NGC 3372). The method was also employed in \citet{buckner_spatial_2020} to investigate the clustering tendencies of young stellar objects (YSOs), and thus the star formation history of NGC 2264 (see also \citet{nony_mass_2021}). \par
In this paper we further investigate the ability of INDICATE to quantify overall structure and its ability to detect and quantify mass segregation. We then apply INDICATE to pre-main sequence stars in a selection of nearby star-forming regions for the first time. \par
The paper is organised as follows. In $\S$~\ref{sec:methods} we describe current methods in more detail. In $\S$~\ref{sec:making_synth_sfr} we describe how the synthetic star-forming regions are constructed. In $\S$~\ref{sec:can_indicate_detect_structure} we test the ability of INDICATE to quantify the overall structure of star-forming regions and in $\S$~\ref{sec:can_indicate_detect_mass_segregation} we test the ability of INDICATE to detect and quantify mass segregation in synthetic star-forming regions. In $\S$~\ref{section:observational_data_indicate_results} we present results of applying INDICATE to real star-forming regions. We conclude in $\S$~\ref{sec:conclusions}.

\section{Methods}
\label{sec:methods}
In this section we will describe the INDICATE method along with other methods for comparing the spatial distributions of stars and the detection of mass segregation in star-forming regions.

\subsection{INDICATE}
\label{sec:INDICATE}
%INDICATE is a tool introduced by \citet{buckner_spatial_2019} to quantify the clustering tendencies of individual stars, where each star is assigned its own index from which a star's spatial distribution is determined to be random or spatially clustered and the higher this index the greater the affiliation of a star to others. 
Previous methods of defining structure and clustering tendencies, such as the $Q$-parameter \citep[e.g.][]{cartwright_statistical_2004} involve calculating a value for the entire region which quantifies the amount of substructure present. INDICATE is different in that it assigns a degree of clustering to each star. This allows INDICATE to determine the significance of any clustering on a star by star basis. 

The INDICATE algorithm proceeds as follows. First, an evenly spaced control field (i.e. a regular, evenly spaced grid of points) is generated with the same number density as the dataset. The number density is calculated by dividing the number of points in the dataset by the rectangular area covered by the data. We use the minimum and maximum values of the x and y coordinates to define the edges of the rectangle. For each point \textit{j} in the dataset, the Euclidean distance to the $N^{\rm{th}}$ nearest neighbour in the control field is measured (i.e. for $N = 5$ the distance from the data point $j$ to its $5^{\rm{th}}$ nearest neighbour in the control field) and then the mean of those distances, $\Bar{r}$, is calculated. Then the algorithm counts how many other points $N_{\rm{\Bar{r}}}$ from the dataset are within a radius of $\Bar{r}$ of point $j$. The (unit-less) index $I_{\rm{j}}$ for point $j$ is then defined as,
\begin{equation}
    I_{\rm j} = \frac{N_{\Bar{\rm r}}}{N},
    \label{eq:index}
\end{equation}
where $N$ is the nearest neighbour number. The index is independent of the shape, size and density of the dataset \citep{buckner_spatial_2019}.

To determine the index, above which stars are considered to be clustered and not randomly distributed, \mbox{INDICATE} is applied to a uniformly distributed set of points (see \textit{Appendix~\ref{ab:poisson_ctrl_field}}), with the same number density as the dataset, and using the same control field. 
%Adding to account for feedback
In \citet{buckner_spatial_2020} this is repeated 100 times to remove any statistical fluctuations. In this work we present the results for running this once but we also show results for 100 repeats in \textit{Appendix}~\ref{ab:sig_repeats}. 

The significant index is defined as $I_{\rm{sig}} = \Bar{I} + 3\sigma$ where $\Bar{I}$ is the mean index of the uniform distributions and $\sigma$ is the standard deviation of this mean. Any star in the dataset with an index greater than $I_{\rm{sig}}$ is considered to have a degree of clustering above random. 

%%%%%%%%%

We pick the 50 most massive stars when testing for the presence of mass segregation as it was the minimum sample size that was tested in \citet{buckner_spatial_2019}. For this sample of 50 stars we run 100 repeats to account for statistical fluctuations that can significantly alter the number of stars with indexes above the significant index.

%%%%%%%%%

Following \citet{allison_using_2009} and \citet{parker_dynamical_2014} we define a subset of the 10 most massive stars to compare the median index against the entire regions median index.

To ensure that the correct distance to the nearest neighbour is found for points on the outskirts of the regions the control grid is extended beyond the dataset. If the control grid is not extended then edge effects can make a very small change to the index of those stars (see \textit{Appendix~B} in \citet{buckner_spatial_2019}). 

\subsection{Local Stellar Surface Density Ratio $\Sigma_{\rm{LDR}}$}
\label{sec:local_stellar_surface_density_ratio}
\citet{maschberger_global_2011} developed the local stellar surface density ratio $\Sigma_{\rm{LDR}}$ to quantify the relative surface density of the most massive stars compared to all stars in a region. This method makes use of the local stellar surface density defined as, 
\begin{equation}
\Sigma = \frac{N-1}{\pi R_{\rm{N}}^{2}},
\end{equation}
where \textit{N} is the nearest neighbour number (here we take $N\,=\,5$) and $R_{\rm{N}}$ is the distance to the $N^{\rm{th}}$ nearest neighbour \citep{casertano_core_1985}.
This allows the local surface density of a subset of stars to be quantified by taking the median surface density of that subset. 

The same can be done for all of the stars in the region to find the ratio,
\begin{equation}
    \Sigma_{\rm{LDR}} = \frac{\Tilde{\Sigma}_{\rm{subset}}}{\Tilde{\Sigma}_{\rm{all}}},
\end{equation}
where $\Tilde{\Sigma}_{\rm{subset}}$ is the median local stellar surface density of the subset (in this paper the 10 most massive stars) and $\Tilde{\Sigma}_{\rm{all}}$ is the median local stellar surface density of the entire star-forming region. 

Following \citet{kupper_mass_2011} and \citet{parker_dynamical_2014} we use $\Sigma_{\rm{LDR}}$ to compare the local surface density of the most massive stars to all stars. If the $\Sigma_{\rm{LDR}}\,>\, 1$ then the most massive stars are located in areas of higher than average surface density. If $\Sigma_{\rm{LDR}}\,<\,1$ then the most massive stars are found in areas of lower then average surface density. To determine the significance of this difference a two sample Kolmogorov-Smirnov (KS) test is used. In this work we have chosen an arbitrary threshold value of 0.01, below which the null hypothesis that the two distributions share the same underlying distribution is rejected.  

\subsection{Mass Segregation Ratio $\Lambda_{\rm{MSR}}$}
\label{sec:mass_segregation_ratio}
We quantify the mass segregation of stars using the mass segregation ratio $\Lambda_{\rm{MSR}}$ \citep[][]{allison_using_2009}. This method makes use of minimum spanning trees (MSTs); these are graphs where each point is connected to at least one other point such that the total edge length is minimised with no closed loops. 

First a minimum spanning tree is generated for the 10 most massive stars and this is compared to the average length of a set of MSTs made by randomly picking 10 stars from the distribution. If the average edge length of the 10 most massive stars' MST $l_{\rm{10}}$ is significantly smaller than the average edge length of the random MSTs $l_{\rm average}$, then the most massive stars are said to be mass segregated. 
This is quantified by the ratio,
\begin{equation}
\Lambda_{\rm{MSR}} = \frac{\left<l_{\rm{average}}\right>}{l_{\rm{10}}}_{-\sigma_{1/6}/l_{\rm{10}}}^{+\sigma_{5/6}/l_{\rm{10}}}.
\end{equation}
For this work the 10 most massive stars are chosen, and then we successively add the next 10 most massive stars to the subset group and repeat the method for the new larger subsets.

The uncertainty is found the same way as in \citet{parker_spatial_2018}, where the upper and lower errors are the lengths of random MSTs which lie 5/6 and 1/6 of the way through an ordered list of all random MST lengths, respectively. These values correspond to a 66 per cent deviation from the median value and prevent any single outlying star from heavily influencing the uncertainty, which would be an issue if, like in \citet{allison_using_2009}, a Gaussian dispersion is used as a uncertainty estimator instead.

If $\Lambda_{\rm{MSR}} \gg 1$ then the subset of the most massive stars are mass segregated, if $\Lambda_{\rm{MSR}} \ll 1$ then the subset of the most massive stars are inversely mass segregated and if $\Lambda_{\rm{MSR}} \approx 1$ then the subset of the most massive stars are not mass segregated and are at similar distances from each other as the average stars in the star-forming region.

\subsection{Radial Distribution}
A straightforward way to quantify mass segregation is to compare the cumulative distributions of the most massive stars' positions to the cumulative distributions of all stellar positions. 

To do this a central position needs to be defined. We follow \citet{parker_comparisons_2015} and use the origin (0,0) of the regions as the centre, as they find that the origin of the distribution is a reasonably robust estimation of the centre in centrally concentrated and fractal distributions. We then find the distance from the origin of the region to each star and plot the cumulative distribution functions of all stars in the region and the 10 most massive stars.

\section{Making Synthetic Star-Forming Regions}
\label{sec:making_synth_sfr}
To investigate the performance of INDICATE we make use of synthetic star-forming regions of different idealised geometries (substructured, smooth centrally concentrated, and uniform). We create synthetic star-forming regions of 1000 stars each and the set-up of these regions is described in the following subsections. Our choice of 1000 stars is motivated by the observation that star clusters and star-forming regions in the Galaxy follow a $N_{\rm{cl}} \propto M_{\rm{cl}}^{-2}$ power law (where $N_{\rm{cl}}$ is the number of regions and $M_{\rm{cl}}$ is the mass of the region) between $10 < M_{\rm{cl}}/M_{\rm{\odot}} < 10^{5}$ \citep{lada_embedded_2003}. A star-forming region containing 1000 stars therefore sits somewhere in the middle of this distribution. We note that many star-forming regions in the Solar neighbourhood contain fewer stars (e.g. 100s); INDICATE (and other methods to quantify structure such as the $Q$-parameter) are affected by statistical noise when applied to regions with fewer than 50 stars. In each case we randomly assign masses to the synthetic data using the initial mass function from \mbox{\citet{maschberger_function_2013}}, with the lower mass, upper mass and mean stellar mass set as $0.01\, M_{\rm{\odot}}$, $150\, M_{\rm{\odot}}$ and $0.2\, M_{\rm{\odot}}$ respectively.  The probability distribution is as follows,

\begin{equation}
    p(m) \propto \left(\frac{m}{\mu} \right)^{-\alpha} \left(1 + \left(\frac{m}{\mu}\right)^{1 - \alpha}\right)^{-\beta},
    \label{eq:maschberger_imf}
\end{equation}
where $\mu$ is the mean stellar mass, $\alpha = 2.3$ is the high mass index and $\beta = 1.4$ is the low mass index for the power law. 

\subsection{Fractal Star-Forming Regions}
\begin{figure*}
  \subfigure[A typical fractal distribution of 1000 stars with D = 1.6. ]{\includegraphics[width=0.45\linewidth]{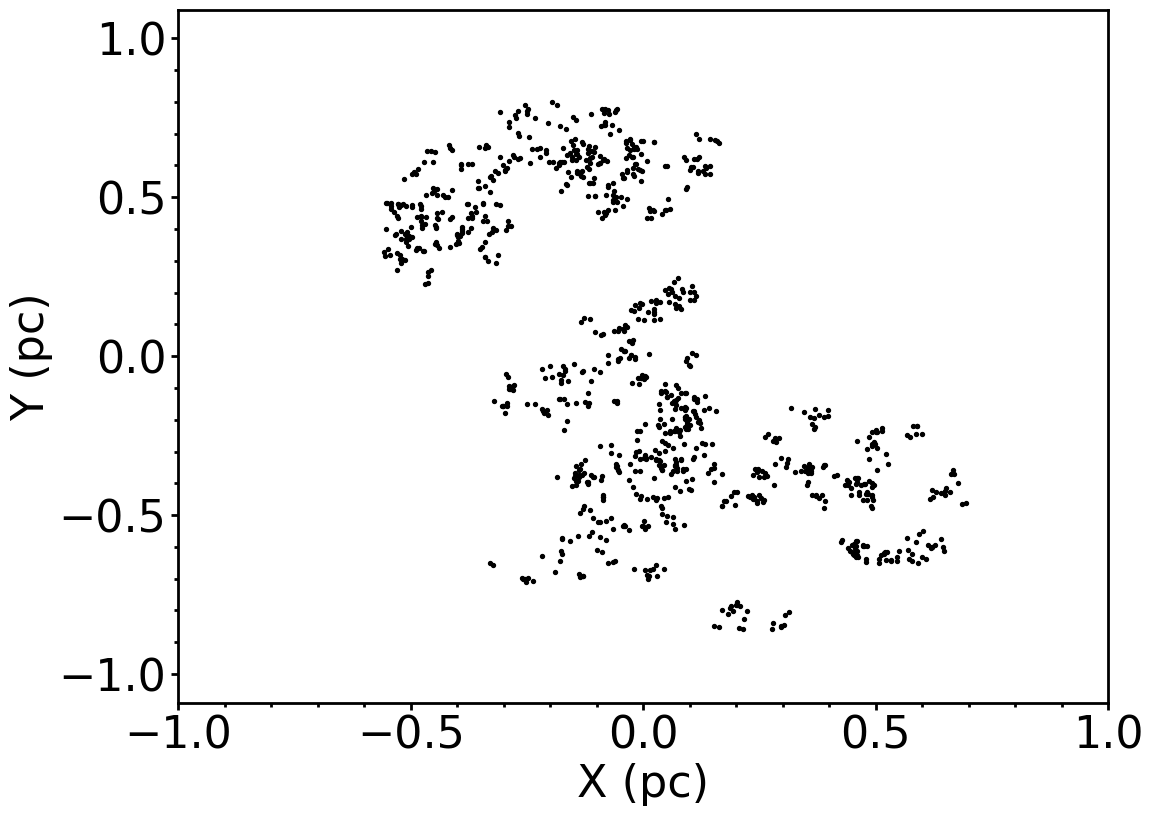}}
  \hspace{10pt}
  \subfigure[A typical smooth, centrally concentrated distribution of 1000 stars with a density index $\alpha = 2.0$]{\includegraphics[width=0.45\linewidth]{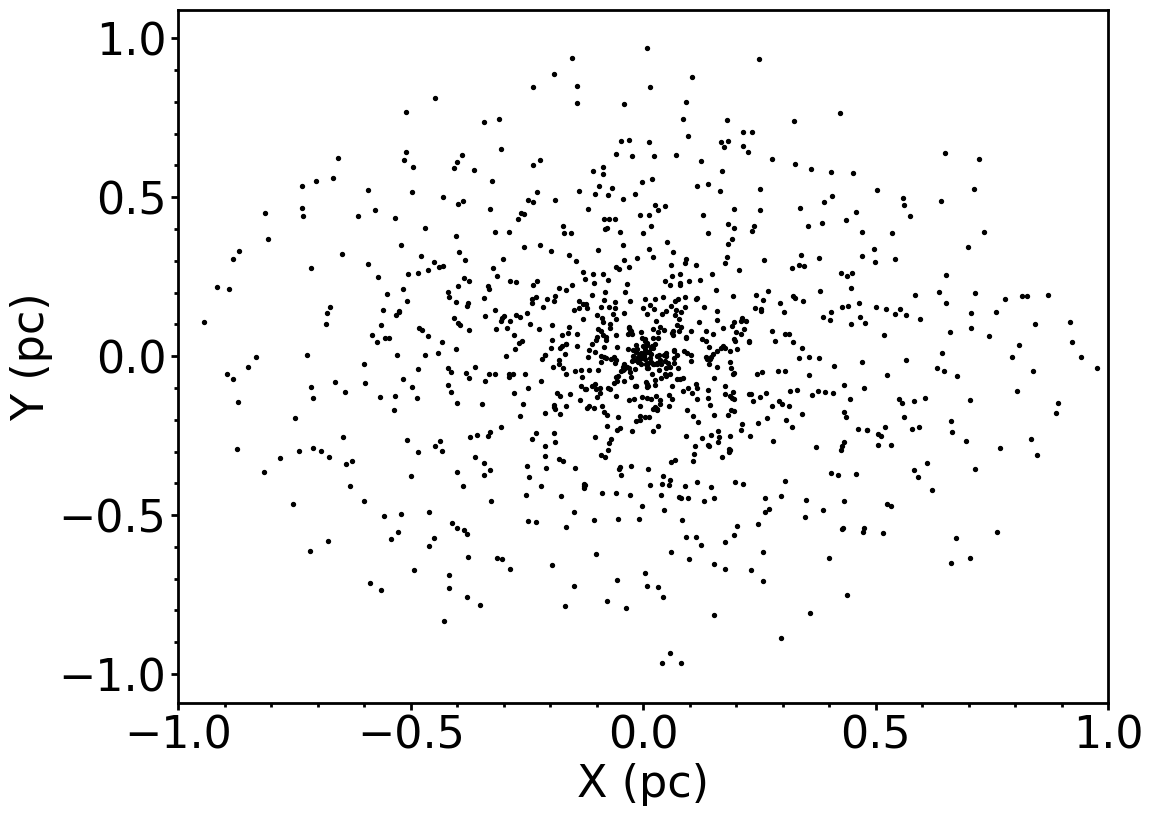}}
  \caption{Typical examples of our synthetic star-forming regions: (a) a substructured star-forming region with fractal dimension $D = 1.6$, (b) a radial smooth centrally concentrated star-forming region with density index $\alpha = 2.0$.}
  \label{fig:example_clusters_fractal_radial}
\end{figure*}

We generate a substructured star-forming region using the box fractal method \citep{goodwin_dynamical_2004,cartwright_statistical_2004}. This method has also been used in previous works \citep[e.g.][]{allison_using_2009, parker_comparisons_2015, daffern-powell_dynamical_2020}. An example of a star-forming region that was generated using this method is presented in figure~\ref{fig:example_clusters_fractal_radial}(a), consisting of 1000 stars with a fractal dimension $D = 1.6$.

The method works as follows. A single star is placed at the centre of a cube of side length $N_{\rm{Div}} = 2$. This cube is then subdivided down into $N_{\rm{Div}}^{3}$ (in this case it is 8) sub-cubes. A star is placed at the centre of each sub-cube and each of the star's corresponding cubes has a probability of being subdivided again into 8 more sub-cubes given by $N_{\rm{Div}}^{(3 - D)}$ where $D$ is the fractal dimension of the region. 

Stars that do not have their cubes subdivided are removed from the region along with any previous generations of stars that preceded them. A small amount of noise is added to each of the stars to stop them having a regular looking structure. The process of subdivision and adding new stars is done until the target number of stars is reached or exceeded in the last iteration of the method. Only the last generation of stars added are kept in the region meaning all previous stars are removed and then any excess stars are removed at random so that the target number of stars lie inside a spherical boundary \citep[][]{daffern-powell_dynamical_2020}.

\subsection{Smooth Centrally Concentrated Star-Forming Regions}
The stars are distributed using the following expression,
\begin{equation}
    n\propto r^{-\alpha}, 
\end{equation}
where \textit{n} is the number density, \textit{r} is the radial distance from the origin of the region and $\alpha$ is the radial density exponent, where higher values of $\alpha$ will produce more centrally concentrated regions \citep{cartwright_statistical_2004}.

\subsection{Uniformly Distributed Star-Forming Regions}
A uniform distribution of 1000 points is generated to test INDICATE when there is no initial structure. Like all other synthetic star-forming regions in this work all points lie within $-1<x<1$ and $-1<y<1$.

\section{Applying INDICATE to synthetic star-forming regions}
\subsection{Can INDICATE determine the structure of star-forming regions?}
\label{sec:can_indicate_detect_structure}
To see how INDICATE performs on star-forming regions with different structures and to test if it can differentiate between them we run it over different sets of star-forming regions with each set corresponding to a different structural parameter that is used in its creation. Each set contains 100 different realisations of regions made with the same parameter. The following fractal dimensions of $D = 1.6, 2.0, 2.6$ and $3.0$ for example would make up 4 sets of star-forming regions, each set containing 100 realisations of substructured regions with each realisation containing 1000 stars. The same is done for smooth, centrally concentrated star-forming regions were 7 sets of regions are made using the following radial density exponents of $\alpha = 0.0,\, 0.5, \,1.0, \,1.5, \,2.0, \,2.5$ and $2.9$, with each realisation also containing 1000 stars. The results of applying INDICATE to all these different sets is shown in figure~\ref{fig:structure_indicate}, which shows the mean, median, mean median and mean maximum index for each set of these star-forming regions. 

The mean and median INDICATE index is calculated for each set by finding the indexes of all stars across all 100 different realisations and then simply finding the mean and median of these values. The mean median INDICATE index is calculated by finding the median INDICATE index for each individual region in a given set resulting in 100 median values for which the mean can be calculated. The mean maximum INDICATE index is calculated by finding the maximum INDICATE index in each individual region for a given set and then finding the mean of these 100 values.

Figure~\ref{fig:structure_indicate} clearly shows that the index is degenerate because different smooth and fractal distributions can have the same INDICATE index. We present the index distributions for the synthetic regions in \textit{Appendix}~\ref{appsec:index_distribution_synth_sfr}. Because the index is similar across star-forming regions with different geometries and levels of substructure, we suggest that INDICATE cannot be used to quantify the type of morphology in the same way that the $Q$-Parameter can.

\subsection{Can INDICATE be used to quantify mass segregation?}
\label{sec:can_indicate_detect_mass_segregation}

To test the ability of INDICATE to detect and quantify if the most massive stars are  in regions of localised above-average stellar surface density we apply it to all 1000 stars in our substructured, smooth centrally concentrated and uniform synthetic star-forming regions where the masses are randomly assigned to stars using the IMF from \citet{maschberger_function_2013}. We change the mass configurations of these regions by swapping the 10 most massive stars with the 10 stars of highest INDICATE index, \textit{high mass high index} configuration (\textit{hmhi}), and by swapping the 10 most massive stars with the 10 most central stars, \textit{high mass centre} configuration \textit{(hmc)}. Table~\ref{tab:indicate repeats all stars} shows how many times across 100 realisations for a substructured region ($D=1.6$), a smooth, centrally concentrated region ($\alpha=2.0$) and a uniform distribution the median index of the entire region of 1000 stars, the 10 most massive stars and 10 random stars are above the significant index. Table~\ref{tab:mass seg tests MSR LDR} shows the results of applying $\Lambda_{\rm{MSR}}$ and $\Sigma_{\rm{LDR}}$ to all stars in the same 100 realisations of each morphology.

% To see how often INDICATE detects mass segregation ($\Tilde{I}_{10} > I_{\rm{sig}}$) of the 10 most massive stars we apply INDICATE to only the 50 most massive stars across the 100 realisations of each morphology. We present these results in table~\ref{tab:indicate mass segregation tests}. For tables~\ref{tab:indicate repeats all stars} and \ref{tab:indicate mass segregation tests} the median indexes are calculated by finding the median index in each region for all stars, the 10 most massive stars and the 10 random stars for each individual realisation then finding the median of these 100 values. For the tests in table~\ref{tab:indicate mass segregation tests} the swap is performed in a full region of 1000 stars then the 50 most massive stars are taken out. For the ranges of indexes across all realisations for the different morphologies see \textit{Appendix}~\ref{app:testing indicate 100 realisations}.} 

\begin{table}
    \setlength{\tabcolsep}{4pt}
    \centering
    \caption{INDICATE results of 100 different realisations for each of the presented morhpologies. From left to right the columns are: the median of the median indexes found across all 100 realisations for all stars, the number of times a realisation's median index for all stars is above its significant index, the median of the median indexes of the 10 most massive stars, the number of times a realisation's median index for the 10 most massive stars is above its significant index, the median of the median index for 10 randomly chosen stars and the number of times a realisation's median index for 10 random stars is above its significant index.}
    \begin{tabular}{l|cccccc}
        \hline
       Region & $\Tilde{I}_{\rm{all}}$ & $\#>I_{\rm{sig}}$ & $\Tilde{I}_{\rm{10,mm}}$ & $\# > I_{\rm{sig}}$ & $\Tilde{I}_{\rm{10,ran}}$ & $\# > I_{\rm{sig}}$\\ 
        \hline
        $D=1.6$, m         & 4.4     & 100     &  4.4     &  97    & 4.5    & 98  \\
        $D=1.6$, hmhi      & 4.4     & 100     & 11.8     & 100    & 4.6    & 98  \\
        $D=1.6$, hmc       & 4.4     & 100     &  3.5     &  84    & 4.5    & 97  \\
        $\alpha=2.0$, m    & 1.8     &   2     &  2.0     &  38    & 2.0    & 41  \\
        $\alpha=2.0$, hmhi & 1.8     &   2     & 22.7     & 100    & 2.0    & 32  \\
        $\alpha=2.0$, hmc  & 1.8     &   2     & 22.2     & 100    & 2.1    & 32  \\
        Uniform, m         & 1.0     &   0     &  0.9     &   0    & 0.9    &  0  \\
        Uniform, hmhi      & 1.0     &   0     &  2.2     &  34    & 0.9    &  0  \\
        Uniform, hmc       & 1.0     &   0     &  0.9     &   0    & 0.9    &  0  \\
        
        \hline
    \end{tabular}
   
    \label{tab:indicate repeats all stars}
\end{table}

\begin{table}
    \setlength{\tabcolsep}{4pt}
    \centering
    \caption{Results of applying $\Lambda_{\rm{MSR}}$ and $\Sigma_{\rm{LDR}}$ to all 1000 stars in 100 different realisations of each morphology and mass configuration. From left to the right the columns are the number of times $\Lambda_{\rm{MSR}} < 0.5$ (which would indicate significant inverse mass segregation, as is observed in the Taurus star-forming region), the number of times $\Lambda_{\rm{MSR}} > 2$ which counts how many times $\Lambda_{\rm{MSR}}$ detects strong signals of mass segregation and the number of times the ratio $\Sigma_{\rm{LDR}}$ is found to be $> 1$ and significant according to a KS test with a threshold p-value < 0.01.}
    \begin{tabular}{l|ccc}
        \hline
        Region & $\# \Lambda_{\rm{MSR}}<0.5$ & $\# \Lambda_{\rm{MSR}}>2$ & $\#\Sigma_{\rm{LDR, Sig}} > 1$  \\
        \hline
        D=1.6, m           &  0           &   0    &   1      \\
        D=1.6, hmhi        &  0           &  94    &  71      \\
        D=1.6, hmc         &  0           & 100    &   6      \\
        $\alpha=2.0$, m    &  0           &   1    &   2      \\
        $\alpha=2.0$, hmhi &  0           & 100    & 100      \\ 
        $\alpha=2.0$, hmc  &  0           & 100    & 100      \\
        Uniform, m         &  0           &   0    &   0      \\
        Uniform, hmhi      &  0           &  34    & 100      \\
        Uniform, hmc       &  0           & 100    &   9      \\
        \hline
    \end{tabular}
   
    \label{tab:mass seg tests MSR LDR}
\end{table}

\begin{table}
    \setlength{\tabcolsep}{4pt}
    \centering
    \caption{INDICATE results when applying only to the 50 most massive stars across 100 realisations of each morphology. From left to right the columns are: the median of the median indexes found for all 50 stars across all 100 realisations, the number of times the median index for the 50 most massive stars is above the significant index in a realisation, the median of the median index of the 10 most massive stars found across all regions, the number of times the median index for the 10 most massive stars is greater than the significant index for a realisation, the number of times that $\Lambda_{\rm{MSR}}$ detects mass segregation in the realisations that INDICATE has detected mass segregation, the median of the median indexes found for 10 randomly chosen stars across all regions, the number of times the median index of a realisation is greater than the significant index.}
    \begin{tabular}{l|ccccccc}
        \hline
       Region & $\Tilde{I}_{\rm{50}}$ & $\#>I_{\rm{sig}}$ & $\Tilde{I}_{\rm{10,mm}}$ & $\# > I_{\rm{sig}}$ & $\#$ MS &$\Tilde{I}_{\rm{10,ran}}$ & $\# > I_{\rm{sig}}$\\ 
        \hline
        $D=1.6$, m         & 1.5    & 12    & 1.5    &  14    &    0    & 1.5    & 16  \\
        $D=1.6$, hmhi      & 1.7    & 28    & 3.6    &  94    &   90    & 1.7    & 35  \\
        $D=1.6$, hmc       & 1.6    & 23    & 3.0    &  95    &   95    & 1.8    & 29  \\
        $\alpha=2.0$, m    & 1.4    & 14    & 1.5    &  29    &    1    & 1.6    & 30  \\
        $\alpha=2.0$, hmhi & 2.6    & 54    & 4.8    & 100    &  100    & 2.7    & 65  \\
        $\alpha=2.0$, hmc  & 2.6    & 57    & 4.8    & 100    &  100    & 2.8    & 62  \\
        Uniform, m         & 0.8    &  0    & 0.7    &   0    &    0    & 0.7    &  0  \\
        Uniform, hmhi      & 0.8    &  0    & 1.2    &  15    &   11    & 0.8    &  0  \\
        Uniform, hmc       & 0.8    &  0    & 2.4    &  87    &   87    & 0.8    &  3  \\
        \hline
    \end{tabular}
    \label{tab:indicate mass segregation tests}
\end{table}

Comparing the median INDICATE index of all the stars in the region with the median index of the 10 most massive stars allows the relative clustering tendencies of the most massive stars to be determined. If the median index of the most massive stars is greater than the median index for the entire region then the most massive stars are more clustered than the typical star in a region, and consequently are found in locations of higher than average local surface density. If the opposite is true then the most massive stars are less clustered than the typical star and are found in areas of lower than average local surface density. To determine if any detected difference in the spatial distribution, according to INDICATE, of the most massive stars in these tests is significant we use a 2 sample KS test with a significance threshold of 0.01, below which we reject the null hypothesis that the 10 most massive stars and the entire population of stars are spatially distributed the same way. If the p-value $\ll 0.01$ then there is a significant difference in the index distributions (and therefore clustering tendencies) of the 10 most massive stars compared to the entire region. %It is important to note that the KS test is not determining is INDICATE is finding mass segregation or not but whether or not it is picking up any correlation between the mass of the stars and the INDICATE indexes they have.}

To test the ability of INDICATE to detect and quantify mass segregation we apply INDICATE to just the 50 most massive stars in each of the regions. The criteria used for INDICATE to detect mass segregation are from \citet{buckner_spatial_2019} and require that the 10 most massive stars are non-randomly clustered with respect to all 50 most massive stars (i.e. $\Tilde{I}_{\rm{10}} > I_{\rm{sig}}$). 

To see how often INDICATE detects mass segregation of the 10 most massive stars we apply INDICATE to only the 50 most massive stars across the 100 realisations of each morphology. We present these results in table~\ref{tab:indicate mass segregation tests}. INDICATE detects mass segregation in many of these realisations, specifically for the realisations with \textit{hmhi} and \textit{hmc} mass configurations. We apply $\Lambda_{\rm{MSR}}$ to these realisations and count how many times $\Lambda_{\rm{MSR}}$ finds the 10 most massive stars to be mass segregated in the realisations that INDICATE has detected mass segregation. For centrally concentrated regions with the mass configurations \textit{hmhi} and \textit{hmc} both INDICATE and $\Lambda_{\rm{MSR}}$ find mass segregation in all 100 realisations. For the centrally concentrated region with randomly assigned masses INDICATE finds 29 realisations with mass segregation but $\Lambda_{\rm{MSR}}$ detects mass segregation in only one of these. A similar result is seen for the substructured regions where INDICATE finds that 14 of the realisations are mass segregated whereas $\Lambda_{\rm{MSR}}$ finds no mass segregation in these realisations. INDICATE detects mass segregation in the uniform distribution \textit{hmhi} and \textit{hmc} mass configurations in 15 and 87 realisations, respectively. $\Lambda_{\rm{MSR}}$ detects mass segregation in 11 of the 15 realisations and all 87 realisations for the \textit{hmhi} and \textit{hmc} mass configurations, respectively.

For tables~\ref{tab:indicate repeats all stars} and \ref{tab:indicate mass segregation tests} the median indexes are calculated by finding the median index in each region for all stars, the 10 most massive stars and the 10 random stars for each individual realisation then finding the median of these 100 values. For the tests in table~\ref{tab:indicate mass segregation tests} the swap is performed in a full region of 1000 stars then the 50 most massive stars are taken out. For the ranges of indexes across all realisations for the different morphologies see \textit{Appendix}~\ref{app:testing indicate 100 realisations}.

We use the 50 most massive stars as this is the minimum sample size that was tested in \citet{buckner_spatial_2019}. As long as a subset is larger than this value, any subset may be selected. For example \citet{buckner_spatial_2019} used a subset of 121 OB stars. 

To see if the 10 most massive stars are spatially clustered differently than the entire subset we run a 2 sample KS test with significance threshold of 0.01. The INDICATE plots for the 50 most massive stars in each of the synthetic regions are shown in \textit{Appendix}~\ref{appsec:classical_massseg_synth_data}.
For all of these tests the $\Sigma_{\rm{LDR}}$, $\Lambda_{\rm{MSR}}$ and cumulative distribution of stellar position methods are also applied for comparison. We further employ the 2 sample KS test to determine if any $\Sigma_{\rm{LDR}}$ or CDF results are significantly different between the 10 most massive stars and the entire population in a region.

\subsection{Fractal Star-Forming Regions}
\label{sec:indicate fractal results}
\begin{figure*}
  \subfigure[Mean Indexes]{\includegraphics[width=0.49\linewidth]{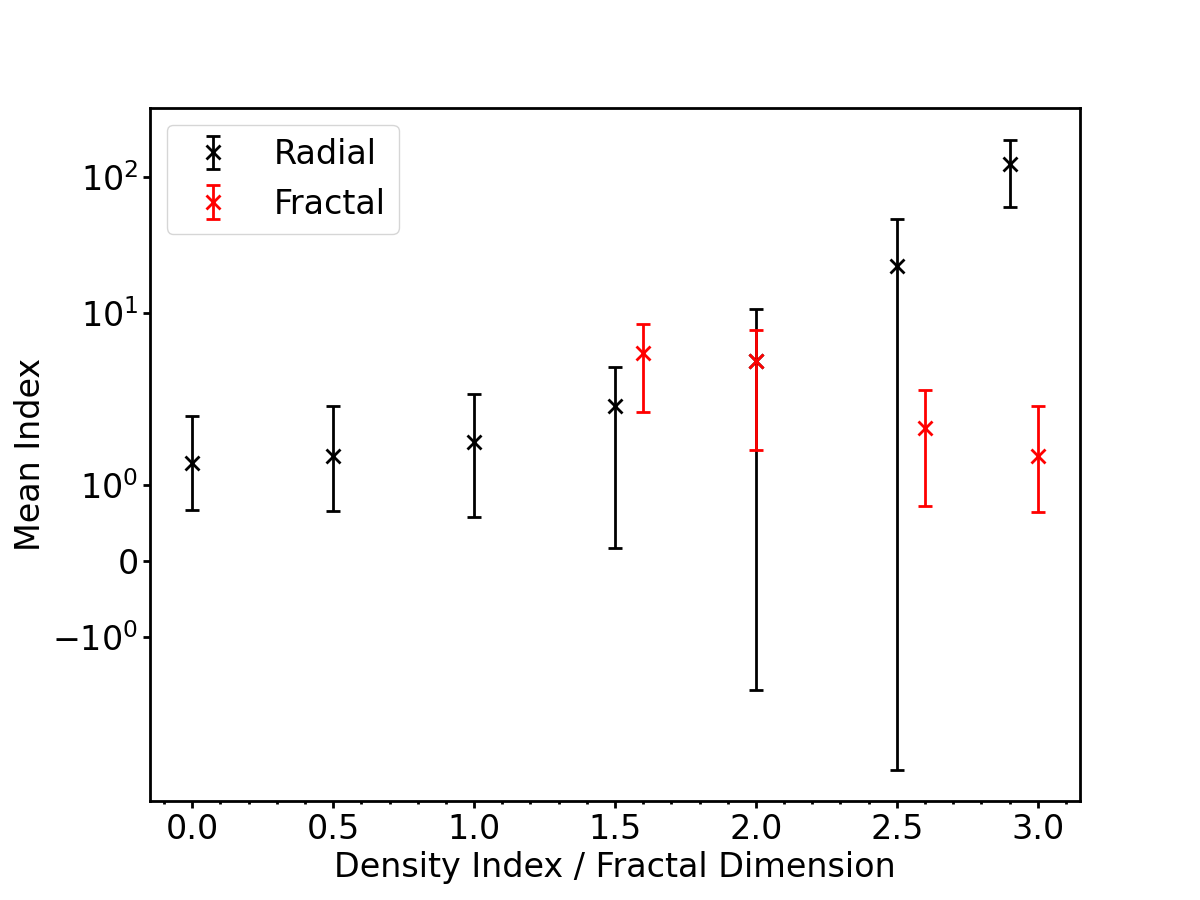}}
   \hspace{0.8pt}
  \subfigure[Median Indexes]{\includegraphics[width=0.49\linewidth]{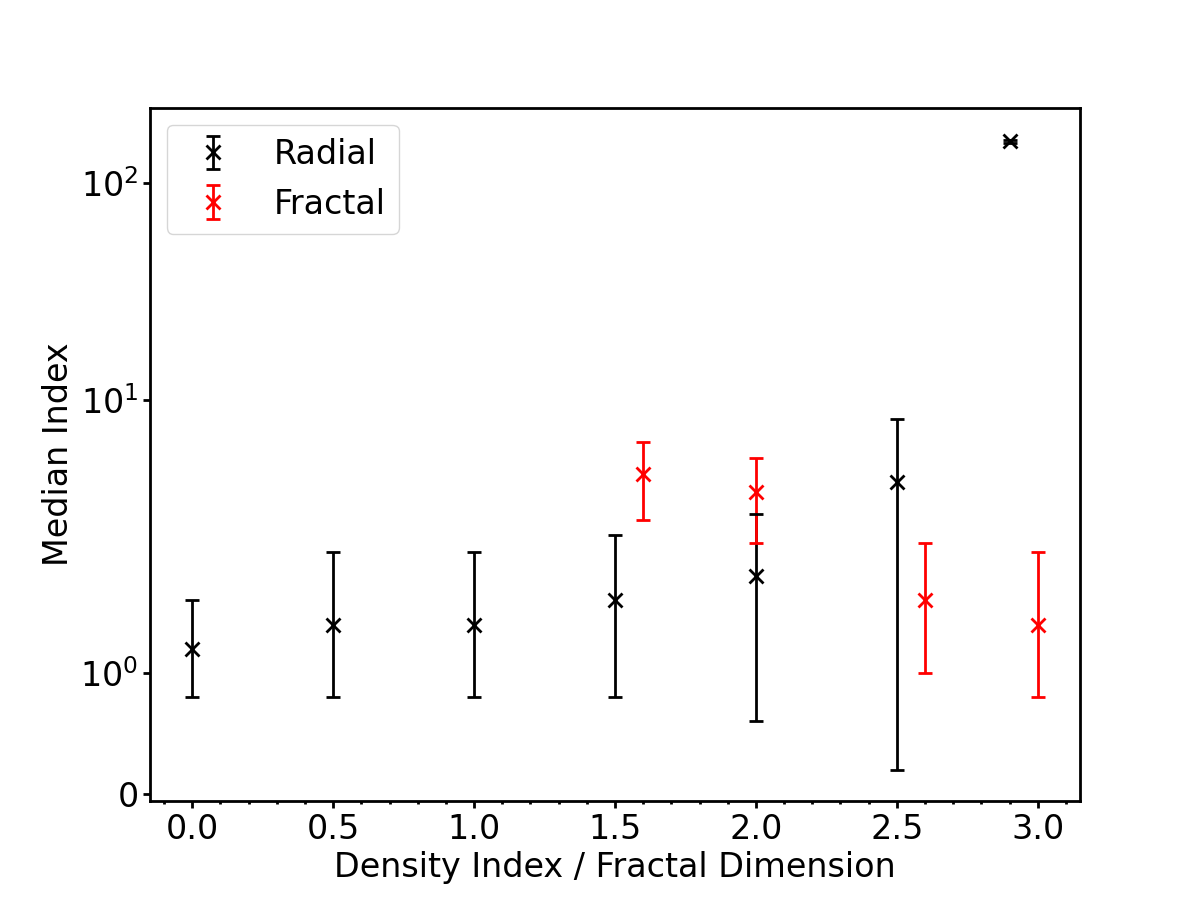}}
   \hspace{0.8pt}
  \subfigure[Mean of Median Indexes]{\includegraphics[width=0.49\linewidth]{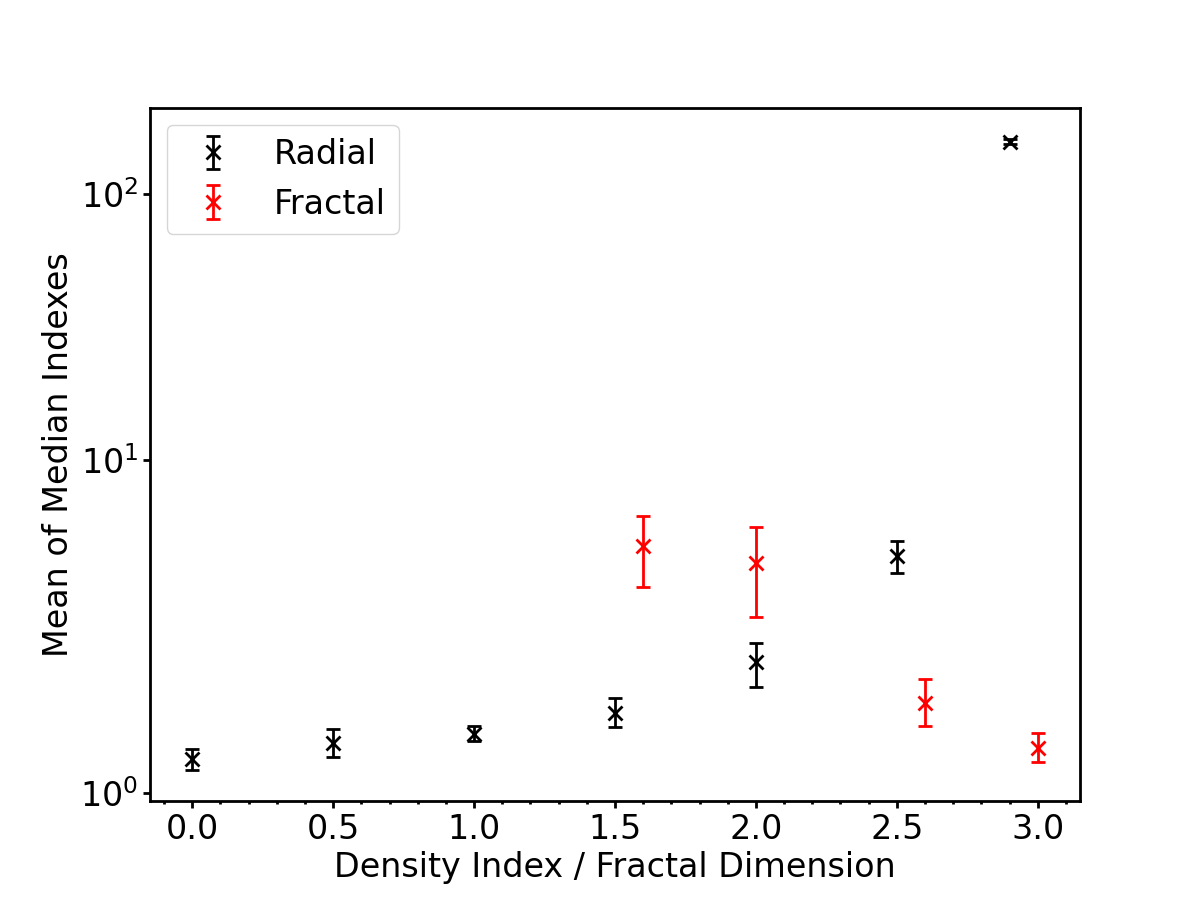}}
   \hspace{0.8pt}
  \subfigure[Mean Maximum Indexes]{\includegraphics[width=0.49\linewidth]{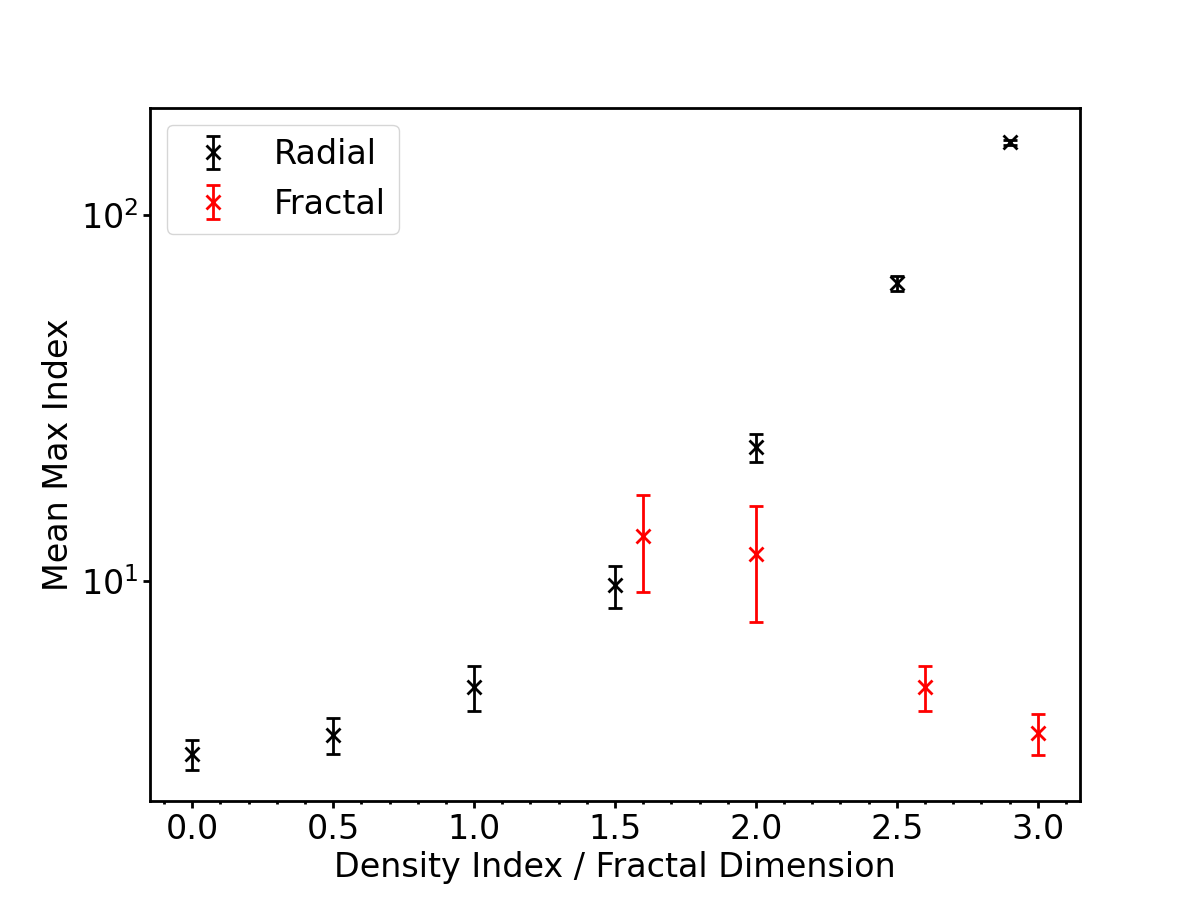}}
  
  \caption{INDICATE results for regions with different spatial distributions. Panel (a) shows the mean indexes found for 100 different realisations of ideal star-forming regions of differing fractal dimension and radial densities and the error bars represent the standard deviation of the mean indexes. Panel (b) shows the median indexes for the same regions and the error bars here represent the median absolute deviation. Panel (c) shows the mean of the median found for each of the 100 regions with the error bars representing the standard deviation of the mean. In panel (d) the mean maximum index is shown with the error bars being the standard deviation of this value.}
  \label{fig:structure_indicate}
\end{figure*}

\subsubsection{Random masses}
Figure~\ref{fig:results_fractal}(a) shows that INDICATE clearly identifies areas of high spatial clustering, and finds that 82.2 per cent  of stars have an index greater than the significant index of 2.3. The median index of the significantly clustered stars is $5.2_{-1.4}^{+2.6}$ where the sub and superscript numbers show the uncertainty defined by the 25$^{\rm{th}}$ and 75$^{\rm{th}}$ quantiles respectively (this value is the same for \textit{hmhi} and \textit{hmc} configurations as while the masses have been swapped each star will have the same position). The maximum index for all stars is 15.8 (which is also the same for \textit{hmhi} and \textit{hmc}) with a median index for the entire region of $4.4_{-1.4}^{+2.6}$ and for the 10 most massive stars it is $4.5_{-0.6}^{+3.3}$. As both of the medians are above the significant index both the entire region and the 10 most massive stars are clustered above random. A KS test returns a p-value of 0.9, suggesting that the difference in clustering tendencies of the most massive stars and all stars is not significant and that both high and low mass stars share similar non-random clustering tendencies. 
Applying INDICATE to the 50 most massive stars we find that 22 per cent of stars in the subset have indexes above the significant index of 2.1. For significantly clustered stars the median index is $2.2_{-0.0}^{+0.2}$ and the maximum index across all 50 stars is 2.6. The median index for the subset is $1.4_{-0.6}^{+0.6}$ and the median index for the 10 most massive stars is $1.6_{-0.8}^{+0.4}$, which is below the significant index meaning that INDICATE is not detecting mass segregation. A KS test returns a p-value $= 1.00$, correctly identifying that there is no difference in clustering tendencies of the 10 most massive stars and the entire subset.

Figure~\ref{fig:results_fractal}(d) shows local stellar surface density against mass, with $\Sigma_{\rm{LDR}} = 1.3$, and a p-value of 0.69, meaning no significant difference between the local stellar surface density of high mass stars and the entire region. This is in agreement with the INDICATE result that the most massive stars are distributed in a similar way to the other stars in the region. 

Figure~\ref{fig:results_fractal}(g) shows the $\Lambda_{\rm{MSR}}$ result, with $\Lambda_{\rm{MSR}} = {0.97}_{-\,0.11}^{+\,0.09}$ for the 10 most massive stars which is consistent with no significant mass segregation and is in agreement with INDICATE. 

Figure~\ref{fig:results_fractal}(j) shows the cumulative radial distribution of the 10 most massive stars and all the stars in the region. The radial distribution starts at 0.6 pc for the 10 most massive stars; this is due to the randomly assigned stellar masses, which happen to mainly be in clumps a large distance away from the centre (i.e. the origin at (0,0)). When comparing the two distributions using a KS test a p-value $\ll 0.01$ is returned, meaning that the most massive stars could be mistakenly inferred to have been drawn from a different underlying radial distribution than all of the stars.

\subsubsection{High Mass High Index}
We now swap the 10 most massive stars with stars that have the highest INDICATE index and these results are shown in the middle column of figure~\ref{fig:results_fractal}. In figure~\ref{fig:results_fractal}(b) we show the positions of the 10 most massive stars (the black crosses).

The median index of the 10 most massive stars has increased from $4.5_{-0.6}^{+3.3}$ to $15.3_{-0.4}^{+0.2}$, with the median index for the entire region staying the same at $4.4_{-1.4}^{+2.6}$, as does the significant index of 2.3 and the percentage of stars with indexes greater than it. These parameters stay the same as the overall geometry of the region has not changed, just the masses assigned to 20 of the stars have been swapped. Comparing the 10 most massive stars to the rest of the population using a KS test gives a p-value $\ll 0.01$, implying a significant difference in the clustering tendencies of the most massive stars. Figure~\ref{fig:results_fractal}(b) shows this difference clearly, as the 10 most massive stars are now positioned in the most clustered locations according to INDICATE and, consequently, appear visibly more spatially concentrated.

Applying INDICATE to just the 50 most massive stars we find that the percentage of stars with indexes above the significant index of 2.1 is 38 per cent, with a median index for all 50 stars of $1.3_{-0.5}^{+2.3}$ and a median index of $3.6_{-0.0}^{+0.0}$ for the 10 most massive stars. The median index for significantly clustered stars is $3.6_{-0.0}^{+0.0}$. A maximum INDICATE index of 3.8 is found for the 50 most massive stars. As the median index for the 10 most massive stars is above the significant index mass segregation has been detected in the region. The reason the amount of stars with significant indexes has changed is due to the fact that we have made the swap in the full region of 1000 stars and then taken the 50 most massive from that. A KS test returns a p-value $\ll 0.01$ confirming that the tendency of the 10 most massive stars to cluster with high mass stars is significantly different to that of the entire subset of the 50 most massive stars.

$\Sigma_{\rm{LDR}}$ has increased from 1.3 to 2.3, with a p-value $\ll 0.01$. Figure~\ref{fig:results_fractal}(e) shows the 10 most massive stars are now above the median surface density of all of the stars (shown by the horizontal dashed black line). In this case the reason for this is that swapping stars to the most clustered areas as measured using INDICATE results in them also being swapped into areas with higher than average local stellar surface density.

We now measure mass segregation according to $\Lambda_{\rm{MSR}}$ of the region, and find the 10 most massive stars have a mass segregation ratio of  $\Lambda_{\rm{MSR}} = 33.74_{\,-\,5.27}^{\,+\,2.54}$. This implies significant mass segregation. In this particular region this is because all of the most massive stars are located in a single clump with a high INDICATE index.

Because all the most massive stars have been moved to the same region the average distance between them is shorter than when looking at the average distances between random stars in the region. Figure~\ref{fig:results_fractal}(h) shows the peak signal for the 10 most massive stars which then rapidly decreases to $\sim1$, meaning no mass segregation for lower mass stars, which is to be expected because these stars have not been swapped.

When comparing the cumulative distributions of the positions of the 10 most massive stars and all the stars (see figure~\ref{fig:results_fractal}(k)), a clear difference can be seen. The distribution of positions for the 10 most massive stars is very narrow because they are in a very concentrated location therefore they are all a similar distance away from the origin. A KS test between the cumulative distribution of positions for all the stars and the 10 most massive stars returns a p-value $\ll 0.01$.

\subsubsection{High Mass Centre}
\begin{figure*}
  \subfigure[INDICATE, m]{\includegraphics[width=0.32\linewidth]{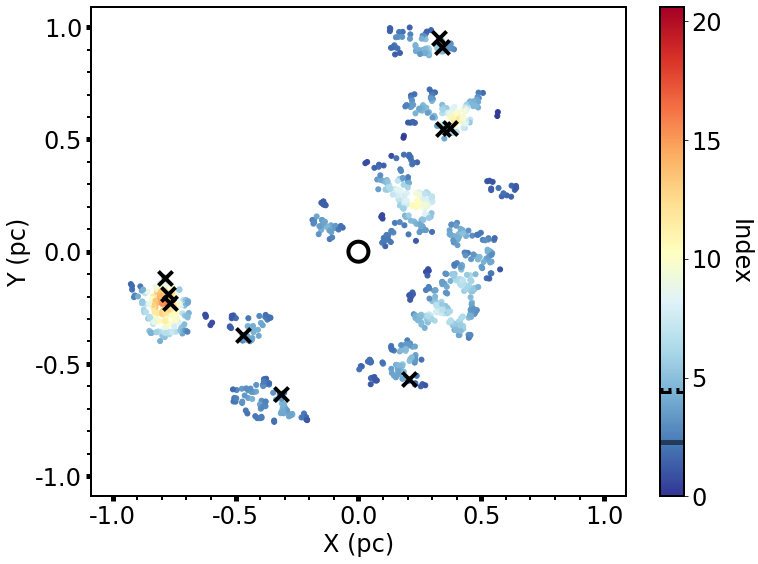}}
  \hspace{0.8pt}
  \subfigure[INDICATE, hmhi]{\includegraphics[width=0.32\linewidth]{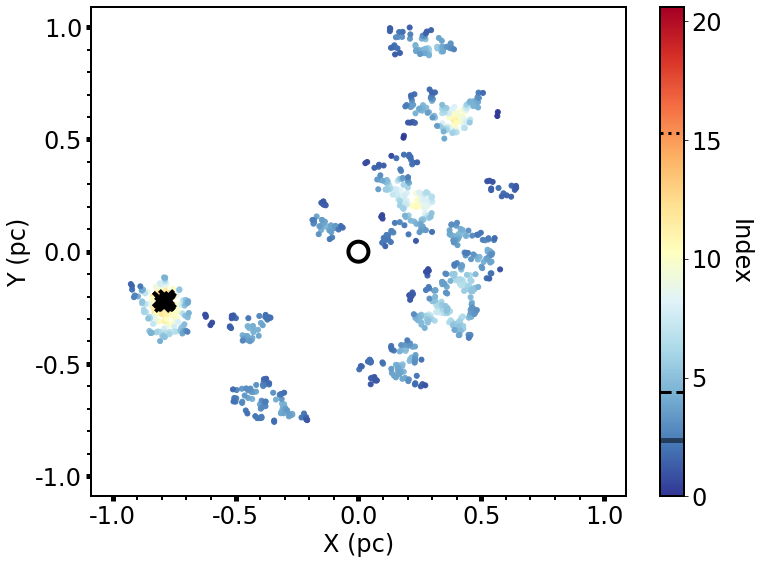}}
  \hspace{0.8pt}
  \subfigure[INDICATE, hmc]{\includegraphics[width=0.32\linewidth]{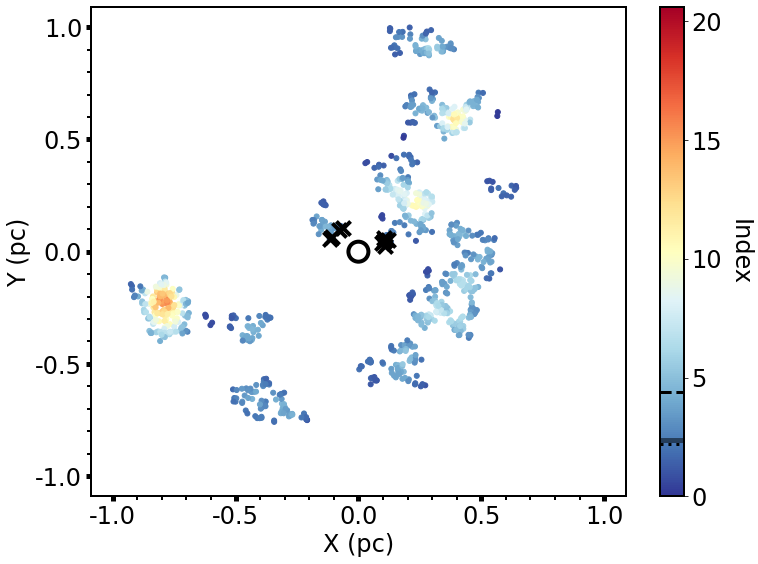}}
  \hspace{0.8pt}
  \subfigure[$\Sigma - m$, m]{\includegraphics[width=0.32\linewidth]{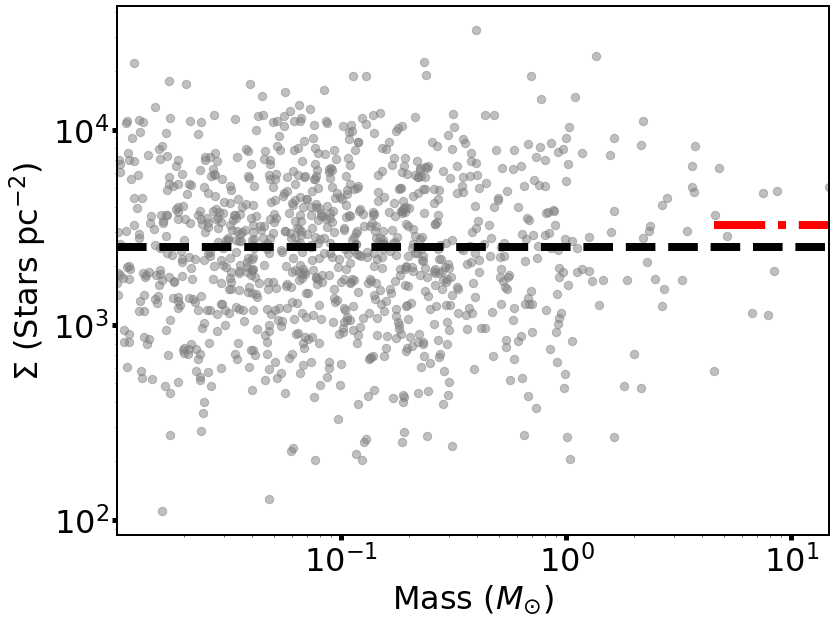}}
  \hspace{0.8pt}
  \subfigure[$\Sigma - m$, hmhi]{\includegraphics[width=0.32\linewidth]{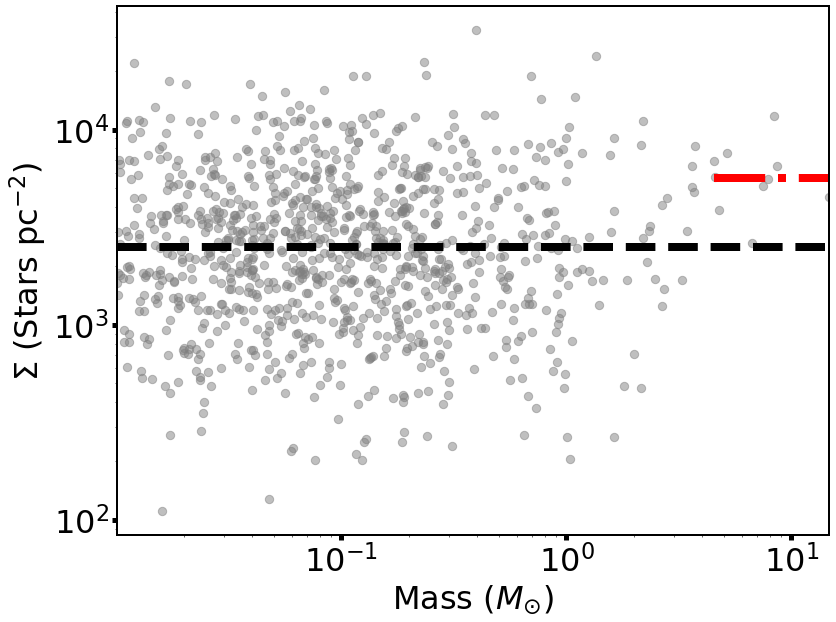}}
  \hspace{0.8pt}
  \subfigure[$\Sigma - m$, hmc]{\includegraphics[width=0.32\linewidth]{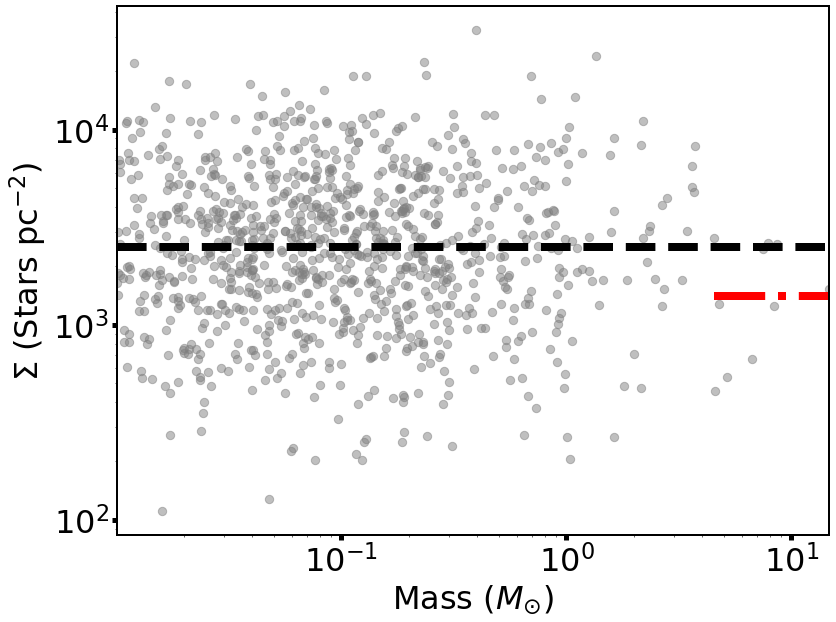}}
  \hspace{0.8pt}
  \subfigure[$\Lambda_{\rm{MSR}}$, m]{\includegraphics[width=0.32\linewidth]{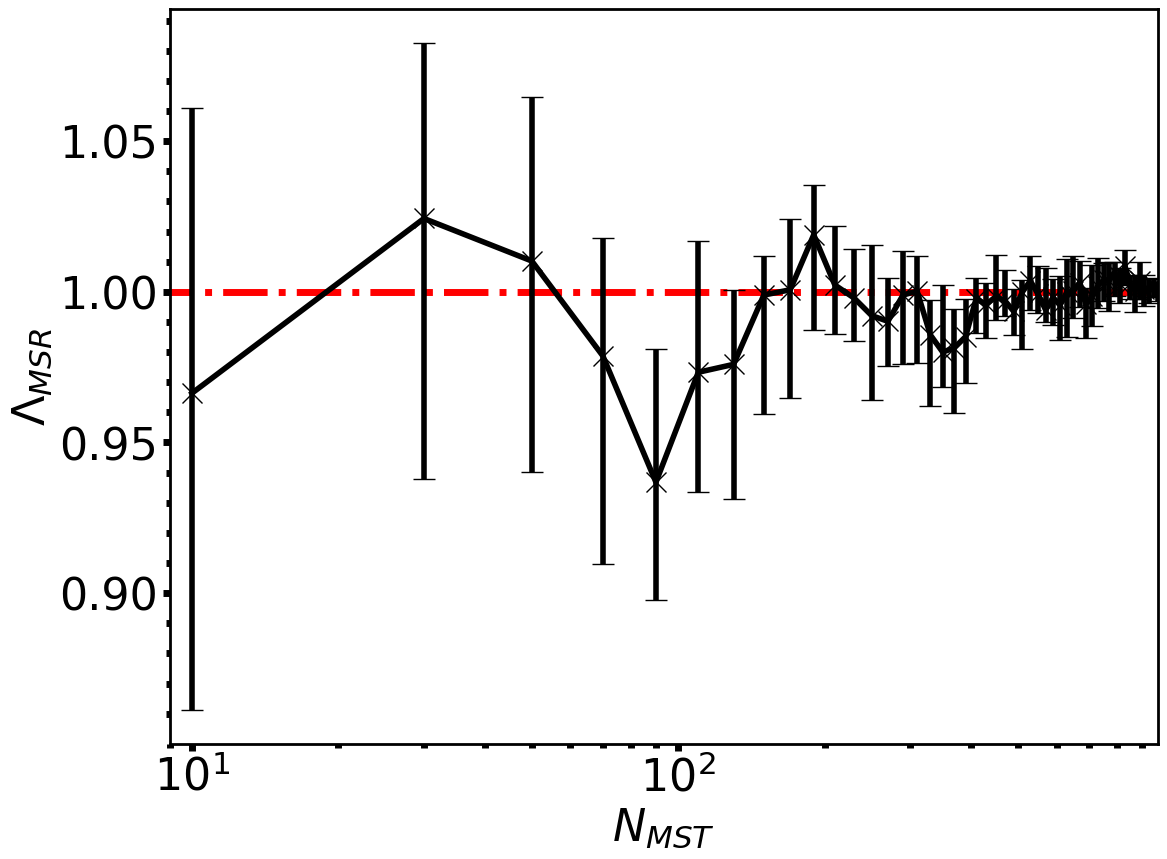}}
  \hspace{0.8pt}
  \subfigure[$\Lambda_{\rm{MSR}}$, hmhi]{\includegraphics[width=0.32\linewidth]{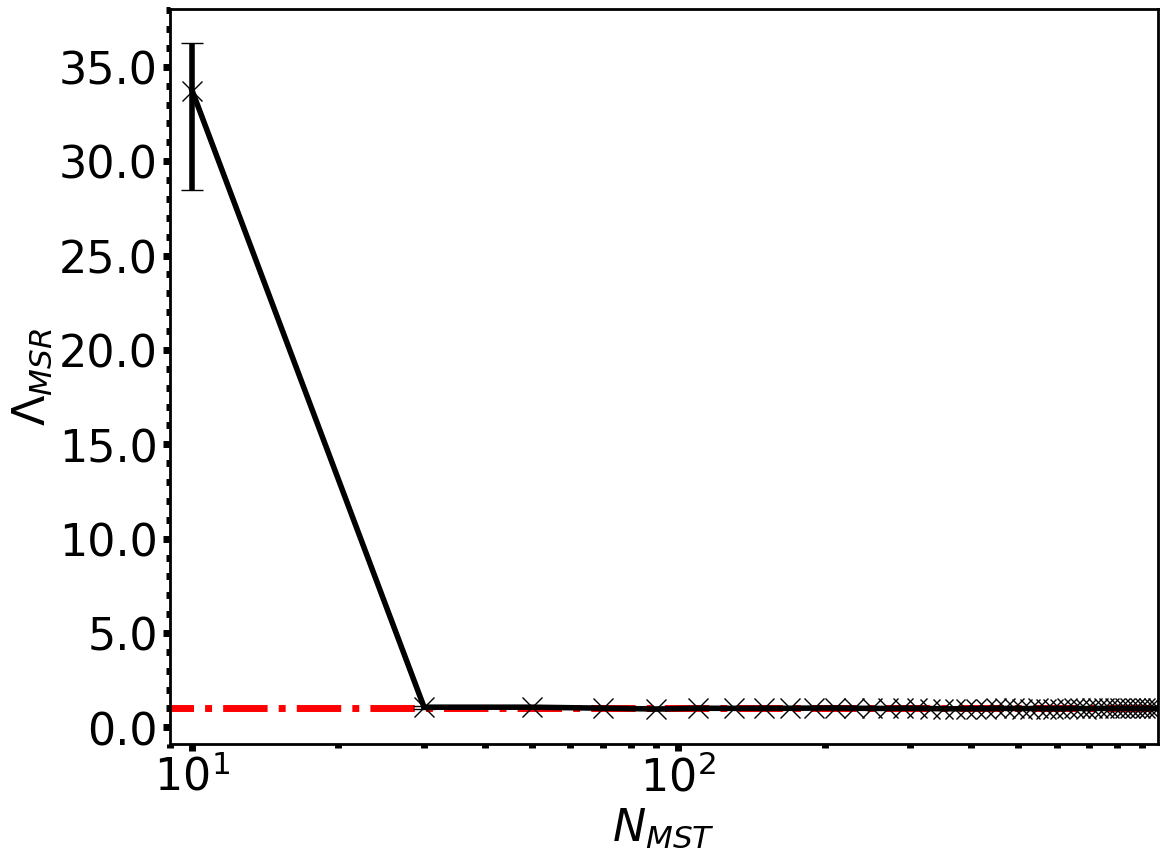}}
  \hspace{0.8pt}
  \subfigure[$\Lambda_{\rm{MSR}}$, hmc]{\includegraphics[width=0.32\linewidth]{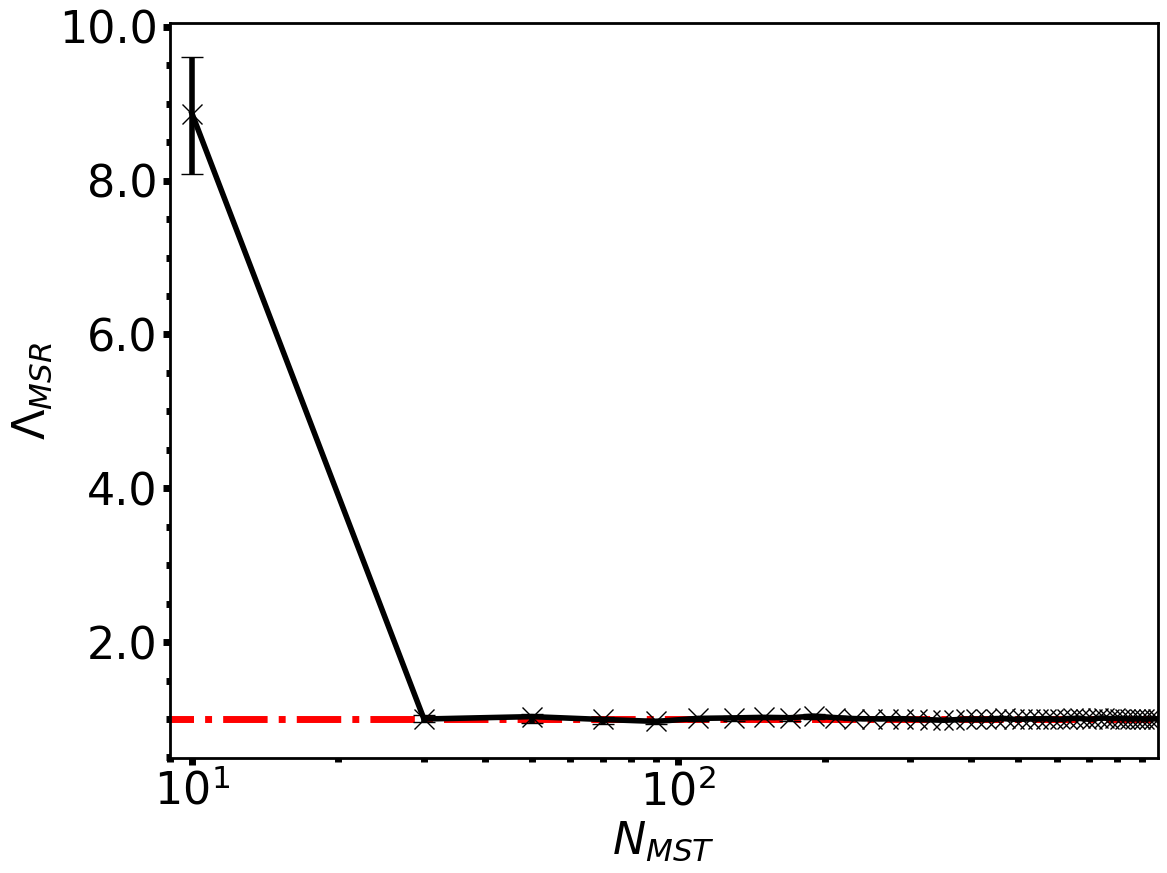}}
  \hspace{0.8pt}
  \subfigure[Radial distribution, m]{\includegraphics[width=0.32\linewidth]{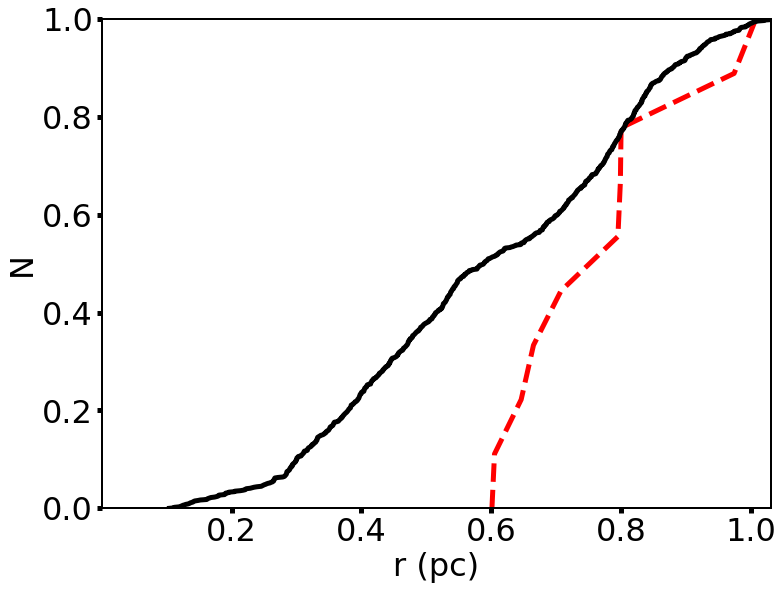}}
  \hspace{0.8pt}
  \subfigure[Radial distribution, hmhi]{\includegraphics[width=0.32\linewidth]{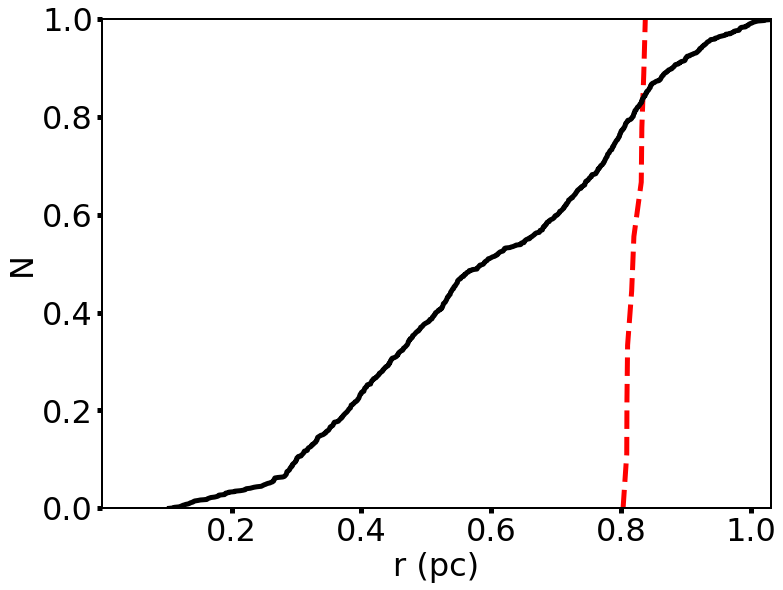}}
  \hspace{0.8pt}
  \subfigure[Radial distribution, hmc]{\includegraphics[width=0.32\linewidth]{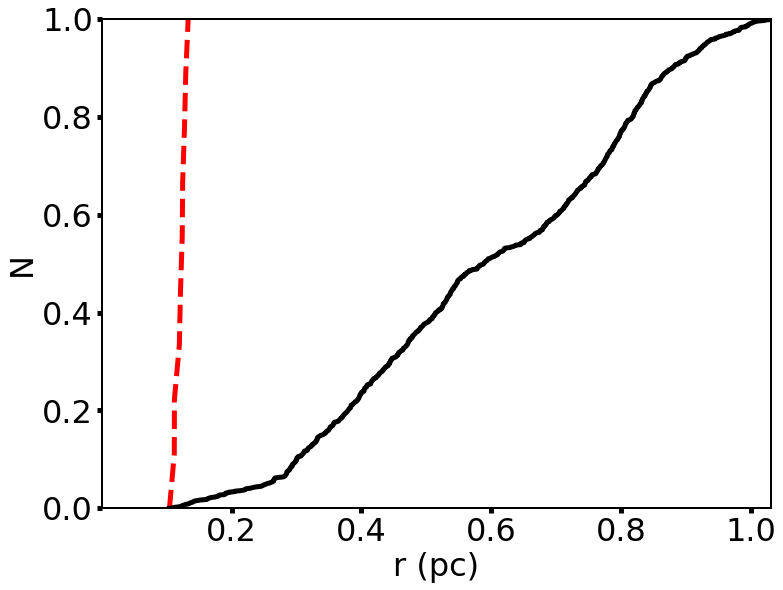}}
  \caption{A synthetic fractal star-forming region of 1000 stars with a fractal dimension of 1.6. The rows are (i) the INDICATE values, (ii) the $\Sigma$-m plots, (iii) the $\Lambda_{\rm{MSR}}$ plots and (iv) the cumulative distributions of radial distances from the centre of the star-forming region. From left to right, the columns are the region with randomly assigned masses (m), highest mass moved to highest INDICATE index (hmhi) and highest masses moved to the centre (hmc). In panels (a), (b) and (c) the stars are colour mapped using the range of indexes found for the centrally concentrated region in figure~\ref{fig:results_radial}, as this region was found to have the greatest INDICATE index. In the colour bar the solid black line represents the significant index for the star-forming region, the dashed black line represents the median index for all of the stars and the dotted black line represents the median index of the 10 most massive stars. The 10 most massive stars in (a), (b) and (c) are highlighted with black crosses. The centre of the region is located in the middle of the black ring. In (d), (e) and (f) the median surface density of the stars is shown by the black dashed line, the median surface density of the 10 most massive stars is shown by the red dash-dotted line. In (g), (h) and (i) the mass segregation ratio is shown by the black line; the horizontal dashed dotted red line shows the value of 1 corresponding to no mass segregation. In (j), (k) and (l) the black line; represents the CDF of radial distance from the centre for all the stars, the red dashed line is the CDF for the 10 most massive stars.}
  \label{fig:results_fractal}
\end{figure*}

The 10 most massive stars are now swapped with the 10 most central stars (the results for this are shown in the right-hand column of figure~\ref{fig:results_fractal}). Because of the box-fractal construction of the region the origin (at (0,0)) is in empty space, so the most massive stars are split into two groups around the origin. 

The median INDICATE index for the 10 most massive stars is $2.2_{-0.0}^{+0.0}$ (decreasing from $4.5_{-0.6}^{+3.3}$ for the 10 most massive stars in the original region with randomly assigned masses), below the median index for the region which is $4.4_{-1.4}^{+2.6}$.  A KS test returns a p-value $\ll$ 0.01 implying a significant difference in the distributions, in this case the most massive stars are in locations of lower clustering than the rest of the stars in the region. Figure~\ref{fig:results_fractal}(c) shows the two main groups of the most massive stars either side of the centre of the star-forming region and the stars here have a relatively low INDICATE index. The central locations of this region happen to be of relatively low index compared to the rest of the star-forming region.

Applying INDICATE to just the 50 most massive stars we find that 28 per cent of stars have indexes above the significant index of 2.1. For significantly clustered stars the median index is $2.6_{-0.2}^{+0.0}$. The median index for the entire subset is $1.4_{-0.4}^{+0.8}$ and is $2.6_{-0.2}^{+0.0}$ for the 10 most massive stars, which is above the significant index of 2.1 meaning the region is mass segregated. A maximum INDICATE index of 2.6 is found for the 50 most massive stars. A KS test returns p-value $\ll 0.01$ confirming that the tendency of the 10 most massive stars to cluster with high mass stars is significantly different to the entire subset.

In figure~\ref{fig:results_fractal}(f) we show the local surface density against mass plot. We find $\Sigma_{\rm{LDR}} = 0.56$ with a p-value $= 0.03$, meaning there is no significant difference in the local surface density of the 10 most massive stars compared to all the stars.
The surface densities therefore display similar behaviour to INDICATE, which also shows a decrease in the measured index. 
%The 10 most massive stars are no longer above the median local stellar surface density (black dashed line) because the points nearest to the centre of this region have a lower local surface density.
When we apply $\Lambda_{\rm{MSR}}$ to this region we detect a significant amount of mass segregation with $\Lambda_{\rm{MSR}}~=~{8.87}_{\,-\,0.78}^{\,+\,0.74}$, (figure~\ref{fig:results_fractal}(i)). This is much lower than when swapping the most massive stars with the most clustered as measured with INDICATE, decreasing from over 30 to 8.87 in figure~\ref{fig:results_fractal}(h) and figure~\ref{fig:results_fractal}(i) respectively. This is due to the areas of highest INDICATE index being highly concentrated in one area, whereas in this case the most central region is in empty space so the most massive stars are spread out around this point. 

The cumulative distribution of the positions of stars is shown in figure~\ref{fig:results_fractal}(l), which also shows the massive stars to be mass segregated and much closer to the centre than the average star. A KS test returns a p-value $\ll 0.01$.

\subsection{Smooth, Centrally Concentrated Star-Forming Regions}
\label{sec:indicate radial results}
\subsubsection{Random Masses}
The region shown in figure~\ref{fig:results_radial}(a) has a clear central region where INDICATE has detected high levels of spatial clustering, finding that 44.1 per cent of stars are spatially clustered above random, with indexes greater than the significant index of 2.3. The median index for stars significantly clustered above random is $5.2_{-1.6}^{+5.2}$. The same values are also found for the \textit{hmhi} and \textit{hmc} configurations. In this case the most massive stars are spread out across the region with none of the 10 most massive stars located in the central area. 
The maximum index for the entire region is 20.4 with a median index of $1.8_{-1.0}^{+3.0}$ for all stars and a median index of $2.0_{-0.8}^{+2.4}$ for the 10 most massive stars: neither the massive stars nor rest of the population is typically in areas of non-random stellar affiliation as both have median indexes below the significant index. A KS test confirms that there is no significant difference in their spatial clustering with a p-value = $0.55$.

Applying INDICATE to just the 50 most massive stars in the region we find that 52 per cent of stars have indexes above the significant index of 2.1, with a median index for the entire subset of 2.6 and 1.1 for the 10 most massive stars. A median index of $4.4_{-1.4}^{+0.4}$ is found for significantly clustered stars. A maximum INDICATE index of 5.2 is found for the 50 most massive stars. As the median index for the 10 most massive stars is below the significant index INDICATE detects no mass segregation in the region. A KS test returns a p-value $= 0.48$ implying no difference in the clustering tendencies of the 10 most massive stars compared to the rest in the subset.

Figure~\ref{fig:results_radial}(d) shows the local stellar surface density against mass plot for this region with the red dashed-dotted line showing the median surface density for the 10 most massive stars and the black dashed line showing the median surface density of all the stars.
A similar result is seen with $\Sigma_{\rm{LDR}} = 1.24$ with a p-value $= 0.63$ from the KS test, indicating the difference is not significant.

Figure~\ref{fig:results_radial}(g) shows the mass segregation ratio for the 10 most massive stars is $\Lambda_{\rm{MSR}} = {1.00}_{\,-\,0.23}^{\,+\,0.11}$ meaning that $\Lambda_{\rm{MSR}}$ finds no significant mass segregation for the 10 most massive stars in this region. 

The cumulative distribution of positions are very similar between the most massive stars and the rest, with p-value $= 0.67$. Figure~\ref{fig:results_radial}(j) shows the radial distribution of the 10 most massive stars in red, it closely matches the radial distribution of the entire region. 

\subsubsection{High Mass High Index}
As before we swap the most massive stars to the areas of greatest clustering as measured by INDICATE. Figure~\ref{fig:results_radial}(b) shows the 10 most massive stars are now located in the centre of the region. The median index for the 10 most massive stars has increased from $2.0_{-0.8}^{+2.4}$ to $19.4_{-0.0}^{+0.8}$, with a p-value $ \ll 0.01$ indicating a significant difference of the clustering tendencies between all the stars (which have a median index of 1.8) and the 10 most massive stars. Therefore, the method correctly detects that the 10 most massive stars are now located in areas of above average stellar affiliation.

We apply INDICATE to just the 50 most massive stars in the region and find that 64 per cent of stars have indexes above the significant index of 2.1, with the subset having a median index of $5.7_{-5.0}^{+0.3}$ and a median index of $6.0_{-0.0}^{+0.2}$ for the 10 most massive stars. The median index for the significantly clustered stars is $6.0_{-0.2}^{+0.0}$. A maximum INDICATE index of 6.2 is found for the 50 most massive stars. As the median index for the 10 most massive stars is larger than significant index INDICATE has detected mass segregation. A KS test returns a p-value $\ll 0.01$ implying a significant difference in the clustering tendencies of the 10 most massive stars despite the median indexes being similar.

Figure~\ref{fig:results_radial}(e) shows the clear difference in the median surface density of the entire region (black dashed line) compared to the median surface density of the 10 most massive stars (red dashed-dotted line). $\Sigma_{\rm{LDR}} = 17$ (increasing from 1.24) with p $\ll$ 0.01, in agreement with INDICATE that the 10 most massive stars are found in areas of greater stellar clustering compared to the rest of the population.

$\Lambda_{\rm{MSR}}$ detects significant mass segregation with $\Lambda_{\rm{MSR}} = {28.23}_{\,-\,7.20}^{\,+\,3.05}$, increasing from 1.00 when compared to this region with randomly assigned masses and is in agreement with \mbox{INDICATE} when applied to the 50 most massive stars. Like in the fractal region in figure~\ref{fig:results_fractal}, there is one area of high clustering so the most massive stars are moved closer together as a result. 
This is also reflected in the cumulative distribution of positions (figure~\ref{fig:results_radial}(k)) with a much steeper function for the 10 most massive stars with p-value $\ll$ 0.01, this is very similar to the results in figure~\ref{fig:results_fractal}(k) for a fractal distribution.

\subsubsection{High Mass Centre}
\begin{figure*}
  \subfigure[INDICATE, m]{\includegraphics[width=0.32\linewidth]{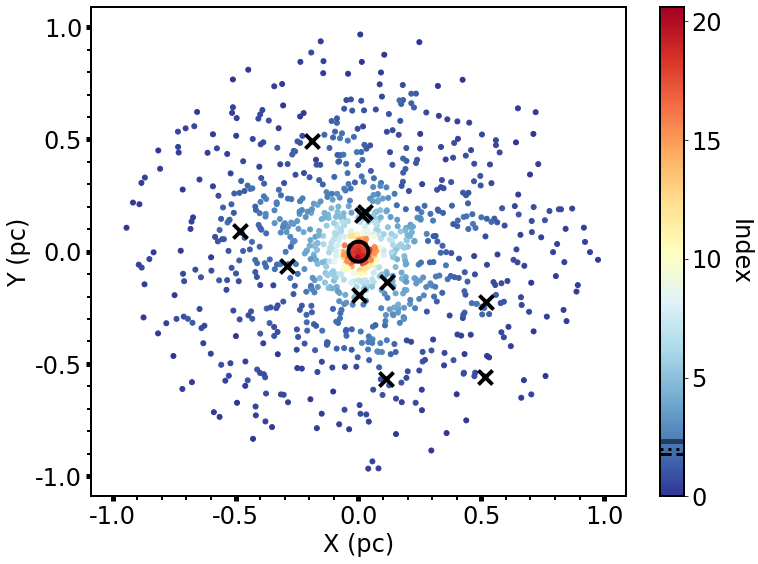}}
    \hspace{0.8pt}
  \subfigure[INDICATE, hmhi]{\includegraphics[width=0.32\linewidth]{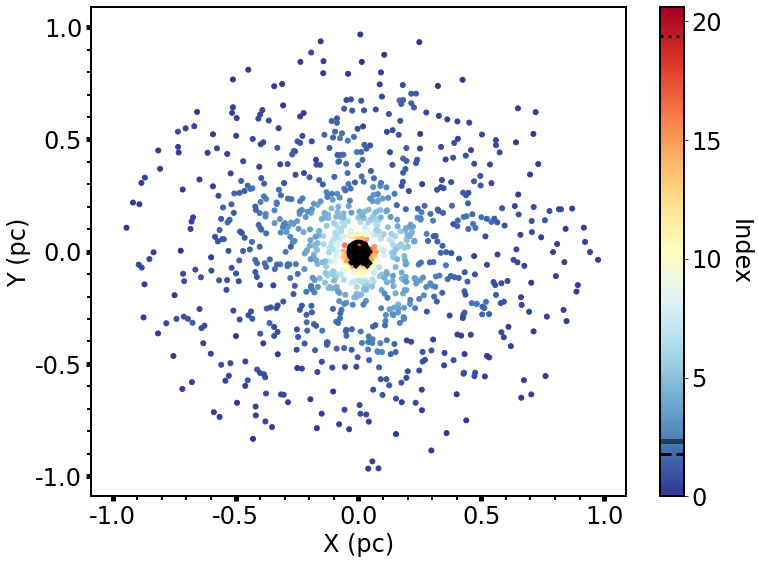}}
    \hspace{0.8pt}
  \subfigure[INDICATE, hmc]{\includegraphics[width=0.32\linewidth]{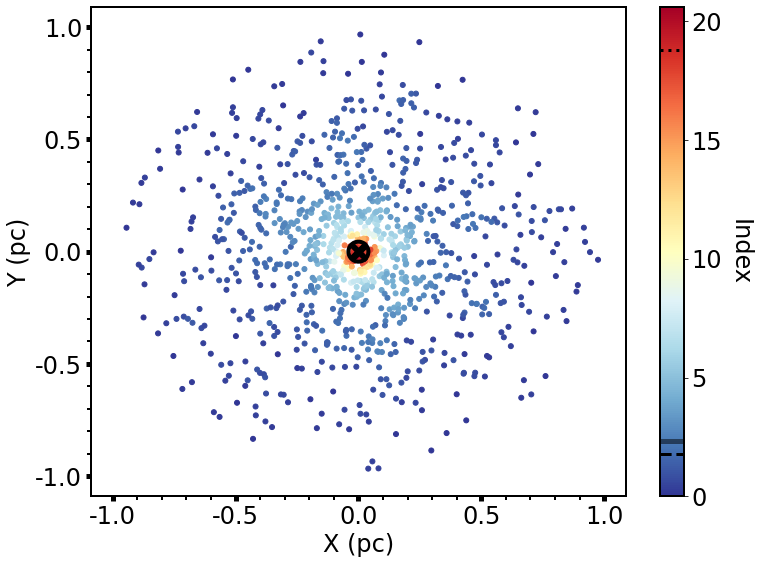}}
    \hspace{0.8pt}
  \subfigure[$\Sigma - m$, m]{\includegraphics[width=0.32\linewidth]{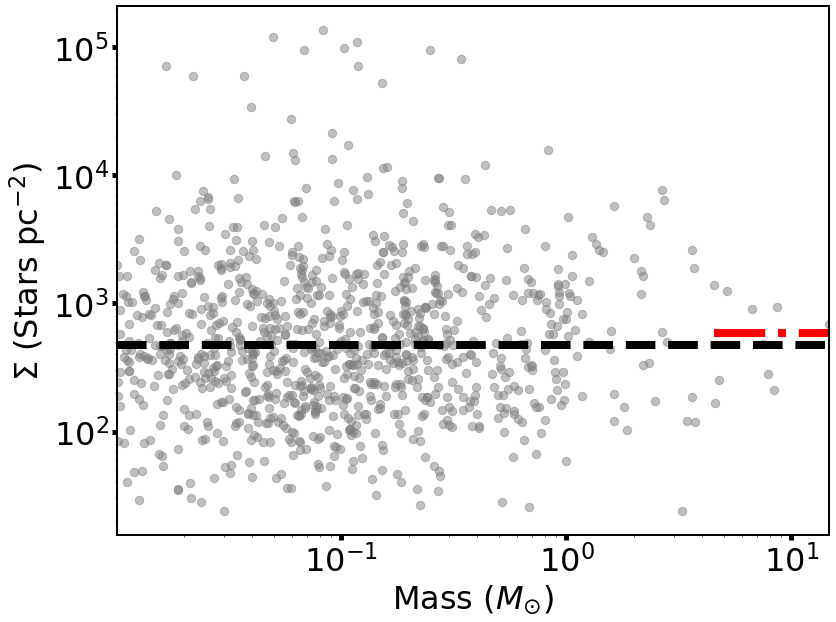}}
   \hspace{0.8pt}
  \subfigure[$\Sigma - m$, hmhi]{\includegraphics[width=0.32\linewidth]{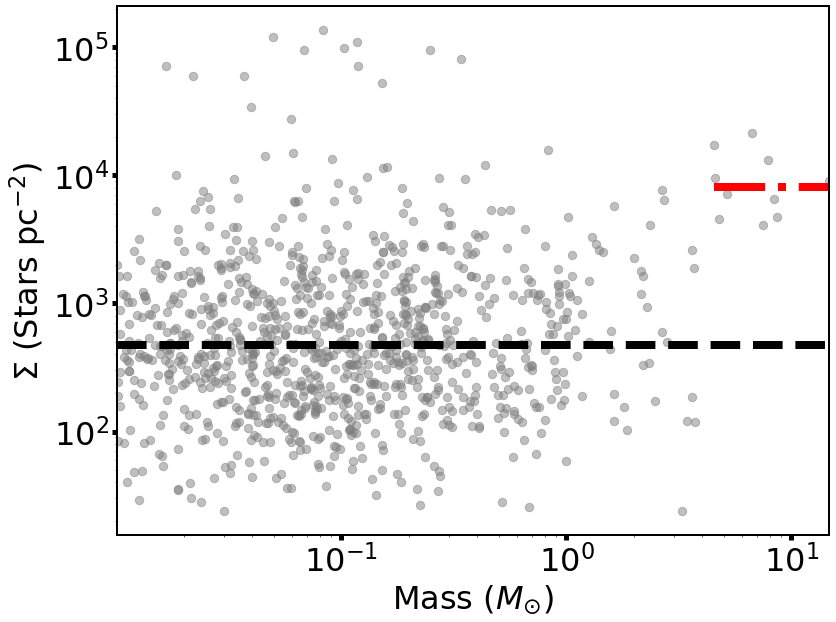}}
   \hspace{0.8pt}
  \subfigure[$\Sigma - m$, hmc]{\includegraphics[width=0.32\linewidth]{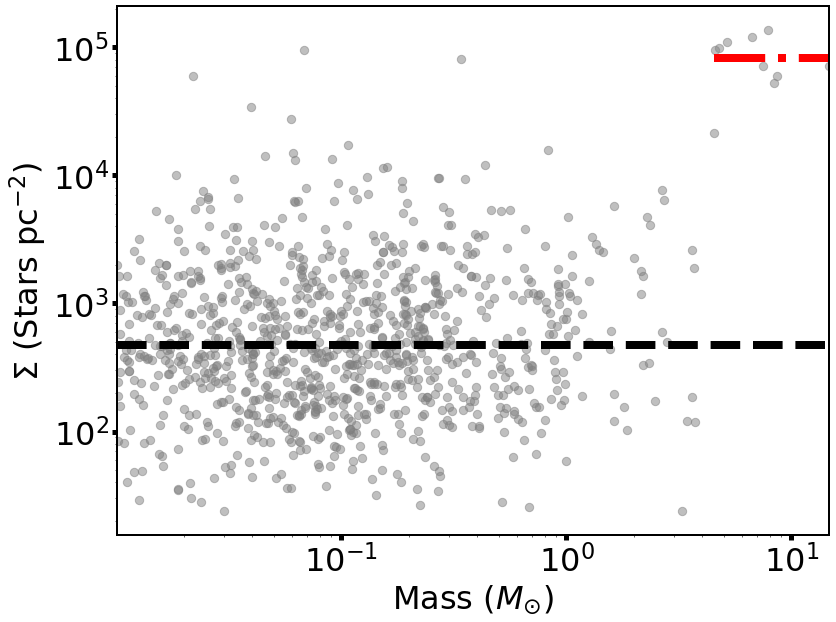}}
   \hspace{0.8pt}
  \subfigure[$\Lambda_{\rm{MSR}}$, m]{\includegraphics[width=0.32\linewidth]{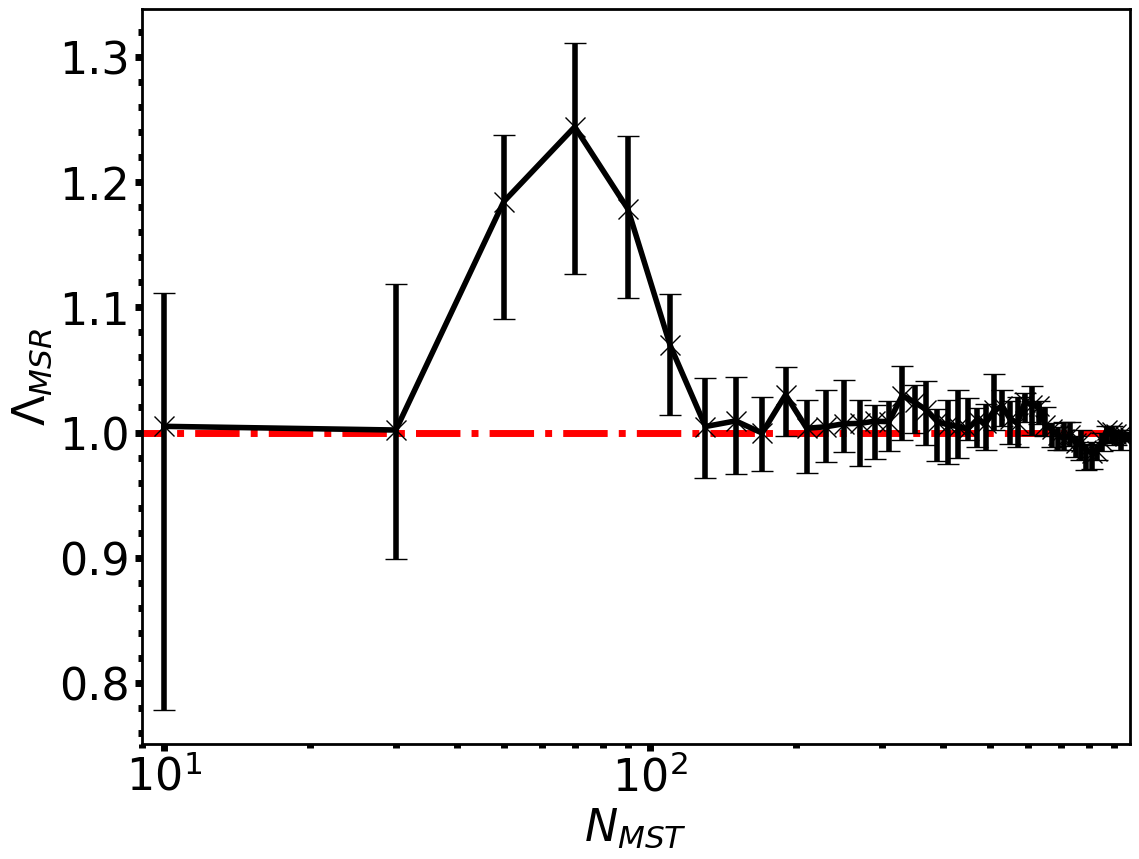}}
   \hspace{0.8pt}
  \subfigure[$\Lambda_{\rm{MSR}}$, hmhi]{\includegraphics[width=0.32\linewidth]{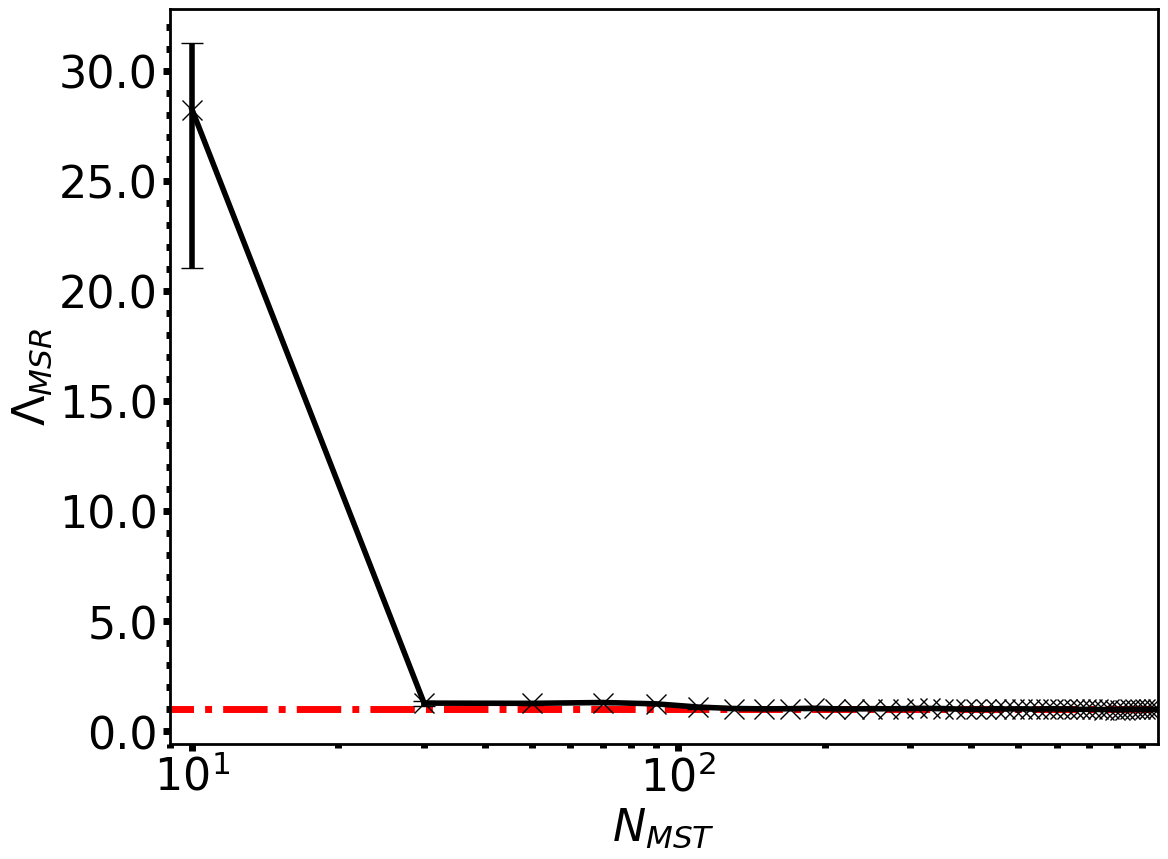}}
   \hspace{0.8pt}
  \subfigure[$\Lambda_{\rm{MSR}}$, hmc]{\includegraphics[width=0.32\linewidth]{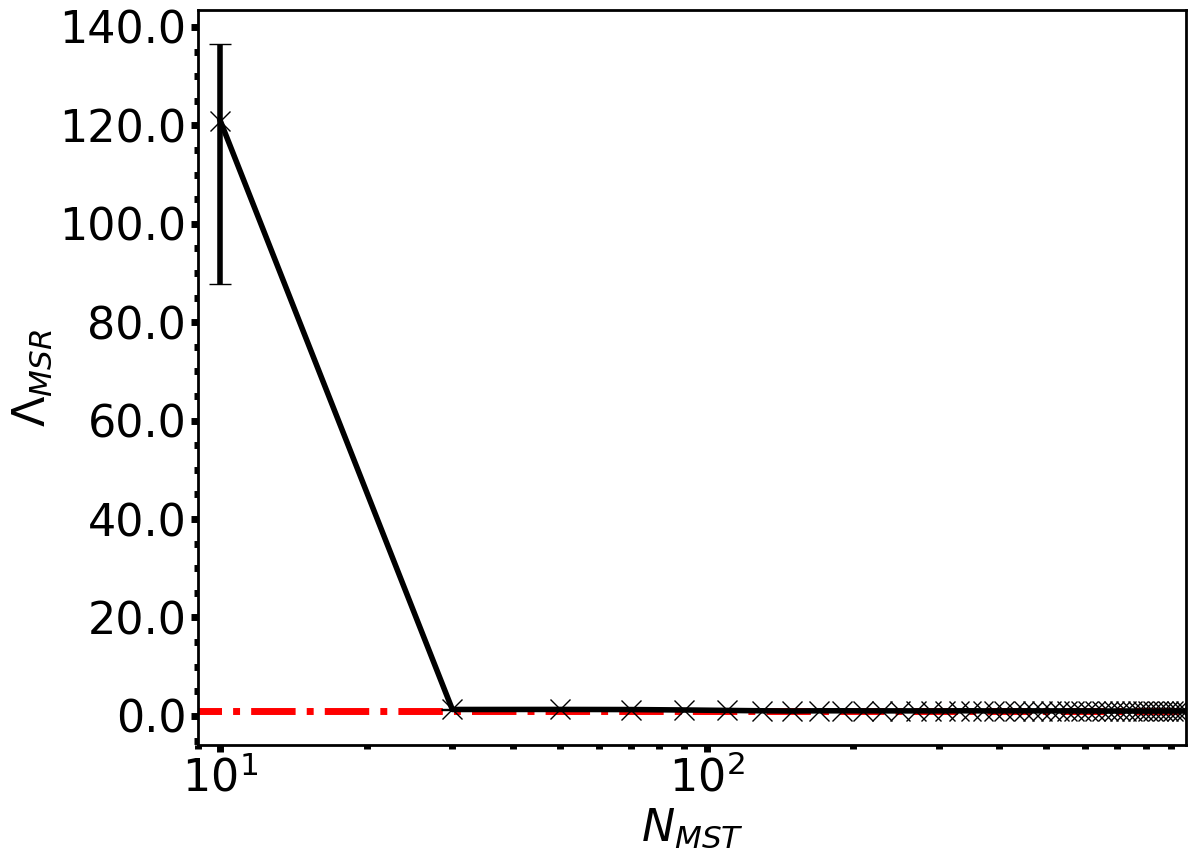}}
   \hspace{0.8pt}
  \subfigure[Radial distribution, m]{\includegraphics[width=0.32\linewidth]{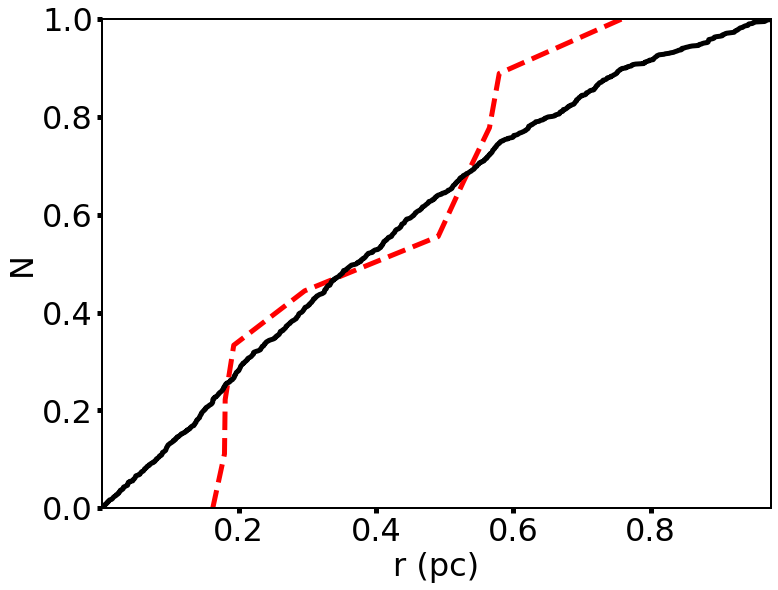}}
   \hspace{0.8pt}
  \subfigure[Radial distribtuion, hmhi]{\includegraphics[width=0.32\linewidth]{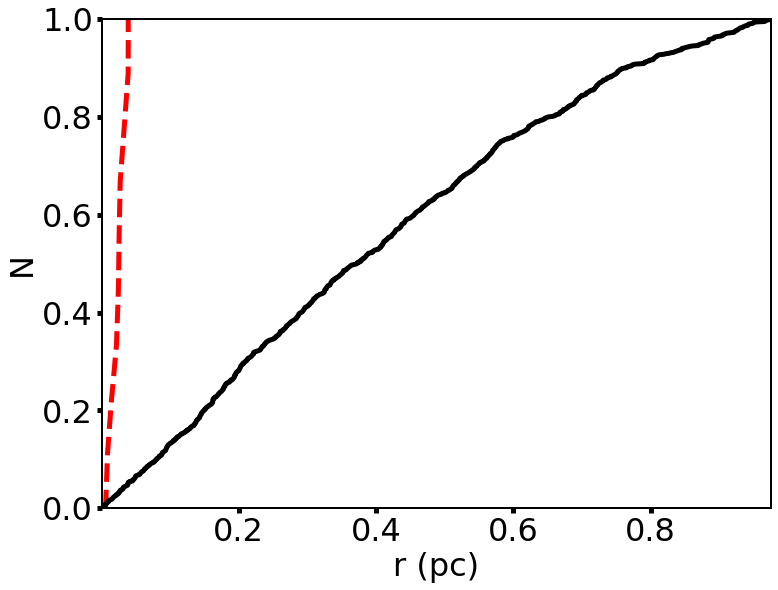}}
   \hspace{0.8pt}
  \subfigure[Radial distribution, hmc]{\includegraphics[width=0.32\linewidth]{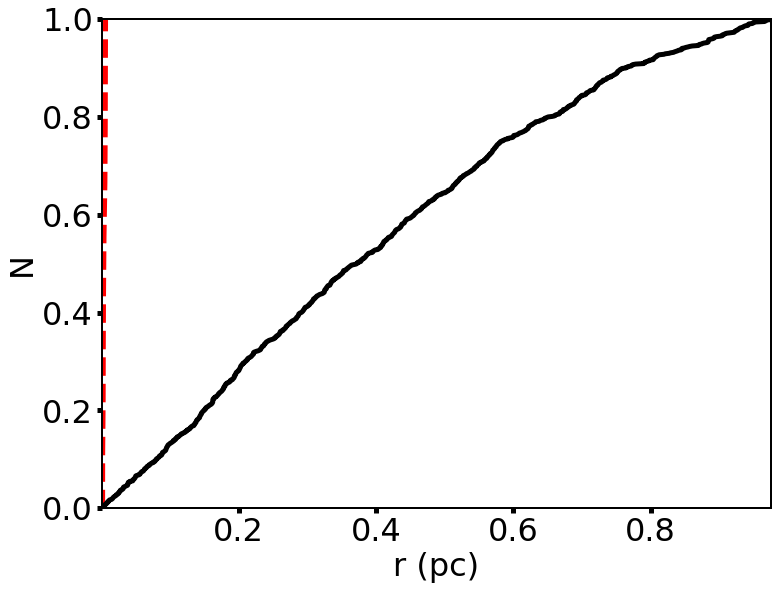}}
  \caption{A synthetic, centrally concentrated star-forming region of 1000 stars with radial density exponent $\alpha = 2.0$. The rows are (i) the INDICATE values, (ii) the $\Sigma$-m plots, (iii) the $\Lambda_{\rm{MSR}}$ plots and (iv) the cumulative distributions of radial distances from the centre of the star-forming region. From left to right the columns are the region with randomly assigned masses (m), highest mass stars moved to locations of highest INDICATE index (hmhi) and highest mass stars moved to centre of the region (hmc). Panels (a), (b) and (c) show the indexes found for each star, in the colour bar the significant INDICATE index is shown with the solid black line, the median index for the entire region is shown by the dashed black line and the median index for the 10 most massive stars is shown with the dotted black line. The 10 most massive stars in (a), (b) and (c) are highlighted with black crosses. The centre of the region is located in the middle of the black ring. In (d), (e) and (f) the median surface density of the region is shown by the black dashed line, the median surface density of the 10 most massive stars is shown by the red dash-dotted line. In (g), (h) and (i) the mass segregation ratio is shown by the black line and the horizontal dashed dotted red line shows the value of 1 corresponding to no mass segregation. In (j), (k) and (l) the black line and represents the CDF of radial distances from the centre of the region for all the stars, the red dashed line is the CDF for the 10 most massive stars.}
  \label{fig:results_radial}
\end{figure*}

Figure~\ref{fig:results_radial}(c) highlights that the most massive stars are now closer to each other after being swapped with the stars closest to the origin. The median index for the 10 most massive stars is $18.8_{-0.0}^{+0.2}$, larger than the median index for the region of $1.8_{-1.0}^{+3.0}$, meaning the 10 most massive stars find themselves in locations of greater than average stellar affiliation. A KS test returns a p-value $\ll 0.01$ finding a significant difference in the clustering tendencies of the 10 most massive stars compared to the rest. 

We apply INDICATE to the 50 most massive stars and find that 62 per cent of them have indexes above the significant index of 2.1, with a median index of $5.6_{-4.9}^{+0.4}$ for the entire region and $6.0_{-0.0}^{+0.0}$ for the 10 most massive stars. The median index found for the significantly clustered stars is again $6.0_{-0.2}^{+0.0}$. A maximum INDICATE index of 6.2 is found for the 50 most massive stars. The median index for the 10 most massive stars is above the significant index and so the region is found to be mass segregated by INDICATE. A KS test returns a p-value $\ll 0.01$ implying a significant difference between the 10 most massive stars' spatial clustering and the 50 most massive stars.

In figure~\ref{fig:results_radial}(f) the most massive stars find themselves in areas of much higher surface density than when they were moved to areas of greatest INDICATE index. $\Sigma_{\rm{LDR}}$ increases from 17 to $\Sigma_{\rm{LDR}} = 174.72$ with a p-value $\ll$ 0.01. As the median INDICATE index for the 10 most massive stars has decreased to 18.8 from 19.4 it demonstrates that INDICATE is not measuring the exact same quantity as the local stellar surface density, because if it was we would expect the median index of the 10 most massive stars to be larger than 19.4. This may be because the origin of the region also happens to be located in the area of greatest local stellar density, which is not the case for the substructured regions. 

$\Lambda_{\rm{MSR}} = {120.90}_{\,-\,33.05}^{\,+\,15.71}$, signifying significant mass segregation for the 10 most massive stars (see figure~\ref{fig:results_radial}(i)).

Figure~\ref{fig:results_radial}(l) shows that the cumulative distribution of the positions of the 10 most massive stars are now much closer to the centre of the region than before. A KS test returns a p-value $\ll 0.01$, implying that there is a significant difference in the spatial distribution of the 10 most massive stars compared to the rest.

\subsection{Uniform Star-Forming Regions}
\label{sec:indicate uniform results}

\subsubsection{Random Masses}

Figure~\ref{fig:results_uniform_cluster}(a) shows the INDICATE results for a uniform distribution with randomly assigned masses. The maximum index is 2.4 with a median INDICATE index for the entire region of $1.0_{-0.4}^{+0.2}$ and $0.8_{-0.0}^{+0.4}$ for the 10 most massive stars. There is no significant difference between the 10 most massive stars and the rest of the region with a p-value $= 0.68$. In this region one star has an INDICATE index equal to the significant index of 2.4.

Applying INDICATE to just the 50 most massive stars we find that no stars have an index above the significant index of 2.1. The median index for the region is $0.8_{-0.4}^{+0.2}$ and the median index of the 10 most massive stars is $0.8_{-0.2}^{+0.3}$. A maximum INDICATE index of 1.4 is found for the 50 most massive stars. As the median for the 10 most massive stars is below the significant index INDICATE detects no mass segregation in the region. A KS test returns a p-value $= 1.00$, confirming that the 10 most massive stars and the entire subset have similar clustering tendencies.

In figure~\ref{fig:results_uniform_cluster}(d) the most massive stars find themselves in similar areas of surface density as the rest of the stars in the region with $\Sigma_{\rm{LDR}} = 0.94$. No significant difference is detected with a KS test returning a p-value $= 0.59$.

Figure~\ref{fig:results_uniform_cluster}(g) shows, as expected, a very weak signal of $\Lambda_{\rm{MSR}} = {1.02}_{\,-\,0.09}^{\,+\,0.11}$ meaning no mass segregation. 

Figure~\ref{fig:results_uniform_cluster}(j) shows the cumulative distribution of positions starting further out for the 10 most massive stars than for the entire region but quickly matches the overall distribution. No significant difference is detected with a p-value $= 0.60$. 

\subsubsection{High Mass High Index}
We now move the 10 most massive stars to the areas of highest INDICATE index (figure~\ref{fig:results_uniform_cluster}(b)). Unlike in the centrally concentrated and fractal star-forming regions there is more than one region with relatively high indexes. The median index for the 10 most massive stars has increased from $0.8_{-0.0}^{+0.4}$ to $2.2_{-0.2}^{+0.0}$ with a p-value $\ll 0.01$ when comparing the 10 most massive stars to the entire region, suggesting a significant difference. As both the median index for the region and the median index for the 10 most massive stars are below the significant index of 2.4 all the stars still have a random spatial distribution according to INDICATE.

We apply INDICATE to just the 50 most massive stars and find that no stars have an index above the significant index of 2.1. The median index of the entire subset is $0.7_{-0.3}^{+0.5}$ and for the 10 most massive it is $1.1_{-0.4}^{+0.1}$. A maximum INDICATE index of 1.4 is once again found for the 50 most massive stars. As the median index for the 10 most massive stars is below the significant index INDICATE detects no mass segregation. A KS test returns a p-value $= 0.67$ implying no difference in the distribution of the 10 most massive stars compared to the entire subset.

Figure~\ref{fig:results_uniform_cluster}(e) shows the median local stellar surface density of the 10 most massive stars is greater than the median local stellar surface density of the region, with $\Sigma_{\rm{LDR}} = 2.11$ (increasing from 0.94) and a KS test returns a p-value $\ll 0.01$, meaning a significant difference in the local stellar surface density of the 10 most massive stars and all the stars.

Figure~\ref{fig:results_uniform_cluster}(h) shows that $\Lambda_{\rm{MSR}} = {1.32}_{\,-\,0.07}^{\,+\,0.23}$ for the 10 most massive stars suggesting that weak mass segregation is being detected, before decreasing as more stars are added to the subset. This is due to the random picking of stars for the subset MST. In \citet{parker_comparisons_2015} they suggest ignoring results of $\Lambda_{\rm{MSR}} < 2$ to avoid false positives such as this.

In figure~\ref{fig:results_uniform_cluster}(k) the cumulative distributions of positions are shown for the 10 most massive stars in red and all stars in black. A KS test returns a p-value $= 0.19$ showing no significant difference.

\subsubsection{High Mass Centre}
\begin{figure*}
  \subfigure[INDICATE, m]{\includegraphics[width=0.32\linewidth]{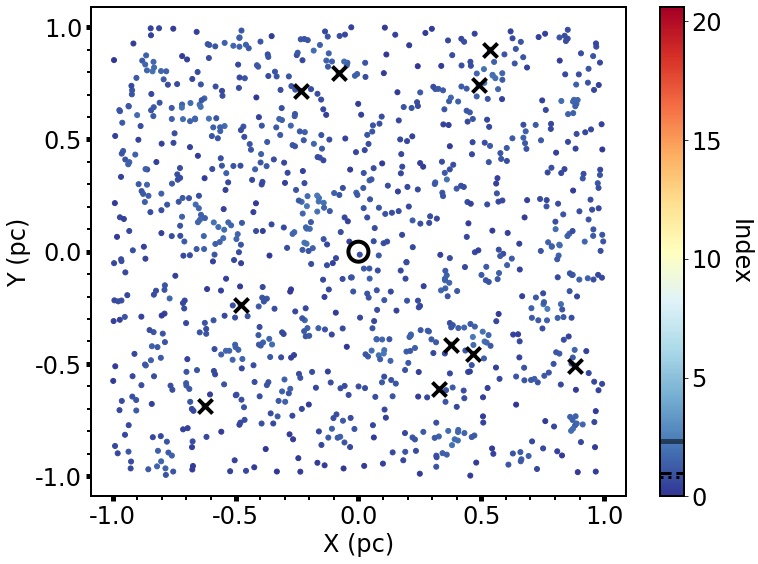}}
   \hspace{0.8pt}
  \subfigure[INDICATE, hmhi]{\includegraphics[width=0.32\linewidth]{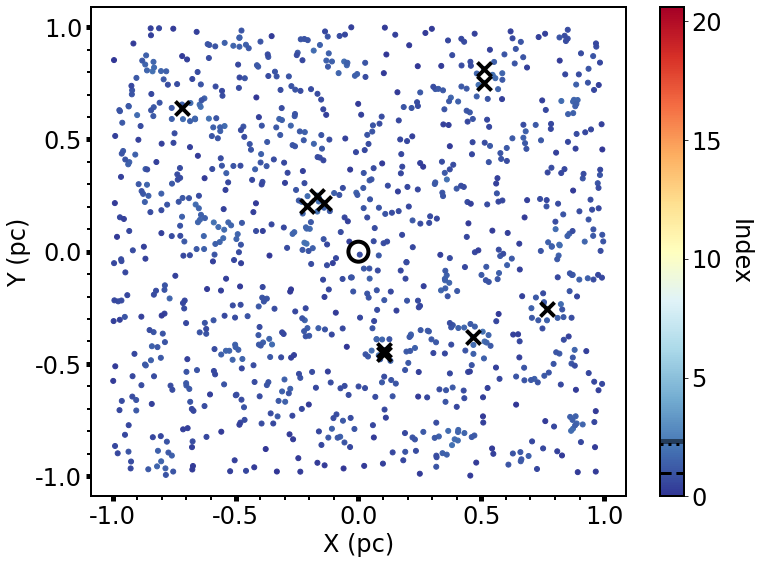}}
   \hspace{0.8pt}
  \subfigure[INDICATE, hmc]{\includegraphics[width=0.32\linewidth]{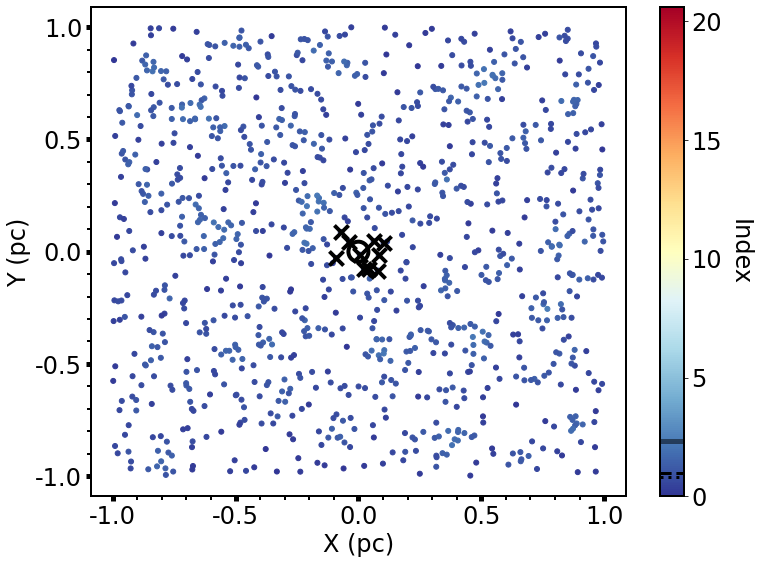}}
   \hspace{0.8pt}
  \subfigure[$\Sigma - m$, m]{\includegraphics[width=0.32\linewidth]{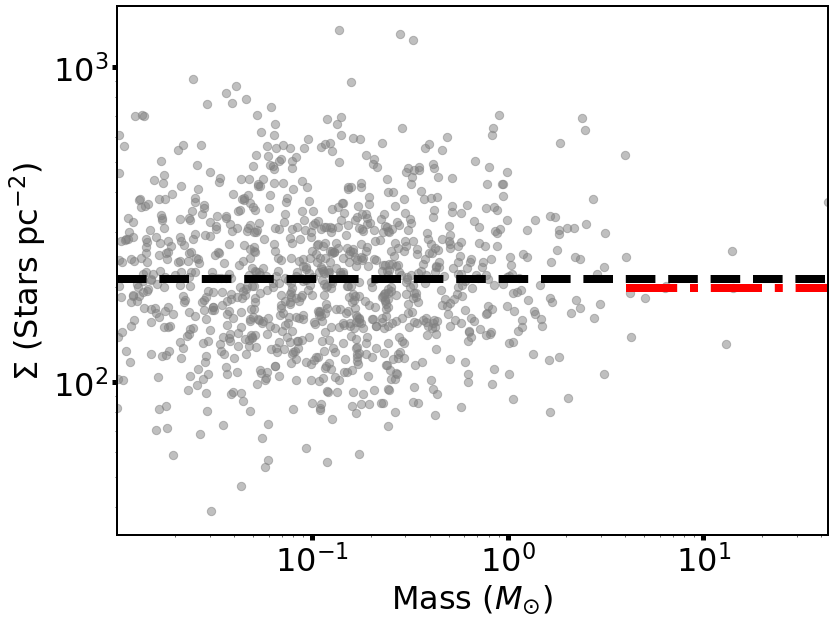}}
   \hspace{0.8pt}
  \subfigure[$\Sigma - m$, hmhi]{\includegraphics[width=0.32\linewidth]{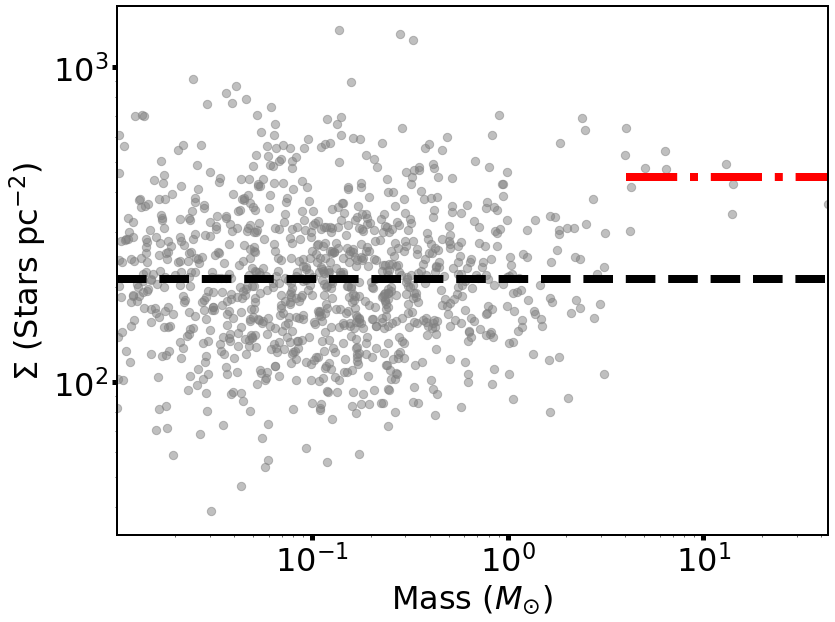}}
   \hspace{0.8pt}
  \subfigure[$\Sigma - m$, hmc]{\includegraphics[width=0.32\linewidth]{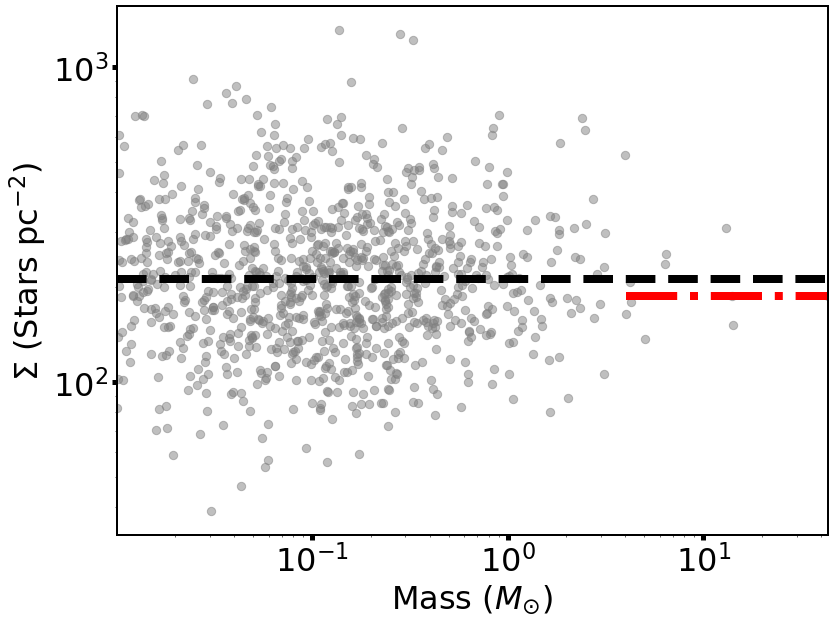}}
   \hspace{0.8pt}
  \subfigure[$\Lambda_{\rm{MSR}}$, m]{\includegraphics[width=0.32\linewidth]{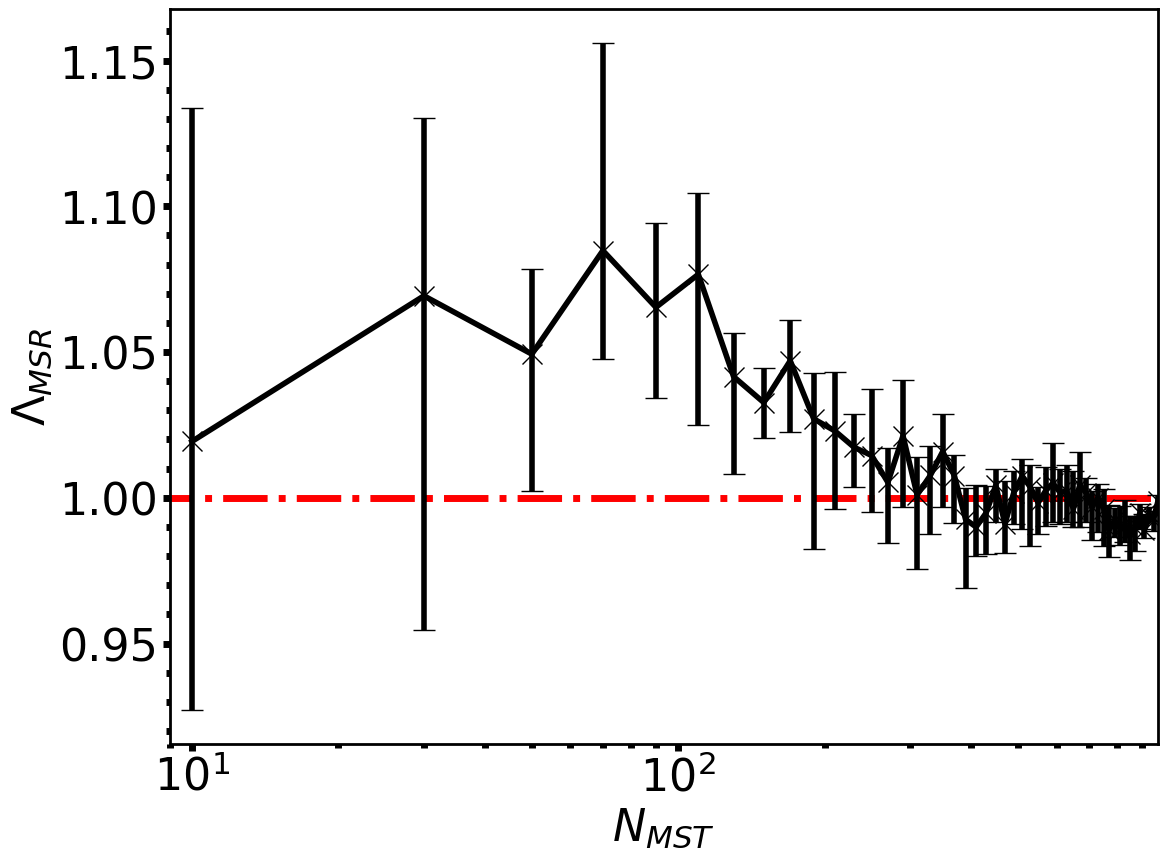}}
   \hspace{0.8pt}
  \subfigure[$\Lambda_{\rm{MSR}}$, hmhi]{\includegraphics[width=0.32\linewidth]{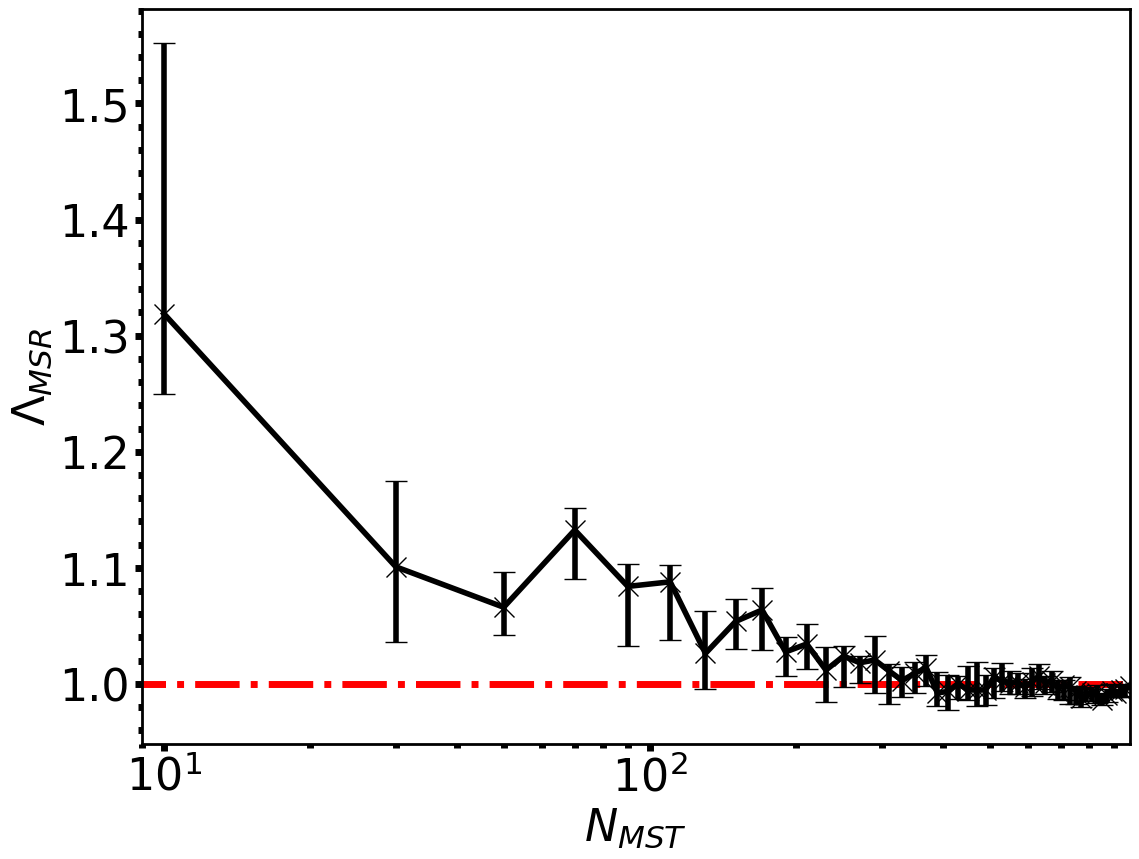}}
   \hspace{0.8pt}
  \subfigure[$\Lambda_{\rm{MSR}}$, hmc]{\includegraphics[width=0.32\linewidth]{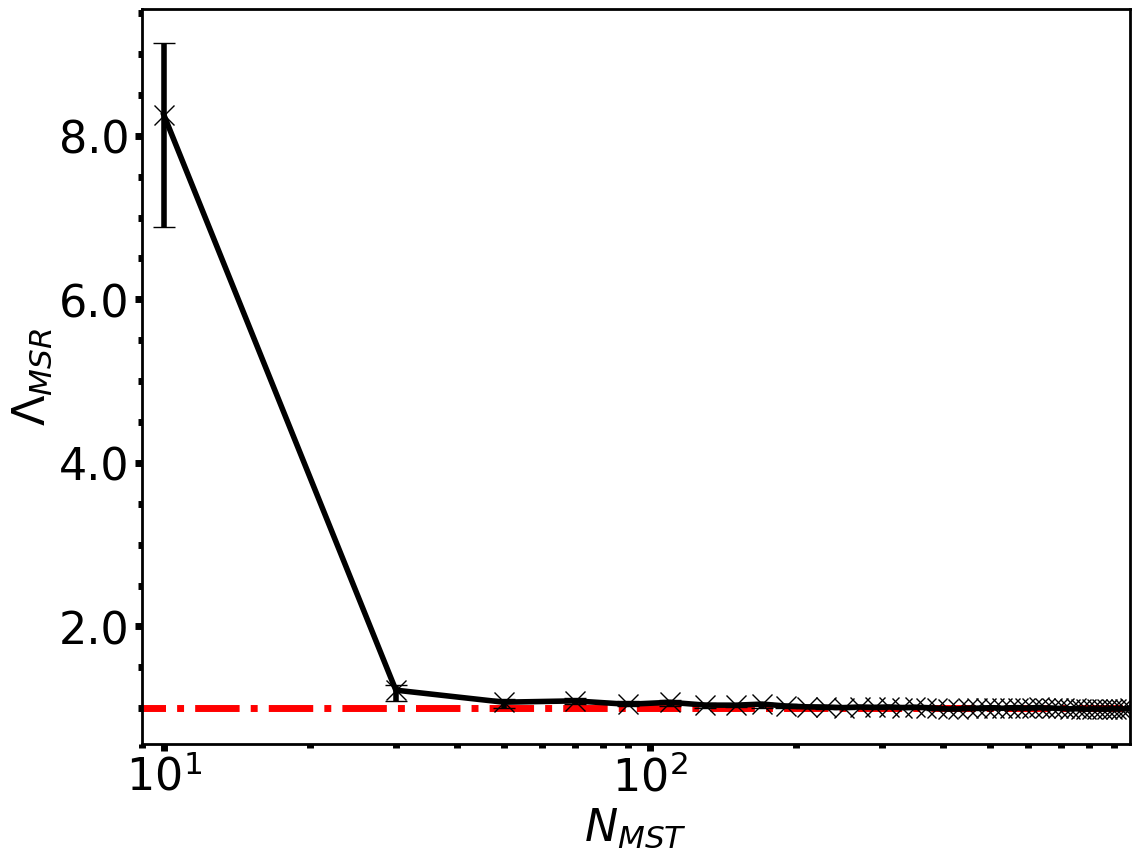}}
  \hspace{0.8pt}
  \subfigure[Radial distributions, m]{\includegraphics[width=0.32\linewidth]{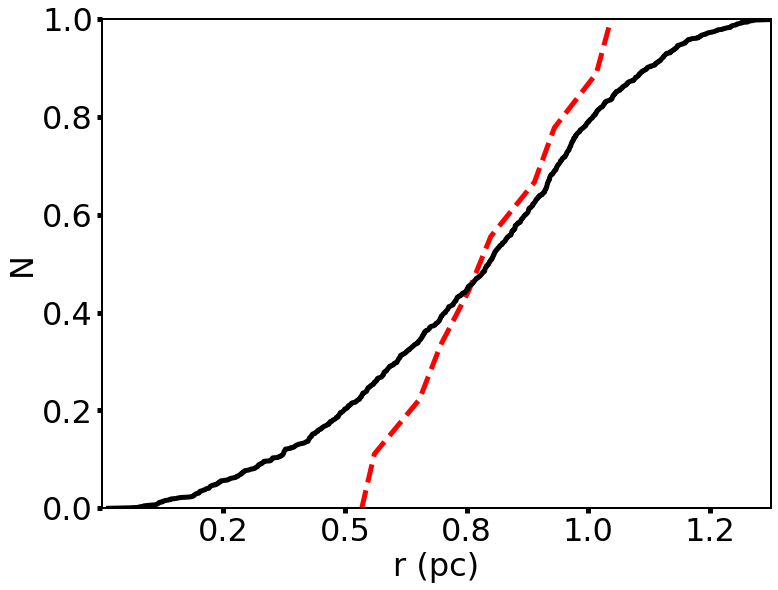}}
   \hspace{0.8pt}
  \subfigure[Radial distributions, hmhi]{\includegraphics[width=0.32\linewidth]{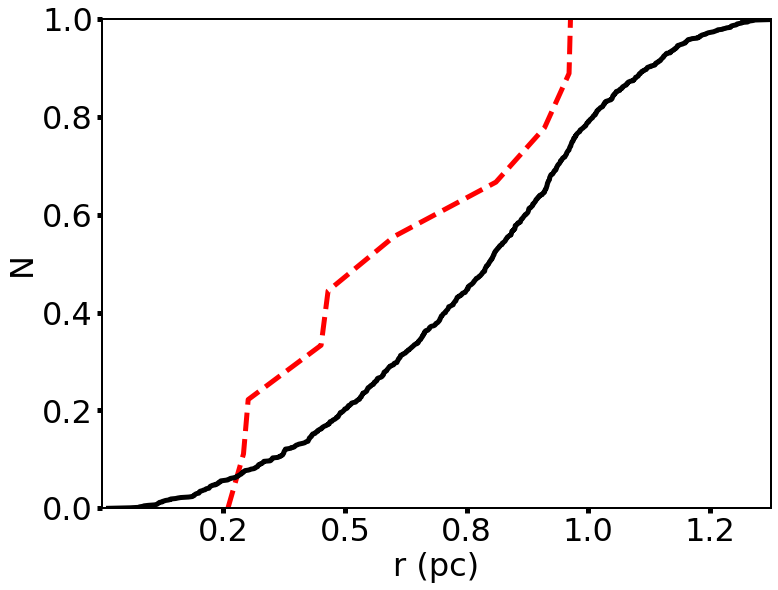}}
   \hspace{0.8pt}
  \subfigure[Radial distributions, hmc]{\includegraphics[width=0.32\linewidth]{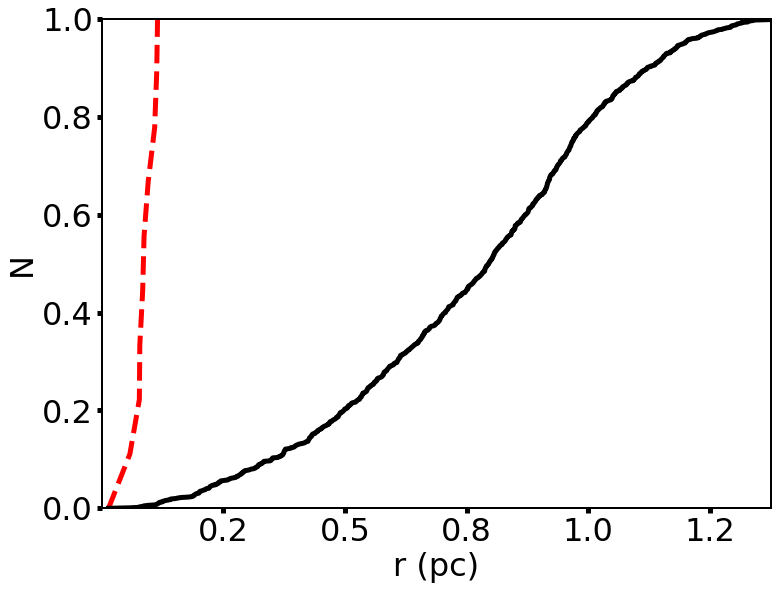}}
  \caption{Uniform distribution of 1000 stars. The rows are (i) the INDICATE values, (ii) the $\Sigma$-m plots, (iii) the $\Lambda_{\rm{MSR}}$ plots and (iv) the cumulative distributions of radial distances from the centre of the star-forming region. From left to right, we show the distribution with randomly assigned masses (m), highest mass moved to highest INDICATE index (hmhi), highest masses moved to centre (hmc). The colour bars in panels (a), (b) and (c) show the INDICATE index, with the solid black line representing the significant index, the dashed black line represents the median index for the region and the dotted black line represents the median index of the 10 most massive stars. The 10 most massive stars in (a),(b) and (c) are highlighted with black crosses. The centre of the region is located in the middle of the black ring. In (d), (e) and (f) the median surface density of all stars is shown by the black dashed line, the median surface density of the 10 most massive stars is shown by the red dash-dotted line. In (g), (h) and (i) the mass segregation ratio is shown by the black line and the horizontal dashed dotted red line shows the value of 1 corresponding to no mass segregation. In (j), (k) and (l) the black line represents the CDF of radial distance from the centre for all of the stars and the red dashed line is the CDF for the 10 most massive stars.}
  \label{fig:results_uniform_cluster}
\end{figure*}

Now the 10 most massive stars are swapped with the 10 most central stars and this is shown in figure~\ref{fig:results_uniform_cluster}(c).
The median index for 10 most massive stars is $0.8_{-0.0}^{+0.1}$ (the same as when masses are randomly assigned) and similarly there is no significant difference in the spatial clustering of the most massive stars and the rest of the stars with a \mbox{p-value $=0.64$}.

We apply INDICATE to just the 50 most massive stars and find that 10 per cent of stars have an index above the significant index of 2.1. The median index for the significantly clustered stars is $2.4_{-0.2}^{+0.0}$. The median index for the region is found to be $0.8_{-0.4}^{+0.4}$ and $2.0_{-0.4}^{+0.4}$ for the 10 most massive stars. A maximum INDICATE index of 2.4 is found for the 50 most massive stars. As the median index for the 10 most massive stars is below the significant index according to INDICATE they are randomly distributed and so no mass segregation has been detected. %A KS test returns a p-value $\ll 0.01$ meaning the 10 most massive stars and the rest are distributed differently as the 10 most massive stars are closer together than would be expected from a random uniform distribution.

Figure~\ref{fig:results_uniform_cluster}(f) shows the local surface density against mass plot. We find $\Sigma_{\rm{LDR}} = 0.88$ (lower than when masses are swapped with stars of greatest INDICATE index) with a KS test giving a p-value $= 0.49$ implying no significant differences in the surface density of the most massive stars compared to all the stars.

Figure~\ref{fig:results_uniform_cluster}(i) shows the $\Lambda_{\rm{MSR}}$ results and a lower value is found than for other examples with the highest masses moved to the centre. $\Lambda_{\rm{MSR}} = {8.25}_{\,-\,1.36}^{\,+\,0.88}$ meaning mass segregation is detected for the 10 most massive stars and this quickly drops off as the rest of the stars are uniformly distributed. This is opposite to the INDICATE result which finds no mass segregation using the given criteria but does find a significant difference in the clustering of the 10 most massive stars compared to the entire subset. 
Figure~\ref{fig:results_uniform_cluster}(l) shows the radial cumulative distributions of positions. The 10 most massive stars show a similar trend as the fractal and smooth star-forming regions, with a much steeper function when the most massive stars are swapped with the most central stars. A KS test returns a p-value $\ll 0.01$.

\subsection{Summary}
The INDICATE method has clearly identified regions of clustering in the synthetic datasets. 
INDICATE gives results that are in agreement with $\Sigma_{\rm{LDR}}$ when applied to the entire region and results that are generally in agreement with $\Lambda_{\rm{MSR}}$ when applied to only the 50 most massive stars in the synthetic star forming regions. The INDICATE results when applied to all 1000 stars in a region are summarised in table~\ref{tab:indicate_results} and results when only applied to the 50 most massive stars are shown in table~\ref{tab:indicate_results_50_most_massive}. The results for the $\Sigma_{\rm{LDR}}$, $\Lambda_{\rm{MSR}}$ and CDF methods are summarised in table~\ref{tab:other_results}.

{\renewcommand{\arraystretch}{1.5}
\begin{table}
    \centering
    \caption{Results of INDICATE being applied to all stars in the synthetic star-forming regions. From left to right the columns are: the median index for all stars in the region, the median index for the 10 most massive stars, the significant index and the p-value returned from a KS test comparing the indexes between the 10 most massive stars and all stars in the region. The null hypothesis is rejected when p-value $\ll 0.01$.}
    \begin{tabular}{l|cccccr}
        \hline
        Region & $\Tilde{I}_{\rm{all}}$ & $\Tilde{I}_{\rm{10}}$ & $I_{\rm{sig}}$ & $\%>I_{\rm{sig}}$ & p\\ \hline
        $D=1.6$, m         & $4.4_{-1.4}^{+2.6}$      & $4.5_{-0.6}^{+3.3}$          & 2.3        & 82.2       & 0.90\\
        $D=1.6$, hmhi      & $4.4_{-1.4}^{+2.6}$     & $15.3_{-0.4}^{+0.2}$         & 2.3        & 82.2       & $\ll 0.01$\\
        $D=1.6$, hmc       & $4.4_{-1.4}^{+2.6}$      & $2.2_{-0.0}^{+0.0}$          & 2.3        & 82.2       & $\ll 0.01$\\
        $\alpha=2.0$, m    & $1.8_{-1.0}^{+3.0}$      & $2.0_{-0.8}^{+2.4}$          & 2.3        & 44.1       & 0.55\\
        $\alpha=2.0$, hmhi & $1.8_{-1.0}^{+3.0}$      & $19.4_{-0.0}^{+0.8}$         & 2.3        & 44.1       & $\ll 0.01$\\
        $\alpha=2.0$, hmc  & $1.8_{-1.0}^{+3.0}$      & $18.8_{-0.0}^{+0.2}$         & 2.3        & 44.1       & $\ll 0.01$\\
        Uniform, m         & $1.0_{-0.4}^{+0.2}$      & $0.8_{-0.0}^{+0.4}$          & 2.4        & 0.0        & 1.00\\
        Uniform, hmhi      & $1.0_{-0.4}^{+0.2}$      & $2.2_{-0.2}^{+0.0}$          & 2.4        & 0.0        & $\ll 0.01$\\
        Uniform, hmc       & $1.0_{-0.4}^{+0.2}$      & $0.8_{-0.0}^{+0.1}$          & 2.4        & 0.0        & 0.64\\
         
        \hline
    \end{tabular}
   
    \label{tab:indicate_results}
\end{table}}

{\renewcommand{\arraystretch}{1.5}
\begin{table}
    \centering
    \caption{Results of applying INDICATE to just the 50 most massive stars in the synthetic regions. From left to right the columns are: the median index for all 50 stars, the median index for the 10 most massive stars, the significant index, the percentage of stars with indexes greater than the significant index and the p-value from a KS test between all 50 stars and the 10 most massive stars.}
    \begin{tabular}{l|ccccc}
        \hline
        Region & $\Tilde{I}_{\rm{50}}$ & $\Tilde{I}_{\rm{10}}$ & $I_{\rm{sig}}$ & $\%>I_{\rm{sig}}$ & p\\ \hline
        $D=1.6$, m         & $1.4_{-0.6}^{+0.6}$     & $1.6_{-0.8}^{+0.4} $      & 2.1        & 22     & 1.00\\
        $D=1.6$, hmhi      & $1.3_{-0.5}^{+2.3}$     & $3.6_{-0.0}^{+0.0} $      & 2.1        & 38     & $\ll 0.01$\\
        $D=1.6$, hmc       & $1.4_{-0.4}^{+0.8}$    & $2.6_{-0.2}^{+0.0}  $     & 2.1        & 28     & $\ll 0.01$\\
        $\alpha=2.0$, m    & $2.6_{-2.0}^{+1.8}$     & $1.1_{-0.7}^{+2.0} $      & 2.1        & 52     & 0.48\\
        $\alpha=2.0$, hmhi & $5.7_{-5.0}^{+0.3}$    & $6.0_{-0.0}^{+0.2}  $     & 2.1        & 64     & $\ll 0.01$\\
        $\alpha=2.0$, hmc  & $5.6_{-4.9}^{+0.4}$     & $6.0_{-0.0}^{+0.0} $      & 2.1        & 62     & $\ll 0.01$\\
        Uniform, m         & $0.8_{-0.4}^{+0.2}$     & $0.8_{-0.2}^{+0.3} $      & 2.1        & 0      & 1.00\\
        Uniform, hmhi      & $0.7_{-0.3}^{+0.5}$    & $1.1_{-0.4}^{+0.1}$       & 2.1        & 0      & 0.67\\
        Uniform, hmc       & $0.8_{-0.4}^{+0.4}$     & $2.0_{-0.4}^{+0.4} $      & 2.1        & 10     & 0.01\\
        \hline
    \end{tabular}
    \label{tab:indicate_results_50_most_massive}
\end{table}}

{\renewcommand{\arraystretch}{1.5}
\begin{table}
    \centering
    \caption{Results of the other methods being applied to all stars in the synthetic star-forming regions. From left to right the columns are: the local stellar surface density ratio, the p-value from a KS comparing the median local stellar surface density of the 10 most massive stars to the median local stellar surface density of the entire region, the mass segregation ratio and the p-value of a KS test comparing the CDF of positions of the 10 most massive stars and all the stars in each region.}
    \begin{tabular}{l|ccccl}
        \hline
        Region               & $\Sigma_{\rm{LDR}}$    & $\Sigma_{\rm{LDR}}$ (p) & $\Lambda_{\rm{MSR}}$ & CDF (p) \\ \hline
        $D=1.6$, m           & 1.30      &      0.69          &  $0.97_{\rm{\,-\,0.11}}^{\rm{\,+\,0.09}}$    & $\ll 0.01$ \\
        $D=1.6$, hmhi        & 2.30      & $\ll 0.01$         & $33.74_{\rm{\,-\,5.27}}^{\rm{\,+\,2.54}}$    & $\ll 0.01$ \\
        $D=1.6$, hmc         & 0.56      &      0.03          &   $8.87_{\rm{\,-\,0.78}}^{\rm{\,+\,0.74}}$   & $\ll 0.01$ \\
        $\alpha=2.0$, m      & 1.24      &      0.63          & $1.00_{\rm{\,-\,0.23}}^{\rm{\,+\,0.11}}$     &       0.67 \\
       $\alpha=2.0$, hmhi    & 17.00     & $\ll 0.01$         & $28.23_{\rm{\,-\,7.20}}^{\rm{\,+\,3.05}}$    & $\ll 0.01$ \\
       $\alpha=2.0$, hmc     & 174.72    & $\ll 0.01$         & $120.90_{\rm{\,-\,33.05}}^{\rm{\,+\,15.71}}$ & $\ll 0.01$ \\
        Uniform, m           & 0.94      &      0.59          &  $1.02_{\rm{\,-\,0.09}}^{\rm{\,+\,0.11}}$    &       0.60 \\
        Uniform, hmhi        & 2.11      & $\ll 0.01$         &  $1.32_{\rm{\,-\,0.07}}^{\rm{\,+\,0.23}}$    &       0.19 \\
        Uniform, hmc         & 0.88      &      0.49          &  $8.25_{\rm{\,-\,1.36}}^{\rm{\,+\,0.88}}$    & $\ll 0.01$ \\
         
        \hline
    \end{tabular}
   \label{tab:other_results}
\end{table}}

\section{Applying INDICATE to Observational Data}
\label{section:observational_data_indicate_results}
We now apply INDICATE to the following star-forming regions: Taurus, Orion Nebula Cluster (ONC), NGC1333, IC348 and $\rho$ Ophiuchi. The ONC may be incomplete due to saturation due to its high stellar density and therefore may be missing the most massive stars, however in \citet{hillenbrand_preliminary_1998} they find that the optical point source component of the ONC is incomplete but when combined with infrared data it is near complete (see $\S$~2 \citet{hillenbrand_preliminary_1998}).
For the ONC, NGC1333 and IC348 we removed stars without known masses when performing KS tests between the 10 most massive stars and all of the stars in the regions. 
The results of applying INDICATE to all points in the observational data sets are presented in table~\ref{tab:results_table_ks_ob}. The results of applying INDICATE to just the 50 most massive stars in these regions is shown in \textit{Appendix}~\ref{sec:classical_massseg_observational_data}.
{\renewcommand{\arraystretch}{1.5}
\begin{table}
	\centering
	\caption{Results of applying INDICATE to all stars in the observational regions. From the left to right the columns are: the median index for the all the stars, the median index for the 10 most massive stars in the region, the percentage of stars that are significantly clustered above random and the p-value from a KS test comparing the 10 most massive stars to all stars in each region.}
	\label{tab:results_table_ks_ob}
	\begin{tabular}{l|ccccl} % four columns, alignment for each
		\hline
		Name & $\Tilde{I}(all)$ & $\Tilde{I}(10)$ & $I_{\rm{sig}}$ & $\% > I_{\rm{sig}}$ & p \\
		\hline
		Taurus           & $6.6_{-1.6}^{+1.8}$    &  $3.6_{-0.2}^{+3.3} $    & 2.1    & 85.9    & 0.07  \\
		ONC              & $1.4_{-0.8}^{+3.0}$    & $10.3_{-5.8}^{+16.5} $   & 2.4    & 46.1    & 0.003 \\
		NGC1333          & $5.1_{-2.9}^{+2.9}$    &  $5.9_{-1.5}^{+2.4}  $   & 2.4    & 73.9    & 0.57  \\
		IC348            & $3.2_{-1.8}^{+3.5}$    &  $2.6_{-0.4}^{+8.3} $    & 2.3    & 59.2    & 0.63  \\
		$\rho$ Ophiuchi  & $1.8_{-1.0}^{+1.2}$    &  $1.3_{-0.7}^{+1.1} $    & 2.2    & 39.6    & 0.82  \\
		\hline
	\end{tabular}
\end{table}}

\subsection{Taurus}
The Taurus star-forming region is located 140 pc away with an estimated age of  around $1$ Myr \citep[][]{bell_pre-main-sequence_2013}. We use the dataset from \citet{parker_mass_2011}, entailing 361 objects. The masses of the 10 most massive stars are between $1.9 \, M_\odot$ and $4.1 \, M_\odot$ (masses are calculated in $\S$~2 of \citet{parker_mass_2011}). 
\begin{figure}
	\includegraphics[width=\columnwidth]{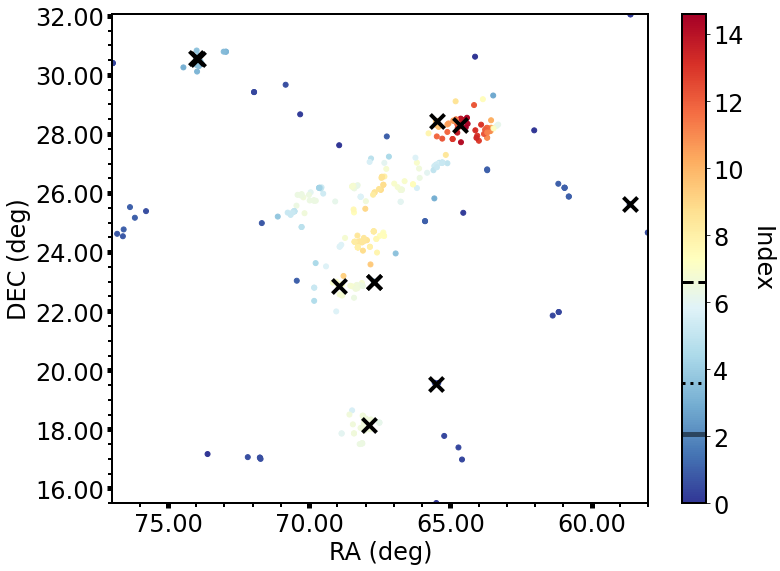}
    \caption{The Taurus star-forming region. The ten most massive stars are highlighted with black crosses. In the colour bar the significant INDICATE index is shown by the solid black line, the median index for all the stars is shown by the dashed black line and the median index for the 10 most massive stars is shown using the dotted black line.}
    \label{fig:fig_taurus_indicate}
\end{figure}
Previous studies of Taurus find a corresponding fractal dimension $D$ of $1.55 \pm 0.25$, inferred from the $Q$-parameter value 0.45 \citep{cartwright_statistical_2004}. In \citet{parker_mass_2011} they find a mass segregation ratio $\Lambda_{\rm{MSR}} = 0.70 \pm 0.10$ for the 20 most massive stars.

Figure~\ref{fig:fig_taurus_indicate} shows Taurus after INDICATE is applied, revealing that 85.9 per cent of stars are spatially clustered above random with indexes greater than the significant index of 2.1. The median index of significantly clustered stars is $6.8_{-0.8}^{+1.8}$. A maximum INDICATE index of 14.6 is found for the region. Taurus has a median index of $6.6_{-1.6}^{+1.8}$ for the entire region and $3.6_{-0.2}^{+3.3}$ for the 10 most massive stars. The most massive stars are highlighted with crosses in figure~\ref{fig:fig_taurus_indicate}, they are spread out across the star-forming region (see also \citet{parker_mass_2011}), with most lying in less clustered regions. INDICATE detects no significant difference in distribution of indexes between the most massive stars and all stars with a \mbox{p-value $= 0.07$}. 

\subsection{ONC}
\begin{figure}
	\includegraphics[width=\columnwidth]{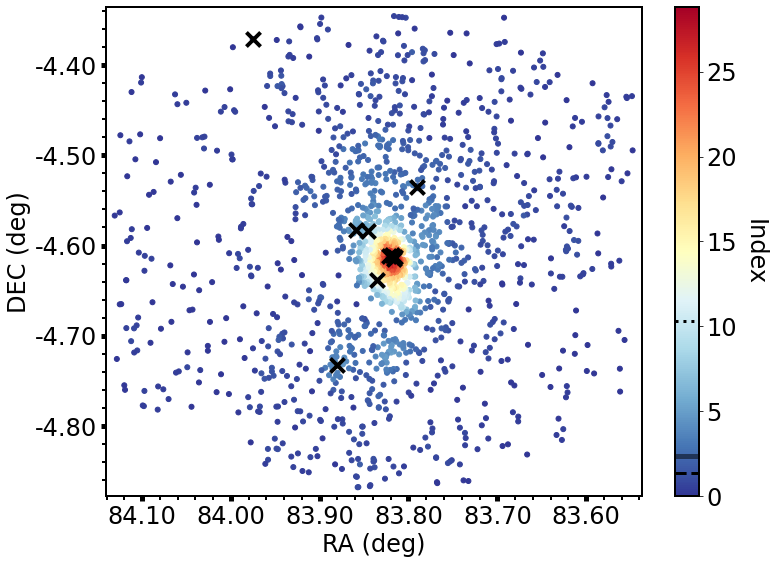}
    \caption{The Orion Nebula Cluster. The ten most massive stars are highlighted with black crosses. In the colour bar the significant index is shown with the solid black line, the median index of all the stars is shown by the dashed black line and the median index for the 10 most massive stars is shown by the dotted black line.}
    \label{fig:fig_onc_indicate}
\end{figure}

We use the dataset from \citet{hillenbrand_preliminary_1998} which contains 1576 objects. The ONC is a very dense centrally concentrated region shown in figure~\ref{fig:fig_onc_indicate}. The line of empty space to the south of the area of highest index that extends from the north-east to the south-west is due to a band of extinction. 641 objects do not have an assigned mass in the dataset and are therefore removed when comparing the indexes between the 10 most massive stars and the entire region. The masses of the 10 most massive stars are between $5.7 \, M_\odot$ and $45.7 \, M_\odot$.  The distance to the ONC is about 400~pc away with an estimated age of around 1~Myr \citep[][]{jeffries_no_2011, reggiani_quantitative_2011}. The mass segregation ratio of the ONC as found using the 4 most massive stars is $\Lambda_{\rm{MSR}} = 8.0 \pm 3.5$.

We individually determine the $Q$-parameter for stars with and without mass measurements, and all sample stars, respectively, and find that there is no significant difference in the $Q$-parameter, suggesting that the three subsets follow the same spatial distribution. The median index for the ONC is $1.4_{-0.8}^{+3.0}$, the median index for the 10 most massive stars is $10.3_{-5.8}^{+16.5}$ with 46.1 per cent of stars spatially clustered above random (the ONC has a significant index of 2.4). A median index of $8.6_{-4.8}^{+8.4}$ is found for significantly clustered stars. A maximum INDICATE index of 28.8 is found for the ONC. These results are similar to the synthetic region shown in figure~\ref{fig:results_radial}. A KS test gives a p-value of 0.003, below our chosen threshold of 0.01 meaning that the 10 most massive stars have different clustering tendencies when compared to the entire region.

\subsection{NGC 1333}
\begin{figure}
	\includegraphics[width=\columnwidth]{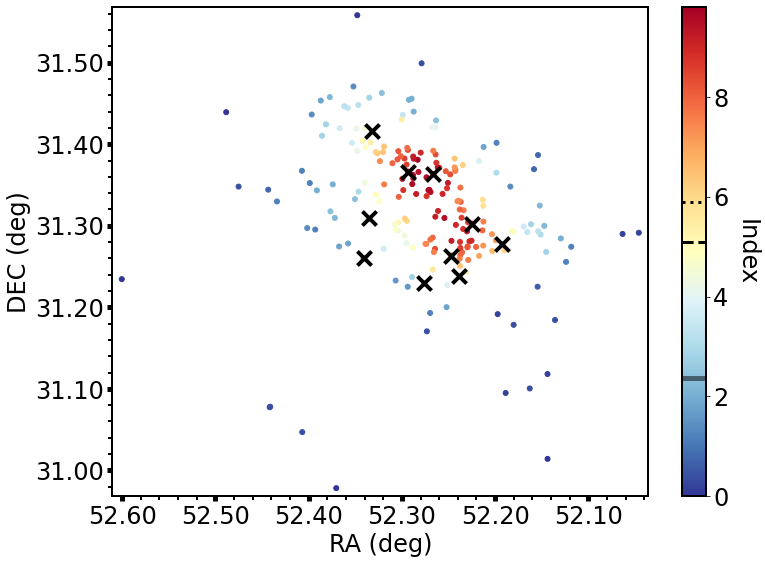}
    \caption{The NGC1333 star-forming region. The ten most massive stars are highlighted with black crosses. The significant index from INDICATE is shown in the colour bar by the solid black line, the median index for all the stars is shown by the dashed black line and the median index for the 10 most massive stars is shown by the dotted black line.}
    \label{fig:fig_NGC1333_indicate}
\end{figure}

The NGC 1333 star-forming region (shown in figure~\ref{fig:fig_NGC1333_indicate}) contains 203 objects, 162 of which have an assigned mass in the dataset used by \citet{parker_dynamical_2017}. The masses of the 10 most massive stars are between $1.1 \, M_\odot$ and $3.3 \, M_\odot$. The distance to the region is 235~pc with an age of around 1~Myr \citep{parker_dynamical_2017,pavlidou_substructure_2021}. In \citet{parker_dynamical_2017} they find a mass segregation ratio of $\Lambda_{\rm{MSR}} = 1.2_{-0.3}^{+0.4}$ implying no mass segregation present when looking at the 10 most massive stars.
INDICATE finds that 74 per cent of stars are spatially clustered above random (the region has a significant index of 2.4). The median index of significantly clustered stars is $7.4_{-2.8}^{+1.2}$. A maximum INDICATE index of 9.8 is found. INDICATE has highlighted an extended central region of relatively high spatial clustering, with the most massive stars spread out around this region. A median index of $5.1_{-2.9}^{+2.9}$ is found for all stars and for the 10 most massive stars a median index of $5.9_{-1.5}^{+2.4}$ is found with a p-value $= 0.57$.  This implies no significant difference in the spatial clustering of the most massive stars compared to all the stars.

\subsection{IC 348}
The data from \citet{parker_dynamical_2017} contains 478 objects for IC 348, 19 of which do not have an assigned mass in the dataset and are ignored when comparing the clustering tendencies of the most massive stars and all stars.

\begin{figure}
	\includegraphics[width=\columnwidth]{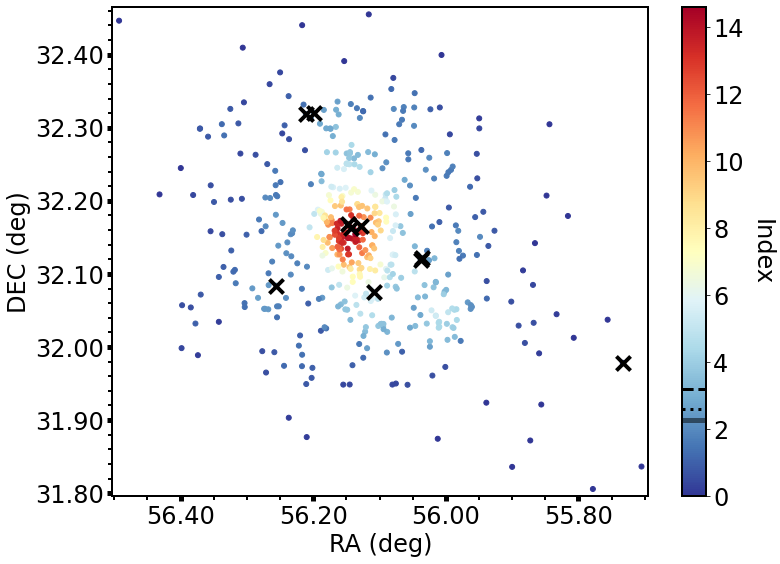}
    \caption{IC348 star-forming region. The ten most massive stars are highlighted with black crosses. The significant index from INDICATE is shown by the solid black line in the colour bar, the median index for all the stars is shown with the dashed black line and the median index for the 10 most massive stars is shown with the dotted black line.}
    \label{fig:fig_ic348_indicate}
\end{figure}

The results of running INDICATE on this region are shown in figure~\ref{fig:fig_ic348_indicate}, clearly showing a central region of relatively higher spatial clustering. The distance to IC 348 is around 300~pc \citep{parker_dynamical_2017} with an age between $2-6$~Myr \citep{cartwright_statistical_2004, bell_pre-main-sequence_2013}. IC 348 has been previously investigated in \citet{parker_dynamical_2017} using the $Q$-parameter to determine overall structure. It was found to have a $Q$-value of 0.85, corresponding to a smooth and centrally concentrated distribution with a radial density exponent of $\alpha = 2.5$. In \citet{parker_dynamical_2017} they find a mass segregation ratio of $\Lambda_{\rm{MSR}} = 1.1_{-0.3}^{+0.2}$ for the 10 most massive stars, meaning no mass segregation is detected.
INDICATE is applied to IC 348, finding 59.2 per cent of stars to be spatially clustered above random and a significant index of 2.3 for the region. The median index for significantly clustered stars is $5.6_{-2.0}^{+4.0}$. A maximum index of 14.6 is found. Median indexes of $3.2_{-1.8}^{+3.5}$ and $2.6_{-0.4}^{+8.3}$ are found for the entire region and the 10 most massive stars, respectively. The masses of the 10 most massive stars are between $2.4 \, M_\odot$ and $4.7 \, M_\odot$. A KS test between the 10 most massive stars and the rest of the stars gives a p-value $= 0.63$, meaning no significant difference in the distribution of the most massive stars and the rest. This is because the most massive stars are spread out over the entire star-forming region, 5 of them are located within the central region, 4 are found between the edge of this region and the outskirts of the region, with one of the most massive stars found right at the edge of the plot.

\subsection{$\rho$ Ophiuchi}
 $\rho$ Ophiuchi is located around 130~pc away with an age of around $0.3 - 2.0$ Myr \citep[][]{parker_search_2012, bontemps_isocam_2001}. We use the dataset from \citet{parker_search_2012} which contains 255 objects. In \citet{parker_search_2012} the mass segregation ratio is found to be $\Lambda_{\rm{MSR}} = 0.89_{-0.13}^{+0.09}$ for the 20 most massive stars, implying no mass segregation is present in the region

\begin{figure}
	\includegraphics[width=\columnwidth]{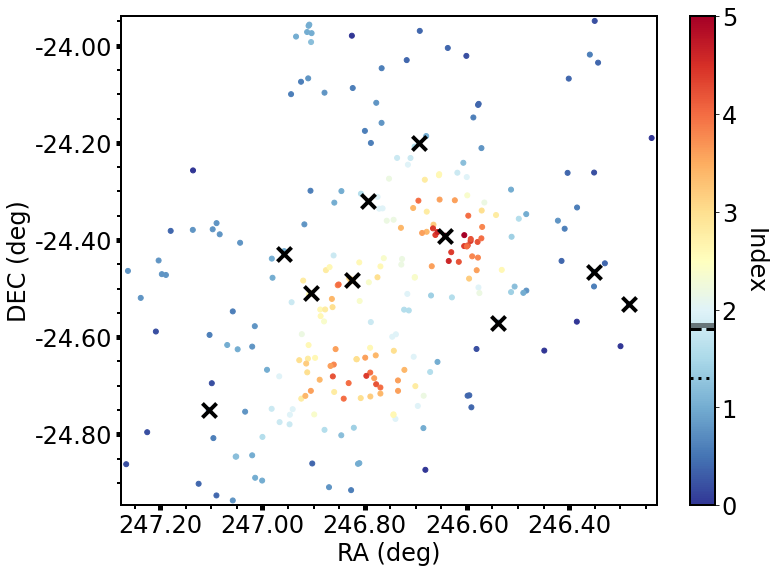}
    \caption{$\rho$~Oph star-forming region. The ten most massive stars are highlighted with black crosses. The significant index from INDICATE is shown by solid black line in the colour bar, the median index for all the stars is shown by the dashed black line and the median index for the 10 most massive stars is shown using the dotted black line.}
    \label{fig:fig_rho_oph_indicate}
\end{figure}
The region is shown in figure~\ref{fig:fig_rho_oph_indicate}.
INDICATE finds that 39.6 per cent of stars are spatially clustered above random, with a significant index of 2.2. The median index for significantly clustered stars is $3.0_{-0.6}^{+0.6}$. The maximum index of the region is 5.0. A median index of $1.8_{-1.0}^{+1.2}$ is found for the entire region and $1.3_{-0.7}^{+1.1}$ is found for the 10 most massive stars, with a p-value $= 0.82$ meaning no significant difference between clustering tendencies of the most massive stars and the rest of the stars. The masses of the 10 most massive stars are between $3.6 \, M_\odot$ and $7.7 \, M_\odot$. These results are similar to IC 348 which also has its most massive stars spread out over the star-forming region. One of the most massive stars is located in an area of high clustering, but the rest have been spread out over relatively lower clustered locations in $\rho$ Ophiuchi.

\section{Conclusions}
\label{sec:conclusions}
We have investigated the performance of the INDICATE method to detect the spatial clustering tendencies in young star-forming regions. We have assessed its ability to quantify mass segregation, and have applied it to pre-main sequence stars in nearby star-forming regions.

We have shown in figure~\ref{fig:structure_indicate} that whilst INDICATE can be used to quantify the clustering tendencies for individual stars in a region it \mbox{cannot} be used to provide any further information on the overall structure of a star-forming region due to the degeneracy of the INDICATE index across different morphologies. 
We confirm that when INDICATE is applied to an entire region it can detect significant differences in the local stellar surface density between the 10 most massive stars and the entire population and will find results that are in agreement with $\Sigma_{\rm{LDR}}$.

When INDICATE is applied to the subset of the 50 most massive stars only it will detect when the 10 most massive stars are more clustered with respect to other massive stars and in most cases will agree with the $\Lambda_{\rm{MSR}}$ method. The two methods are in agreement when applied to regions with \textit{hmhi} and \textit{hmc} mass configurations across 100 realisations of the three different morphologies. The largest discrepancies are found for the substructured realisations and the smooth, centrally concentrated realisations with randomly assigned masses where INDICATE detects 14 and 29 realisations with mass segregation, respectively. When $\Lambda_{\rm{MSR}}$ is applied to the realisations where INDICATE has detected mass segregation it finds no mass segregation in the substructured realisations and 1 in the smooth, centrally concentrated realisations.

We also quantify the clustering tendencies of the most massive stars in these regions compared to all the stars. In the ONC, we find significant differences in the clustering tendencies of the 10 most massive stars when compared to all the stars finding that the 10 most massive stars are in areas of greater local stellar surface density than the average star in the region. 

The other observed regions show no significant differences in the clustering tendencies of the 10 most massive stars as measured using INDICATE compared to all stars in the region. This is due to the 10 most massive stars in these regions being spread out (see figures~\ref{fig:fig_taurus_indicate}, \ref{fig:fig_NGC1333_indicate}, \ref{fig:fig_ic348_indicate} and \ref{fig:fig_rho_oph_indicate}), resulting in a wider range of INDICATE indexes.

In summary, whilst INDICATE can be useful to quantify the degree of affiliation between individual stars and can be used to both detect signals of local stellar surface density and mass segregation depending on whether it is applied to all the stars or just a subset of stars in a region, it does not provide any further information on the type of morphology of a star forming region. 

In a follow-up paper, we will investigate the evolution of $N$-Body simulations with respect to the clustering tendencies of different subsets of stars as measured by INDICATE.
 
\section*{Acknowledgements}
We wish to thank the anonymous referee for their feedback and suggestions which have improved the paper.
RJP acknowledges support from the Royal Society in the form of a Dorothy Hodgkin fellowship.
AB is funded by the European Research Council H2020-EU.1.1 ICYBOB project (Grant No. 818940).

%The Acknowledgements section is not numbered. Here you can thank helpful
%colleagues, acknowledge funding agencies, telescopes and facilities used etc.

%%%%%%%%%%%%%%%%%%%%%%%%%%%%%%%%%%%%%%%%%%%%%%%%%%
\section*{Data Availability}
Data is available on request from the authors.

%The inclusion of a Data Availability Statement is a requirement for articles published in MNRAS. Data Availability Statements provide a standardised format for readers to understand the availability of data underlying the research results described in the article. The statement may refer to original data generated in the course of the study or to third-party data analysed in the article. The statement should describe and provide means of access, where possible, by linking to the data or providing the required accession numbers for the relevant databases or DOIs.



\bibliographystyle{mnras}

\bibliography{indicate_paper_2020.bib} % if your bibtex file is called example.bib

%\begin{thebibliography}{99}
%\end{thebibliography}

%%%%%%%%%%%%%%%%%%%%%%%%%%%%%%%%%%%%%%%%%%%%%%%%%%

%%%%%%%%%%%%%%%%% APPENDICES %%%%%%%%%%%%%%%%%%%%%

\appendix



\section{Poisson Control Field}
\label{ab:poisson_ctrl_field}
\begin{figure*}
\subfigure[INDICATE, fractal]{\includegraphics[width=0.32\linewidth]{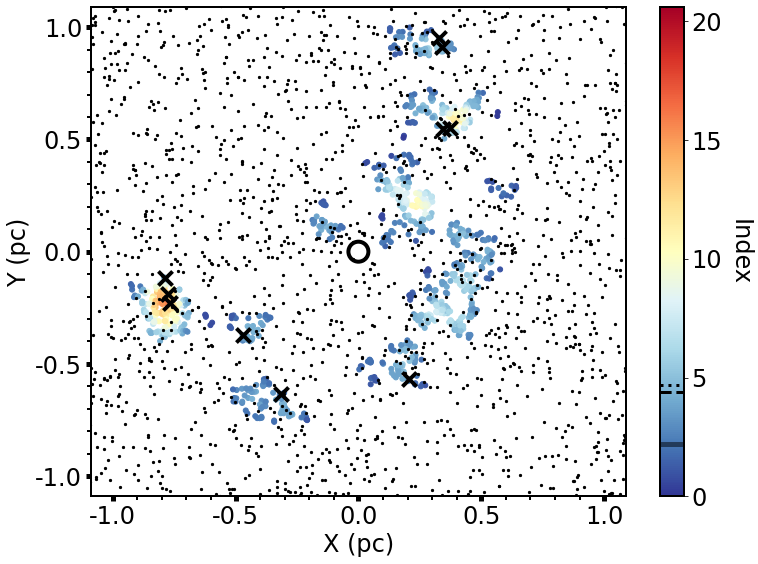}}
    \hspace{0.8pt}
\subfigure[INDICATE, radial]{\includegraphics[width=0.32\linewidth]{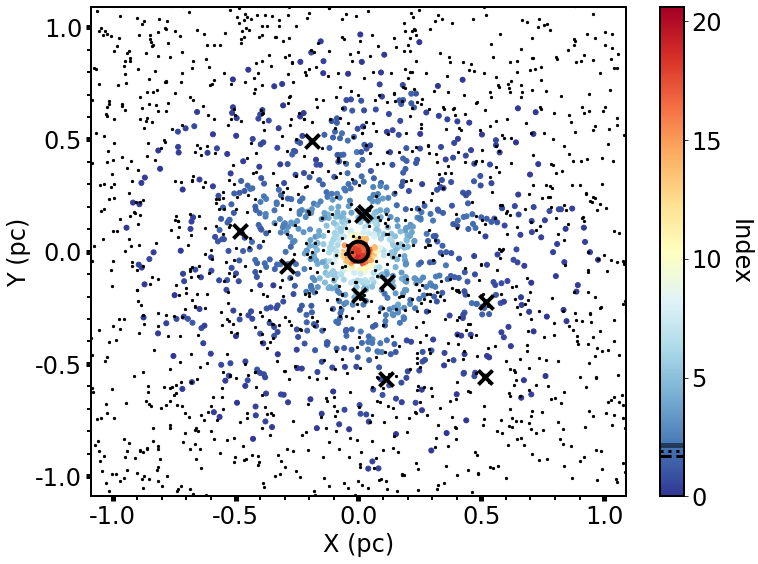}}
    \hspace{0.8pt}
\subfigure[INDICATE, uniform]{\includegraphics[width=0.32\linewidth]{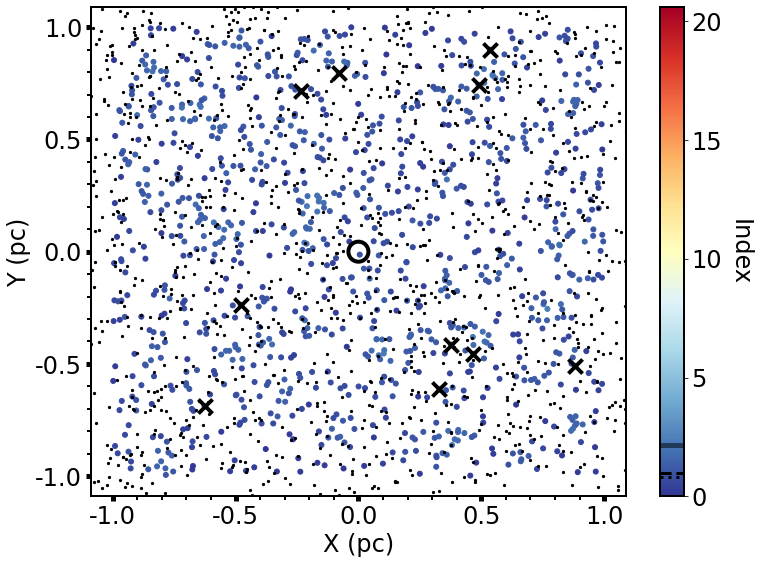}}
\caption{INDICATE using a Poisson distribution as the control field for the synthetic regions used in the main work. From left to right (a) is the substructured region with fractal dimension $D = 1.6$, (b) is the centrally concentrated, smooth distribution with density exponent $\alpha = 2.0$ and (c) shows the uniform distribution. The control field is extended beyond the data to remove edge effects when measuring to the $5^{\rm{th}}$ nearest neighbour. The most massive stars are shown by the black crosses in all the panels and the small black points are the Poisson control field. The colour map has been scaled based on the index results of the smooth, centrally concentrated distribution. The solid black line in the colour bar shows the significant index, the dashed black line is the median INDICATE index for the entire region and the dotted black line is the median INDICATE index for the 10 most massive stars. The centre of each region is located at the middle of the black ring.} 
\label{figapp:poisson_ctrl}
\end{figure*}

The effect of changing the control field is shown in figure~\ref{figapp:poisson_ctrl}. Instead of using an evenly spaced control grid with a uniform field to find the significant index we used a Poisson control field, then found the significant index using a Poisson distribution.

The INDICATE results using a Poisson control field as shown in figure~\ref{figapp:poisson_ctrl} were similar to when a evenly spaced control field is used with a uniform distribution to determine the significant index.
We calculated the significant index for each region using 20 different Poisson distributions of the same number density.

The number of stars clustered above random in the fractal distribution increased to 86.8 per cent from 82.2 per cent.

With the number of stars clustered above random for the radial and uniform distributions now being 46.4 per cent and 0.1 per cent, increasing from 44.4 per cent and 0.0 per cent respectively.

\section{Significant Index Calculations}
\label{ab:sig_repeats}
We calculated each of the synthetic regions significant index using 100 repeats and found that the difference between single run calculations or using repeats is negligible, unless the overall sample size is small such as when we restrict the sample size to 50 to look for classical mass segregation using INDICATE. In these cases we encourage the use of repeats to reduce statistical fluctuations.

For the substructured synthetic region (of fractal dimension $D = 1.6$) with 100 repeats a significant index of 2.3 is found, which is the same as for one run. The percentage of stars clustered above random also stays the same at 82.2 per cent. The median INDICATE index for all the stars in the region is also the same at 4.4, for the median index of the 10 most massive stars it has increased from 4.5 to 4.6 with a p-value $= 0.86$. This is basically the same as for 1 iteration. Swapping the most massive stars with the stars with the highest index we find the same results as for 1 iteration and the same results are also found when the most massive stars are swapped with the most central stars. 

The smooth, centrally concentrated synthetic region (with density exponent $\alpha = 2.0$) has a significant index of 2.3 when calculated using 100 repeats, which is the same as when using one iteration. The percent of stars with indexes above the significant index has also stayed the same at 44.1 per cent, as have the median indexes for the entire region and the 10 most massive stars with respective values of 1.8 and 2.0. The results of the KS test are also the same returning p-value $= 0.55$. We find the same results as a single significance calculation when running repeats after swapping the 10 most massive stars with stars that have the largest INDICATE index and also when we swap the 10 most massive stars with the 10 most central stars.

The uniform synthetic region has a significant index of 2.3 with 100 repeats which is different for the significant index for 1 iteration which was 2.4. The fraction of stars above the significant index has gone up from 0 per cent to 0.1 per cent. The median index for the region is 1.0 and for the 10 most massive stars is 0.8, the same as for 1 iteration. A KS test also returns the same results as for one iteration, i.e. a p-value of $0.68$. Running 100 repeats to calculate the significant index after we have swapped the 10 most massive stars with the stars which have the greatest INDICATE index gives the same result as for one iteration (just with a different significant index of 2.4) and we also find the same results when swapping the 10 most massive stars with the 10 most central stars.

\section{Index Distribution}
\label{appsec:index_distribution_synth_sfr}
\begin{figure*}
\subfigure[Substructured distribution with fractal dimension $D = 1.6$]{\includegraphics[width=0.32\linewidth]{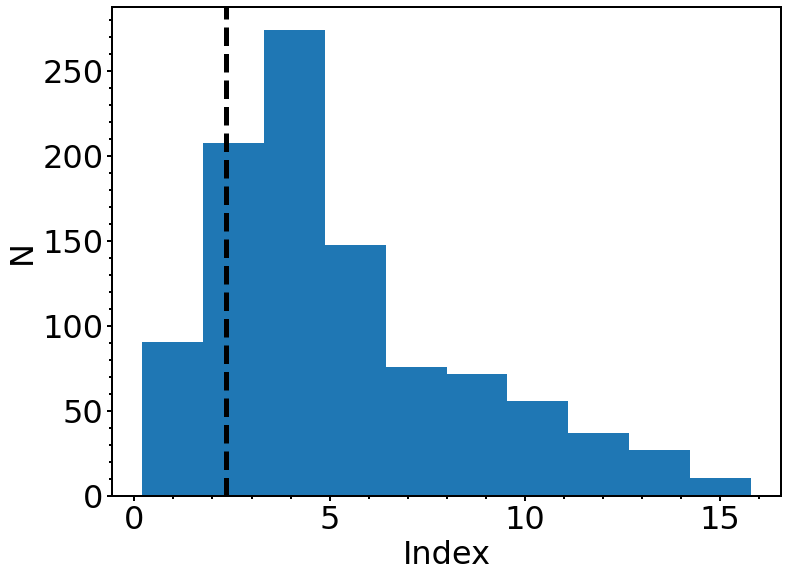}}
    \hspace{0.4pt}
\subfigure[Smooth, centrally concentrated distribution with density exponent $\alpha = 2.0$]{\includegraphics[width=0.32\linewidth]{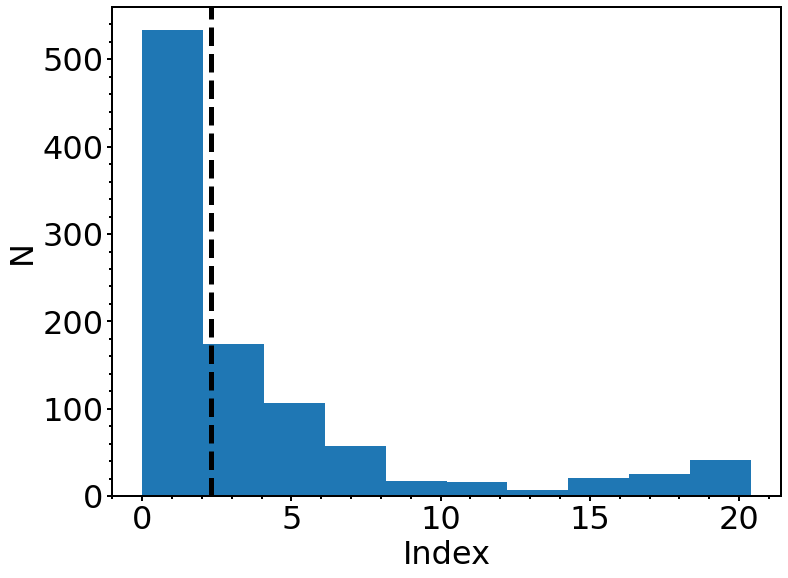}}
    \hspace{0.4pt}
\subfigure[Uniform Distribution]{\includegraphics[width=0.32\linewidth]{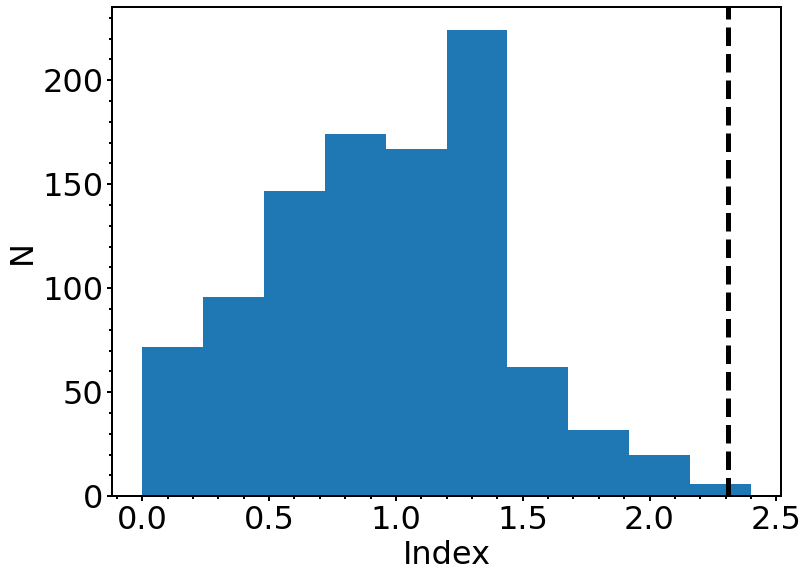}}
\caption{Histograms showing the distribution of the INDICATE indexes for each of the synthetic star-forming regions shown in this work. The vertical dashed line represents the significant index for each region.}
\label{figapp:fractal_index_distributions_single_seeds}
\end{figure*}

%The distribution of the INDICATE index was also investigated. This is shown in figure~\ref{figapp:fractal_index_distributions_3seeds} and shows the index distributions for different fractal dimensions. The more substructured a region the larger their range of index values. 
In figure~\ref{figapp:fractal_index_distributions_single_seeds} we show the distribution of INDICATE indexes for the synthetic star-forming regions presented in $\S$~\ref{sec:indicate fractal results}, $\S$~\ref{sec:indicate radial results} and $\S$~\ref{sec:indicate uniform results} in this paper. 
%The substructured and radial distributions show a wider range of clustering tendencies when compared to the uniform region.

% \begin{figure*}
% \subfigure[$D = 1.6$]{\includegraphics[width=0.49\linewidth]{histograms/frac1d6_3seeds_scaled.png}}
%     \hspace{0.4pt}
% \subfigure[$D = 2.0$]{\includegraphics[width=0.49\linewidth]{histograms/frac2d0_3seeds_scaled.png}}
%     \hspace{0.4pt}
% \subfigure[$D = 2.6$]{\includegraphics[width=0.49\linewidth]{histograms/frac2d6_3seeds_scaled.png}}
%     \hspace{0.4pt}
% \subfigure[$D = 3.0$]{\includegraphics[width=0.49\linewidth]{histograms/frac3d0_3seeds_scaled.png}}
% \caption{Histograms showing the INDICATE index distribution for 3 different realisations of different fractal distributions. (a) shows 3 different $D = 1.6$ distributions, (b) 3 different $D = 2.0$ distributions, (c) 3 different $D = 2.6$ distributions and (d) shows 3 different $D = 3.0$ distributions. The range of index values increases with more substructure as stars are more likely to have a higher INDICATE index in highly substructured regions.}
% \label{figapp:fractal_index_distributions_3seeds}
% \end{figure*}

\section{Testing INDICATE on 100 realisations}
\label{app:testing indicate 100 realisations}
INDICATE was applied to 100 different realisations of the substructured, smooth centrally concentrated and uniform distributions presented in this work. Table~\ref{tab:indicate spreads all stars} shows ranges and interquartile ranges (IQR) of INDICATE indexes of all 1000 stars over 100 different realisations. In table~\ref{tab:indicate spreads all stars} the same results are found for all mass configurations. Table~\ref{tab:indicate spreads 10 most massive stars} shows the INDICATE index ranges for the 10 most massive stars and table~\ref{tab:indicate spreads 10 random stars} shows the INDICATE index ranges for 10 randomly chosen stars.

\begin{table}
    \setlength{\tabcolsep}{3pt}
    \centering
    \caption{INDICATE was applied to all 1000 stars in 100 different realisations of the SFRs presented in this paper. The distribution of INDICATE indexes is summarised here for all stars. From left to right the columns are: the 25${^{\rm{th}}}$ quantile, 75${^{\rm{th}}}$ quantile, the IQR, minimum index, maximum index, the range between the minimum and maximum index and the median significant index found across all realisations.}
    \begin{tabular}{l|ccccccc}
        \hline
        Region & 25${^{\rm th}}$ Quantile & 75${^{\rm th}}$ Quantile & IQR & Min $I$ & Max $I$ & $I$ Range & $\Tilde{I}_{\rm sig}$\\ \hline
       $D=1.6$       & 3.8 & 5.4 & 1.6 & 2.4 & 10.0 & 7.6 & 2.3 \\
       $\alpha=2.0$  & 1.8 & 2.0 & 0.2 & 1.6 & 2.4  & 0.8 & 2.3 \\
       Uniform       & 0.8 & 1.0 & 0.2 & 0.8 & 1.0  & 0.2 & 2.3 \\
       \hline
   \end{tabular}
  
   \label{tab:indicate spreads all stars}
\end{table}

\begin{table}
    \setlength{\tabcolsep}{2pt}
    \centering
    \caption{INDICATE was applied to all 1000 stars in 100 different realisations of the SFRs presented in this paper. The distribution of INDICATE indexes is summarised here for the 10 most massive stars. From left to right the columns are: the 25${^{\rm{th}}}$ quantile, 75${^{\rm{th}}}$ quantile, the IQR, minimum index, maximum index and the range between the minimum and maximum index.}
    \begin{tabular}{l|cccccc}
        \hline
        Region & 25${^{\rm{th}}}$ Quantile & 75${^{\rm{th}}}$ Quantile & IQR & Min $I$ & Max $I$ & $I$ Range \\ \hline
       $D=1.6$, m         & 3.4 &  5.7 & 2.3 & 2.2 &  8.3 &  6.1   \\
       $D=1.6$, hmhi      & 9.8 & 14.5 & 4.8 & 6.0 & 27.0 & 21.0   \\
       $D=1.6$, hmc       & 2.6 &  5.3 & 2.7 & 0.9 & 11.5 & 10.6   \\
       $\alpha=2.0$, m    & 1.4 & 2.6 & 1.2 & 0.7 & 7.7 & 7.0      \\
       $\alpha=2.0$, hmhi & 21.4 & 24.4 & 3.0 & 18.2 & 27.7 & 9.5  \\
       $\alpha=2.0$, hmc  & 20.7 & 23.6 & 2.9 & 17.8 & 27.4 & 9.6  \\
       Uniform, m         & 0.8 & 1.0 & 0.2  & 0.4 & 1.5 & 1.1     \\
       Uniform, hmhi      & 2.1 & 2.4 & 0.3 & 1.8 & 2.8 & 1.0      \\
       Uniform, hmc       & 0.7 & 1.2 & 0.5 & 0.4 & 1.6 & 1.2      \\
       \hline
   \end{tabular}
  
   \label{tab:indicate spreads 10 most massive stars}
\end{table}

\begin{table}
    \setlength{\tabcolsep}{2pt}
    \centering
    \caption{INDICATE was applied to all 1000 stars in 100 different realisations of the SFRs presented in this paper. The distribution of INDICATE indexes is summarised here for 10 randomly chosen stars in each realisation. From left to right the columns are: the 25${^{\rm{th}}}$ quantile, 75${^{\rm{th}}}$ quantile, the IQR, minimum index, maximum index and the range between the minimum and maximum index.}
    \begin{tabular}{l|cccccc}
        \hline
        Region & 25${^{\rm{th}}}$ Quantile & 75${^{\rm{th}}}$ Quantile & IQR & Min $I$ & Max $I$ & $I$ Range \\ \hline
       $D=1.6$, m         & 3.7 & 5.7 & 2.0 & 1.9 & 13.6 & 11.7  \\
       $D=1.6$, hmhi      & 3.6 & 5.6 & 2.0 & 1.9 & 13.6 & 11.7 \\
       $D=1.6$, hmc       & 3.6 & 5.7 & 2.1 & 1.9 & 13.4 & 11.5 \\
       $\alpha=2.0$, m    & 1.5 & 2.5 & 1.0 & 0.4 & 6.6  & 3.2  \\
       $\alpha=2.0$, hmhi & 1.5 & 2.5 & 1.0 & 0.4 & 6.6  & 6.2  \\
       $\alpha=2.0$, hmc  & 1.5 & 2.6 & 1.1 & 0.4 & 8.0  & 7.6  \\
       Uniform, m         & 0.8 & 1.0 & 0.2 & 0.5 & 1.3 & 0.8  \\
       Uniform, hmhi      & 0.8 & 1.0 & 0.2 & 0.5 & 1.4 & 0.9  \\
       Uniform, hmc       & 0.8 & 1.0 & 0.2 & 0.5 & 1.3 & 0.8\\
       \hline
   \end{tabular}
  
   \label{tab:indicate spreads 10 random stars}
\end{table}

\section{Detecting Classical Mass Segregation in Synthetic Data}
\label{appsec:classical_massseg_synth_data}
\begin{figure*}
  \subfigure[$D = 1.6$, m]{\includegraphics[width=0.32\linewidth]{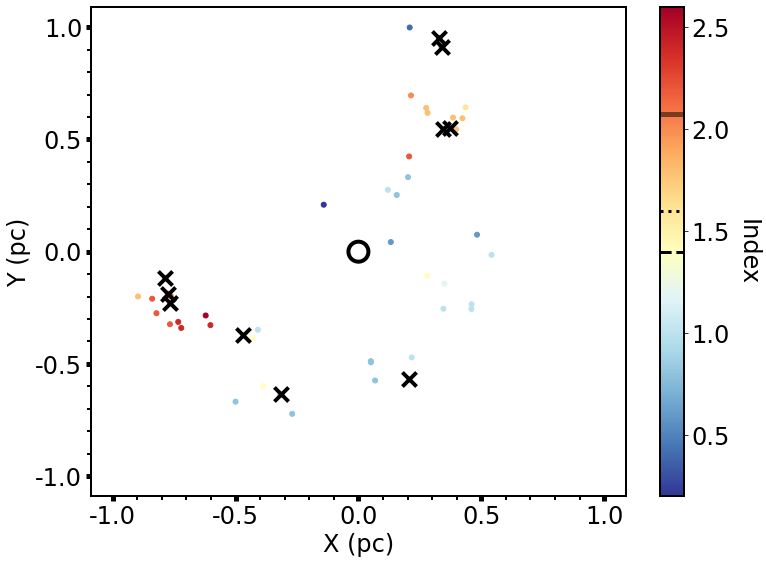}}
   \hspace{0.8pt}
  \subfigure[$D = 1.6$, hmhi]{\includegraphics[width=0.32\linewidth]{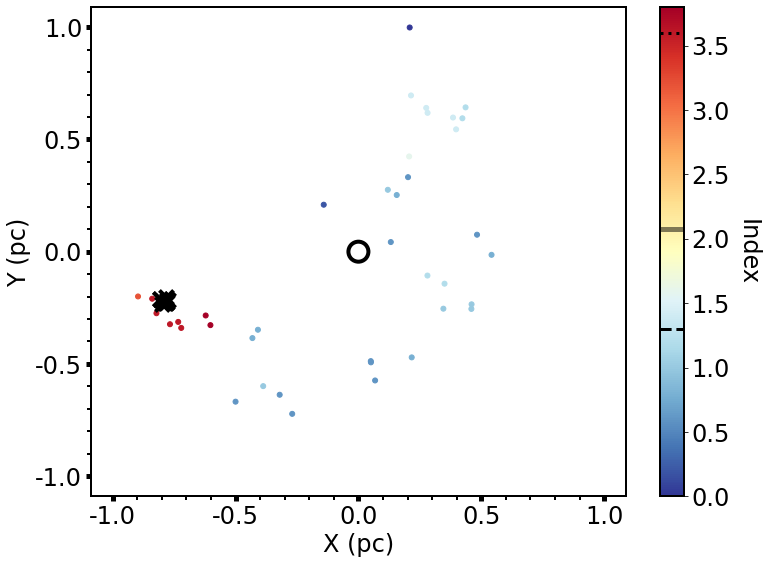}}
   \hspace{0.8pt}
  \subfigure[$D = 1.6$, hmc]{\includegraphics[width=0.32\linewidth]{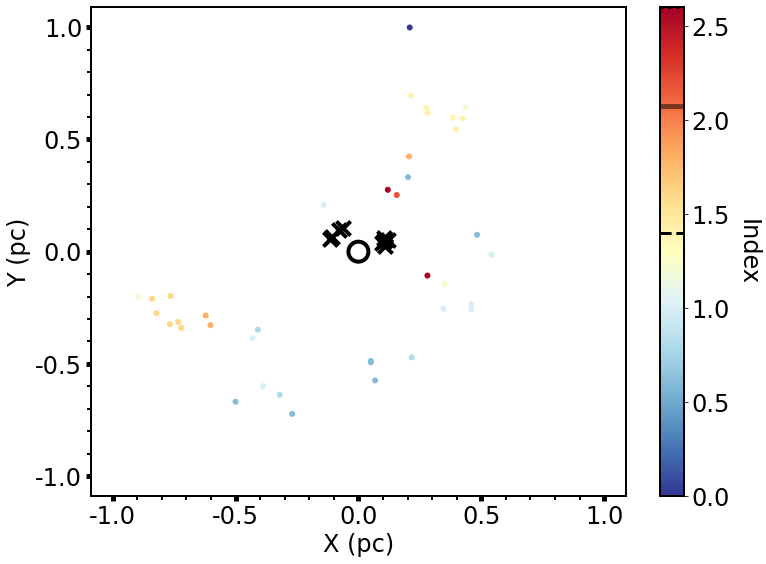}}
   \hspace{0.8pt}
  
  \subfigure[$\alpha = 2.0$, m]{\includegraphics[width=0.32\linewidth]{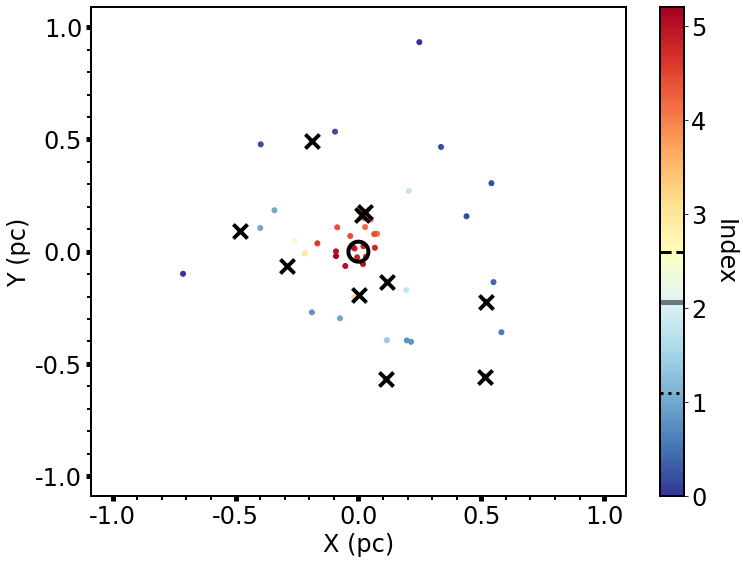}}
   \hspace{0.8pt}
  \subfigure[$\alpha = 2.0$, hmhi]{\includegraphics[width=0.32\linewidth]{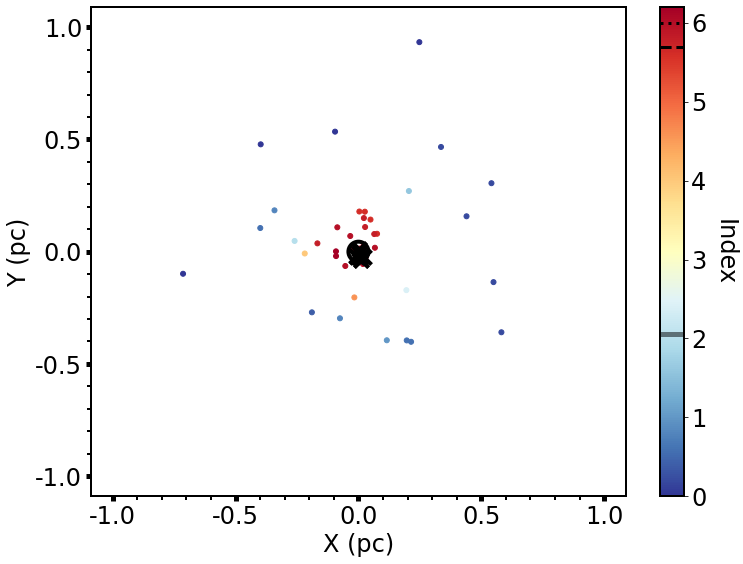}}
   \hspace{0.8pt}
  \subfigure[$\alpha = 2.0$, hmc]{\includegraphics[width=0.32\linewidth]{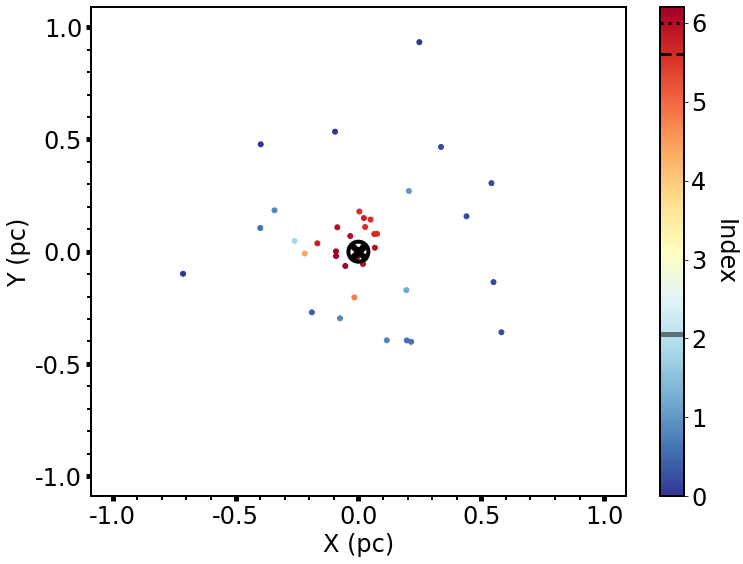}}
   \hspace{0.8pt}
  
  \subfigure[Uniform, m]{\includegraphics[width=0.32\linewidth]{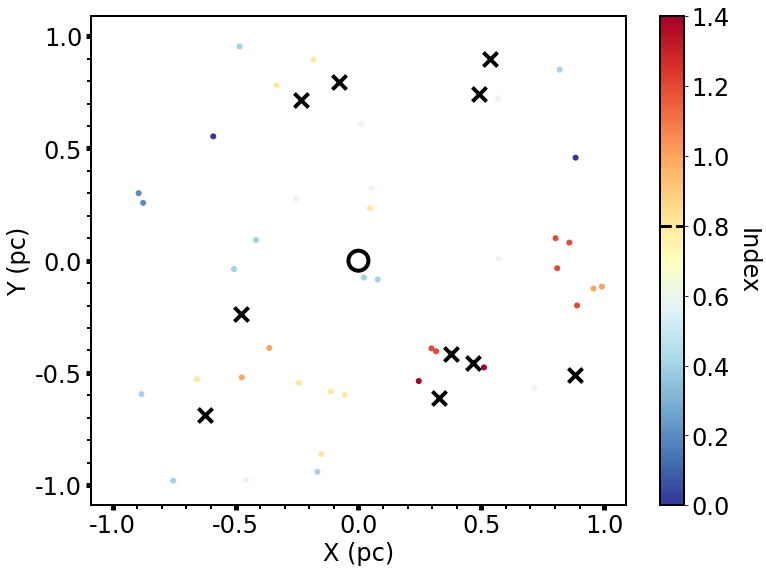}}
   \hspace{0.8pt}
  \subfigure[Uniform, hmhi]{\includegraphics[width=0.32\linewidth]{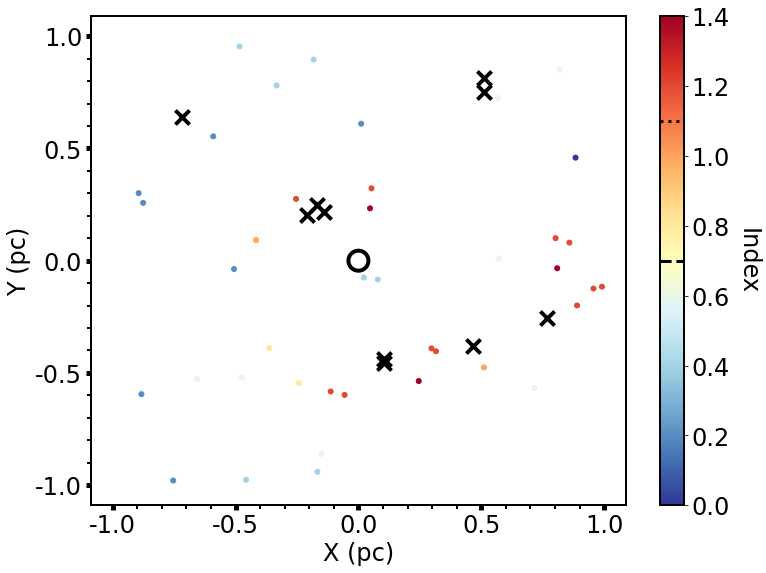}}
   \hspace{0.8pt}
  \subfigure[Uniform, hmc]{\includegraphics[width=0.32\linewidth]{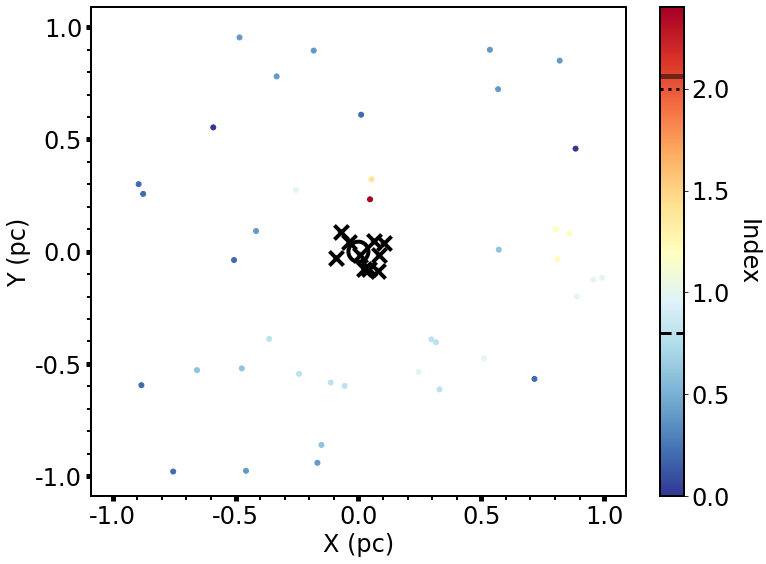}}
   \hspace{0.8pt} 
  
  \caption{Results of applying INDICATE to just the 50 most massive stars in each of the synthetic region. From left to right is the random mass configuration, the \textit{hmhi} and \textit{hmc} configurations. From top to bottom we have the 50 most massive stars from the substructured SFR with fractal dimension of $D = 1.6$, smooth centrally concentrated SFR with density exponent $\alpha = 2.0$ and the uniformly distributed SFR. The solid black line in the colour bar shows the significant index, the dashed black line is the median INDICATE index for the entire subset and the dotted black line is the median INDICATE index for the 10 most massive stars. The centre of each region is located at the middle of the black ring.}
  \label{appfig:50_most_massive}
\end{figure*}

To test if INDICATE can detect classical mass segregation we apply INDICATE to only the 50 most massive stars in each synthetic region and for each regions different mass configurations. Significant index calculations are done 100 times. Figure~\ref{appfig:50_most_massive} shows the INDICATE results for 50 most massive stars in each of the synthetic regions.

\section{Detecting Classical Mass Segregation In Observational Data}
\label{sec:classical_massseg_observational_data}
We run INDICATE over the 50 most massive stars in the observational data, repeating the significance calculation 100 times for each region. We use the same criteria to determine if INDICATE has detected mass segregation as for the synthetic star-forming regions in $\S$~\ref{sec:can_indicate_detect_mass_segregation}. A summary of these results is shown in table~\ref{tab:observational_indicate_results_50_most_massive}.
{\renewcommand{\arraystretch}{1.5}
\begin{table}
	\centering
	\caption{Results of applying INDICATE to only the 50 most massive stars in the observed star-forming regions. From left to right the columns are: the median index for the entire subset of the 50 most massive stars, the median index for the 10 most massive stars in the subset, the significant index, the percentage of stars with indexes above the significant index and the p-value from a KS test between all 50 stars and the 10 most massive stars INDICATE indexes.}
	\begin{tabular}{l|ccccl} % four columns, alignment for each
		\hline
		Name & $\Tilde{I}(50)$ & $\Tilde{I}(10)$ & $I_{\rm{sig}}$ & $\% > I_{\rm{sig}}$ & p \\
		\hline
		Taurus          & $1.3_{-0.5}^{+0.7}$ & $1.2_{-0.2}^{+0.2}$ & 2.5 & 12 & 0.86 \\
		ONC             & $1.4_{-1.0}^{+2.8}$ & $4.3_{-0.8}^{+0.1}$ & 2.1 & 46 & 0.15 \\
		NGC1333         & $2.9_{-1.1}^{+0.9}$ & $2.5_{-0.6}^{+0.9}$ & 2.1 & 66 & 1.00 \\
		IC348           & $4.2_{-3.4}^{+1.0}$ & $2.0_{-1.6}^{+2.6}$ & 2.0 & 62 & 0.67 \\
		$\rho$ Ophiuchi & $1.0_{-0.4}^{+0.6}$ & $0.8_{-0.2}^{+0.3}$ & 2.1 & 0  & 0.39 \\
		\hline
		\label{tab:observational_indicate_results_50_most_massive}
	\end{tabular}
\end{table}}

\subsection{Taurus}
The 50 most massive stars in Taurus have a median index of 1.3 and the 10 most massive a median index of 1.2.

INDICATE finds that 12 per cent of stars have a index above the significant index of 2.5. The median index for significantly clustered stars is $2.8_{-0.5}^{+0.0}$. The maximum index found was 3.0. The median index for the 10 most massive stars is $1.2_{-0.2}^{+0.2}$ which is below the significant index of $2.5$, therefore no mass segregation is detected. This is in agreement with the mass segregation ratio result for Taurus in \citet{parker_mass_2011} for the 20 most massive stars.

A KS test returns a p-value $= 0.86$ meaning no significant difference in the clustering tendencies of the 10 most massive and the entire subset of the 50 most massive stars.

\subsection{ONC}
The median index for just the 50 most massive stars is 1.4 and the median index for the 10 most massive stars is 4.3.

INDICATE finds that 46 per cent of stars have a index above the significant index of 2.1. The median index for significantly clustered stars is $4.2_{-0.2}^{+0.2}$. A maximum index of 4.4 is found.

The median index for the 10 most massive stars is $4.3_{-0.8}^{+0.1}$ which is above the significant index of $2.1$ meaning mass segregation has been detected. This means that the 10 most massive stars are more affiliated with other high mass stars than the typical high mass star. This is in agreement with the mass segregation ratio result found in \citet{allison_using_2009} for the 4 most massive stars.

A KS test returns a p-value $= 0.15$ implying no significant difference between the 10 most massive stars and the 50 most massive stars' clustering tendencies. The only criteria for mass segregation is that $\Tilde{I}_{\rm{10}} > I_{\rm{sig}}$. This results shows that the 10 most massive star are affiliated with other massive stars above random expectation. The KS test may reveal that the way the 10 most massive stars are distributed may have a correlation with their mass.

\subsection{NGC 1333}
A median index of 2.9 is found for all the stars in the subset and a median of 2.5 is found for the 10 most massive stars. 

INDICATE finds that 66 per cent of stars have indexes above the significant index of 2.1. The median index of significantly clustered stars is $3.4_{-0.4}^{+0.4}$. A maximum index of 4.4 is found. The median index for the 10 most massive stars is $2.5_{-0.6}^{+0.9}$ which is above the significant index of $2.1$ meaning INDICATE has detected mass segregation in NGC 1333. This is in disagreement with the mass segregation ratio results found in \citet{parker_dynamical_2017}, where they find no signals of mass segregation for the 10 most massive stars.

A KS test returns a p-value $= 1.00$ implying no significant difference in the clustering tendencies of the 10 most massive stars and the 50 most massive stars.

\subsection{IC 348}
A median index of 4.2 is found for all stars in the subset and a median index of 2.0 is found for the 10 most massive stars.

INDICATE finds that 62 per cent of stars have an index above the significant index of 2.0. The median index of significantly clustered stars is $5.0_{-0.6}^{+0.4}$. The maximum index is 6.0. The median index for the 10 most massive stars is $2.0_{-1.6}^{+2.6}$. The significant index is $2.0$. As the median index for the 10 most massive stars is not well above the significant index INDICATE detects no mass segregation. This results is in agreement with the mass segregation ratio result found for the 10 most massive stars in \citet{parker_no_2017}.

A KS test returns a p-value $= 0.67$ implying no significant difference in the clustering tendencies of the 10 most massive stars and the entire subset.

\subsection{$\rho$ Ophiuchi}
A median index of $1.0_{-0.4}^{+0.6}$ is found for all the stars in the subset and a median index of $0.8_{-0.2}^{+0.3}$ is found for the 10 most massive stars. 

INDICATE finds that no stars have an index above the significant index of 2.1. A maximum index of 2.0 is found. As the median index of the 10 most massive stars is below the significant index no mass segregation is detected. In agreement with the mass segregation result of \citet{parker_characterizing_2012} for the 20 most massive stars.

A KS test returns a p-value $= 0.39$ implying no significant difference in the clustering tendencies of the 10 most massive stars compared to the overall region.

\bsp	% typesetting comment
\label{lastpage}
\end{document}